\documentclass[12pt]{aastex}
\usepackage{graphicx}
\usepackage{natbib}

\begin{document}

\title{A Sample of Intermediate-Mass Star-Forming Regions: Making Stars
at Mass Column Densities $<$1~g~cm$^{-2}$}

\author{K. Arvidsson \and C. R. Kerton}
\affil{Department of Physics \& Astronomy, Iowa State University, Ames, IA 50011; kima@iastate.edu, kerton@iastate.edu}

\author{M. J. Alexander \and H. A. Kobulnicky}
\affil{Department of Physics \& Astronomy, University of Wyoming, 1000 E. University, Laramie, WY 82071; malexan9@uwyo.edu, chipk@uwyo.edu}

\and

\author{B. Uzpen\altaffilmark{1}}
\affil{ITT Technical Institute, 500 E. 84th Avenue, Thornton, CO 80229; buzpen@itt-tech.edu}

\altaffiltext{1}{Department of Physics \& Astronomy, University of Wyoming, 1000 E. University, Laramie, WY 82071; uzpen@uwyo.edu}

\maketitle

\begin{abstract}
In an effort to understand the factors that govern the transition from low- to high-mass star formation, we identify for the first time a sample of intermediate-mass star-forming regions (IM SFRs) where stars up to---but not exceeding---$\sim 8$ M$_{\sun}$ are being produced. We use \emph{IRAS} colors and \emph{Spitzer Space Telescope} mid-IR images, in conjunction with millimeter continuum and $^{13}$CO maps, to compile a sample of 50 IM SFRs in the inner Galaxy. These are likely to be precursors to Herbig AeBe stars and their associated clusters of low-mass stars. IM SFRs constitute embedded clusters at an early evolutionary stage akin to compact \ion{H}{2} regions, but they lack the massive ionizing central star(s). The photodissociation regions that demarcate IM SFRs have typical diameters of $\sim1$ pc and luminosities of $\sim 10^4$ L$_{\sun}$, making them an order of magnitude less luminous than (ultra)compact \ion{H}{2} regions. IM SFRs coincide with molecular clumps of mass $\sim 10^3$ M$_{\sun}$ which, in turn, lie within larger molecular clouds spanning the lower end of the giant molecular cloud mass range, $10^4-10^5$ M$_{\sun}$. The IR luminosity and associated molecular mass of IM SFRs are correlated, consistent with the known luminosity--mass relationship of compact \ion{H}{2} regions. Peak mass column densities within IM SFRs are $\sim$ 0.1--0.5 g~cm$^{-2}$, a factor of several lower than ultra-compact \ion{H}{2} regions, supporting the proposition that there is a threshold for massive star formation at $\sim$ 1 g~cm$^{-2}$.
\end{abstract}


\section{Introduction}
It has become clear that high-mass star formation cannot be viewed simply as a scaled-up version of the low-mass star formation paradigm \citep{zinn07}. High-mass ($M > 8$ M$_{\sun}$) and low-mass ($M <2$ M$_{\sun}$) star formation seem to require some fundamentally different conditions and physical processes that probably involve differences in turbulence and magnetic fields within their formative clouds as much as differences in cloud mass and density \citep{krumkee05}. Stars forming in the 2 to 8 solar mass range (intermediate-mass; IM) are especially interesting as they straddle the boundary between the low- and high-mass paradigms and can provide a window into the processes that govern the transition from low-mass to high-mass star formation. Extensive studies of these IM star-forming regions (IM SFRs), i.e., SFRs where the most massive star is IM, have not been possible in the past due to difficulties in clearly identifying the regions and the lack of angular resolution to investigate them in detail.

Our goals in this investigation of intermediate-mass star-forming regions are to understand what controls or enables the transition from low- to high-mass star formation. Is it simply a question of an enhanced reservoir of material that promotes the formation of a more massive star, or is star formation shut off in some cases because of disruption of the parent molecular cloud? Is there evidence for sequential or triggered star formation in the interstellar medium (ISM) surrounding IM SFRs as around massive SFRs \citep{2005A&A...433..565D}? How common are bipolar outflows? What is the ISM environment of these regions? Are IM SFRs found in locations where high-mass star formation is also occurring or in relative isolation? What are the typical masses and velocity dispersions of IM star-forming clouds? What are the typical sizes and luminosities of IM SFRs?

The aim of the present paper, the first high-resolution study of IM SFRs, is to determine some basic properties of a carefully selected sample of IM SFRs. We present our candidate selection criteria and classification in Section~\ref{section2}. We then, in Section~\ref{section3}, identify CO features corresponding to the candidates and discuss a particularly interesting sample of the candidates. The photometry and spectral energy distribution is examined in Section~\ref{sed}, and the molecular material is discussed in Section~\ref{masssection}. We compare our sample to Ultra-Compact \ion{H}{2} (UC\ion{H}{2}) regions in Section~\ref{uchiisection}. Finally, we present a discussion in Section~\ref{discussion} and conclusions in Section~\ref{conclusions}.

\section{Infrared Analysis}{\label{section2}}
\subsection{Defining a Sample of Candidate IM SFRs}{\label{defininginitial}}
\citet{kerton02}, using 21 cm \ion{H}{1} emission, identified a sample of IM SFRs and determined that they occupy a distinct region of the \emph{IRAS} color-color diagram, see Figure~\ref{cc-plots}, left panel and Figure~12 in \citet{kerton02}. They are separate from \ion{H}{2} regions, and quite distinct from T Tauri and Herbig AeBe stars \citep{barnes1993ASPC...35..102B,weintraub1990ApJS...74..575W}: $-0.18 < \log(F_{\nu}(25)/F_{\nu}(12)) < 0.40$, $0.96 < \log(F_{\nu}(60)/F_{\nu}(25)) < 1.56$ and $0.29 < \log(F_{\nu}(100)/F_{\nu}(60)) < 0.69$. These distinct colors arise from the combined effect of lowered dust temperatures surrounding IM stars and increased survivability of 12 $\micron$-emitting grains in the less harsh radiation environment \citep{giard1994A&A...291..239G}. The color criteria from \citet{kerton02} are probably biased toward the higher end of the intermediate mass range because the color criteria were based on objects that could produce a significant amount of \ion{H}{1} in a photodissociation region.

We queried the \emph{IRAS} Point Source Catalog v2.1 (PSC) with the above color criteria. In all cases, we required the \emph{IRAS} flux density quality to be marked as either high ($F_{\mathrm{qual}} = 3$) or moderate ($F_{\mathrm{qual}} = 2$). Table~\ref{numbers} lists the results of this query over several regions of interest. Over the entire sky there are 984 sources meeting these criteria. In the region within 1 degree ($|b| < 1 \degr$) of the Galactic Plane where most star formation takes place, there are 360 sources. In this paper, we focused on the 70 sources at Galactic longitudes between $14 \degr < \ell <55 \fdg 7$ where complementary data from several multiwavelength Galactic surveys are available, namely the The Boston University Five
College Radio Astronomy Observatory (BU-FCRAO) Galactic
Ring Survey \citep[GRS;][]{grs} in $^{13}$CO, the mid-IR Galactic Legacy Infrared Mid-Plane Survey Extraordinaire I \citep[GLIMPSE;][]{glimpse}, the Multiband Infrared Photometer for \emph{Spitzer} Galactic survey I \citep[MIPSGAL;][]{mipsgal} and the Bolocam Galactic Plane Survey \citep[BGPS;][]{bgps}. Out of the 70 \emph{IRAS} sources, 66 fall within the fully covered GRS field ($18 \degr < \ell <55 \fdg 7$), and 4 objects fall within the partially covered part of the GRS ($14 \degr < \ell < 18 \degr$).

Use of the \emph{IRAS} PSC to define our baseline sample of IM SFRs likely results in a small bias against very nearby sources which would have angular sizes larger than the \emph{IRAS} beam. The \emph{IRAS} PSC contains objects with angular sizes of less than $0.5 \arcmin$, $0.5 \arcmin$, $1 \arcmin$, and $2 \arcmin$ in the scan direction for 12, 25, 60 and 100 $\micron$ respectively. A 1 pc diameter source is larger than $2 \arcmin$ in angular size for a distance $< 1720$ pc. There is an \emph{IRAS} Small Scale Structure Catalog (SSSC) which contains information on structures up to $8 \arcmin$ in angular extent. A search of this catalog over the whole
sky for the same color criteria (but without the flux density quality criteria) returned 25 sources. Most of them are outside of the GLIMPSE/GRS
region and all but one have an associated \emph{IRAS} point source. Thus, we could miss close IM SFRs by using the \emph{IRAS} PSC, but we are likely not missing a large population since that volume is small compared to the covered volume.

\subsection{Infrared Morphological Classification with \emph{Spitzer}}{\label{classification}}
We investigated the 70 selected \emph{IRAS} sources using images from the \emph{Spitzer} GLIMPSE I and MIPSGAL I legacy surveys. The GLIMPSE I images have $\sim 2 \arcsec$ resolution and $5 \sigma$ sensitivities of 0.2, 0.2, 0.4, and 0.4 mJy in the Infrared Array Camera \citep[IRAC;][]{irac} wavebands centered near 3.6, 4.5, 5.8, and 8.0 $\micron$. For this study we used the available $6 \arcsec$ resolution ($5 \sigma$
sensitivity of 1.7 mJy) Multiband Infrared Photometer for
\emph{Spitzer} \citep[MIPS;][]{rieke04} 24 $\micron$ images from the MIPSGAL I
survey. These images provide a vast improvement in angular resolution over the arcminute-scale \emph{IRAS} images previously available.

Based on the GLIMPSE and MIPSGAL images, we classified by eye the 70 sources into three broad classes: {}``Blobs/Shells'', {}``Filamentary'' and {}``Star-like''. Blobs/Shells are characterized by apparent circular or elliptical morphology, predominantly in the 8.0 and 24 $\micron$ bands. Blobs are typically circular (or elliptical) regions of extended emission having similar morphologies at 8.0 and 24 $\micron$ bands, while shells are usually ring-shaped 8.0 $\micron$ features that spatially coincide with 24 $\micron$ extended emission. Filamentary objects usually are emitting regions that occupy a small part of a larger filamentary structure, like a bead on a string. The structure of the region itself is usually dominated by the larger filament it sits on, and can even be irregular. This class also includes emitting regions located where filaments intersect. Finally, objects were classified as star-like where the emission appear to come from one or more point sources in all bands. No extended emission is evident for these objects. These star-like objects are not likely IM SFRs, since a point source diameter of $2 \arcsec$ or less corresponds to a physical size of $< 0.02$ pc at a distance of 2 kpc.

\begin{figure}
\plottwo{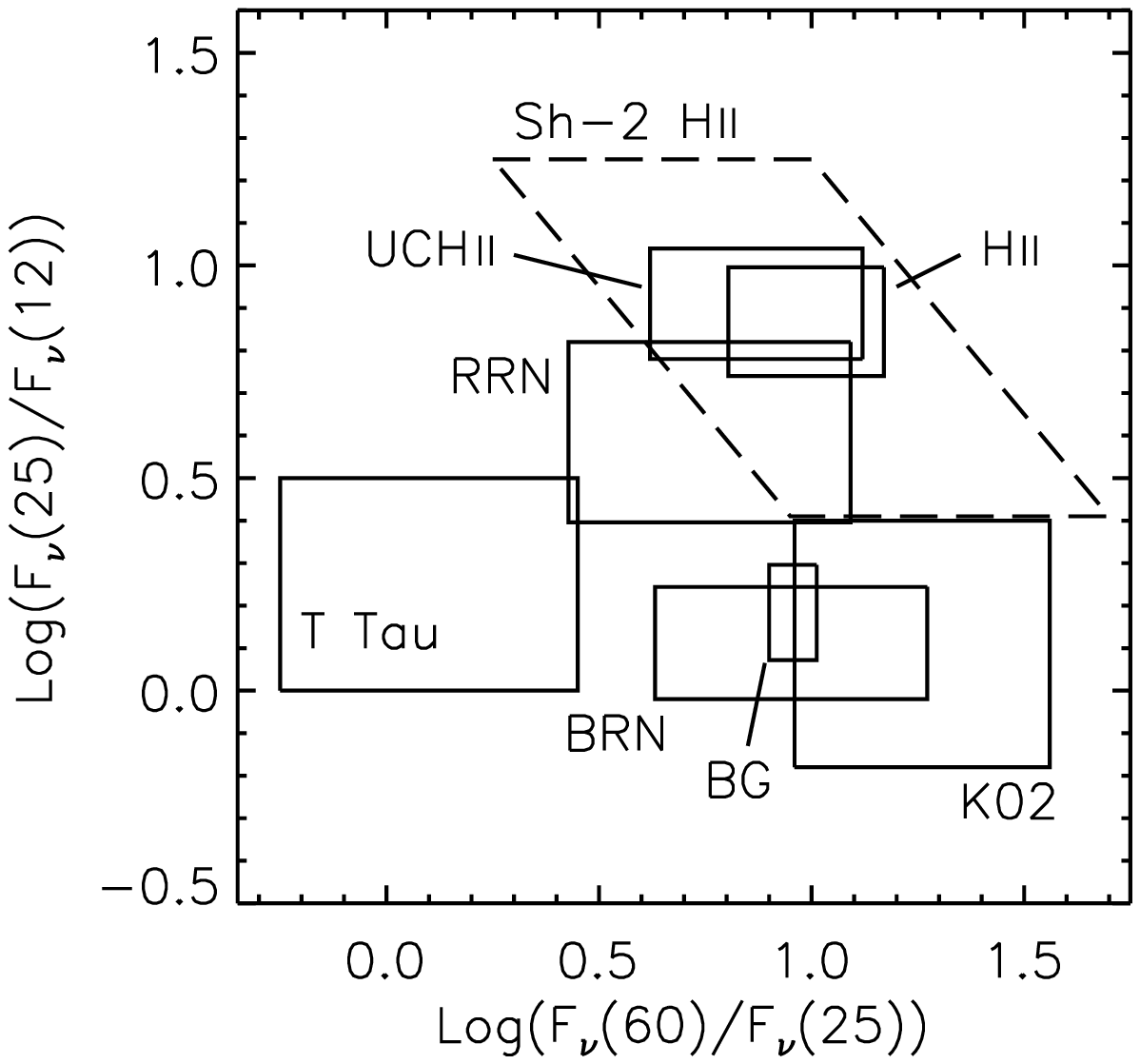}{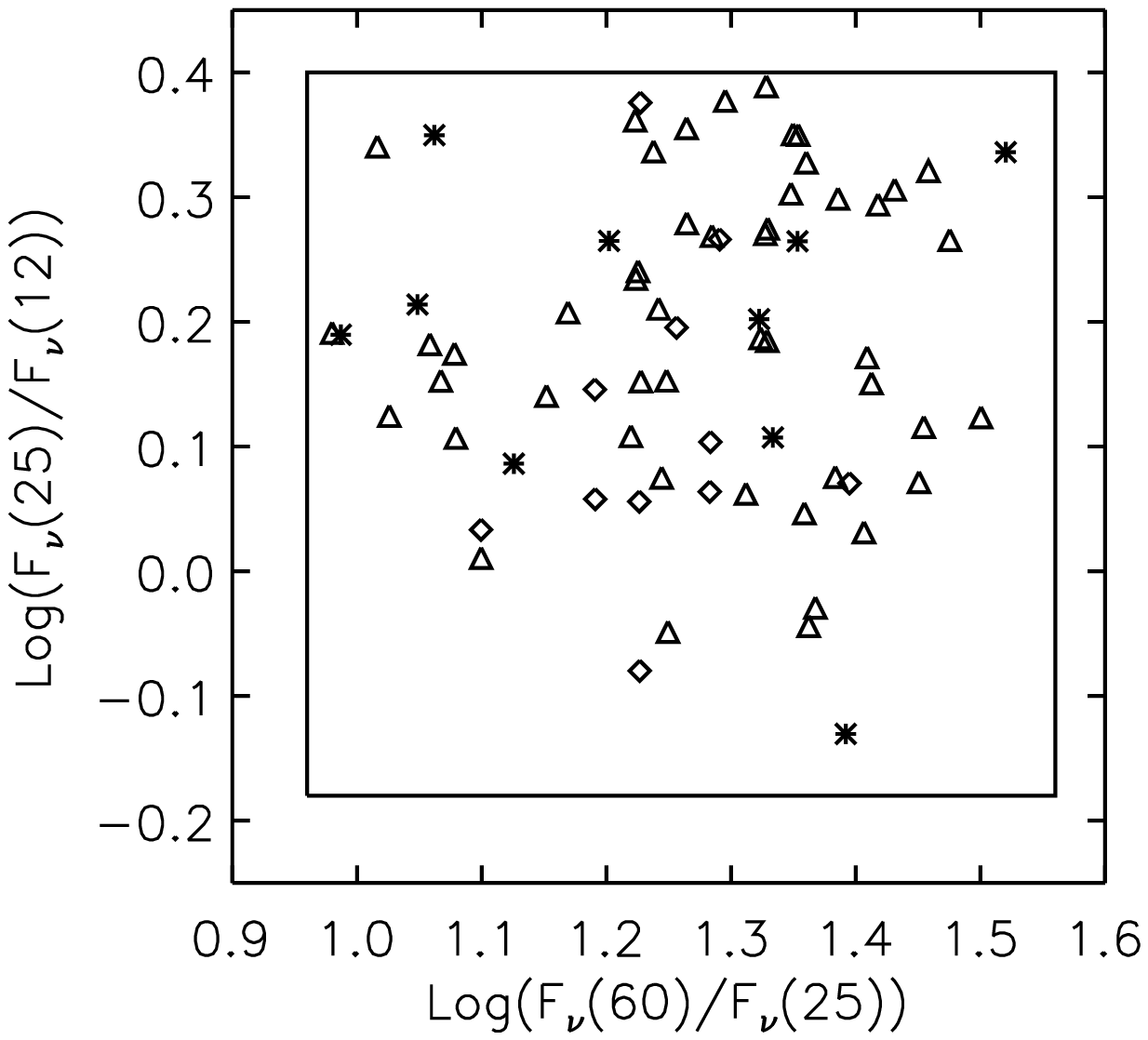}
\caption{(Left) \emph{IRAS} color-color diagram, indicating the extent of various astronomical objects (adapted from Figure 13 in \citet{kerton02}). The box marked K02 corresponds to the color constraints in Section~\ref{defininginitial}. The other boxes show: (Sh--2 \ion{H}{2}) a majority of the \ion{H}{2} regions compiled by \citet{hughesmacleod1989AJ.....97..786H}; (UC\ion{H}{2}) ultra-compact \ion{H}{2} regions from \citet{colors1989ApJ...340..265W}; (\ion{H}{2}) \ion{H}{2} regions; (RRN) red reflection nebulae;
(BRN) blue reflection nebulae; (BG) blue galaxies; (T Tau) T Tauri stars \citep[all from][]{walker1989AJ.....98.2163W}. (Right) Enlarged portion of the \emph{IRAS} color-color diagram. The drawn box corresponds to K02 in the left panel, with the \emph{IRAS} colors of the 70 objects plotted. Triangles correspond to {}``Blobs/Shells'', diamonds to {}``Filamentary'' and stars to {}``Star-like'' objects (Section~\ref{classification}). The three classes of objects show no grouping or preferential location in the \emph{IRAS} color-color diagram.\label{cc-plots}}
\end{figure}

Out of the 70 surveyed \emph{IRAS} objects, 49 ($70 \%$) were classified as blobs/shells, 11 ($16 \%$) as filamentary and 10 ($14 \%$) as star-like. The right panel of Figure~\ref{cc-plots} shows their distribution in \emph{IRAS} color-color space. The three groups show no grouping or preferential location in Figure~\ref{cc-plots}. The \emph{IRAS} names and positions for the filamentary and star-like objects are listed in Table~\ref{rejects}. The 49 blobs/shells were investigated in more detail in the rest of the paper, as their morphology is most consistent with the expected morphology of an IM SFR. Table~\ref{blobsshells} lists the \emph{IRAS} names, positions ($\ell$ and $b$), angular diameters ($\theta$) in arcminutes, velocity associations and widths ($V_{\mathrm{LSR}}$ and $ \Delta V$, discussed in Section~\ref{section3}), calculated near and far distances ($d_{\mathrm{near}}$ and $d_{\mathrm{far}}$), adopted distances, flags noting what distance was adopted, and the calculated diameter in pc of the blobs and shells. Table~\ref{blobsshells} contains 50 entries because one of them, \emph{IRAS} 19012+0505, has been divided into two sources (North and South), that have slightly different velocity associations.

\section{Molecular Counterparts and Distances}{\label{section3}}
\subsection{CO Associations}{\label{co}}
We investigated the GRS CO spectral data cubes visually toward the regions of interest, the 49 objects classified as {}``Blobs/Shells''. This $^{13}$CO molecular line survey has a spectral resolution of $0.2$ km~s$^{-1}$, an angular resolution of $46 \arcsec$, and a sensitivity of $< 0.4$ K. We searched for molecular features similar in morphology to the mid-IR blobs and shells. We found that the majority, 41 objects, have unambiguous CO associations, i.e., either a single CO spectral peak matches the candidate IM SFR, or if multiple peaks appear along the line of sight, the CO morphology for a particular peak can be matched to the mid-IR feature. The width of the morphologically matched feature in each GRS CO spectrum defines a velocity range that we used to construct the zeroth-moment map from the GRS spectral cubes. As a check of the morphological association between CO and the candidate IM SFR, the candidate was required to be spatially associated with a local maximum in the GRS zeroth-moment map. Of the remaining eight objects, seven have several possible associations in the data cube, resulting in an ambiguous radial velocity. These are labeled with the flag {}``A'' in Table~\ref{blobsshells}. The single remaining candidate, \emph{IRAS} 19056+0947, was found to have no CO association at all within a radius of $0 \fdg 3$. This object is labeled with the flag {}``No CO'' and could be a reflection nebula, a known contaminant (see the left panel of Figure~\ref{cc-plots}).

\begin{figure}
\resizebox{6.in}{!}{\includegraphics{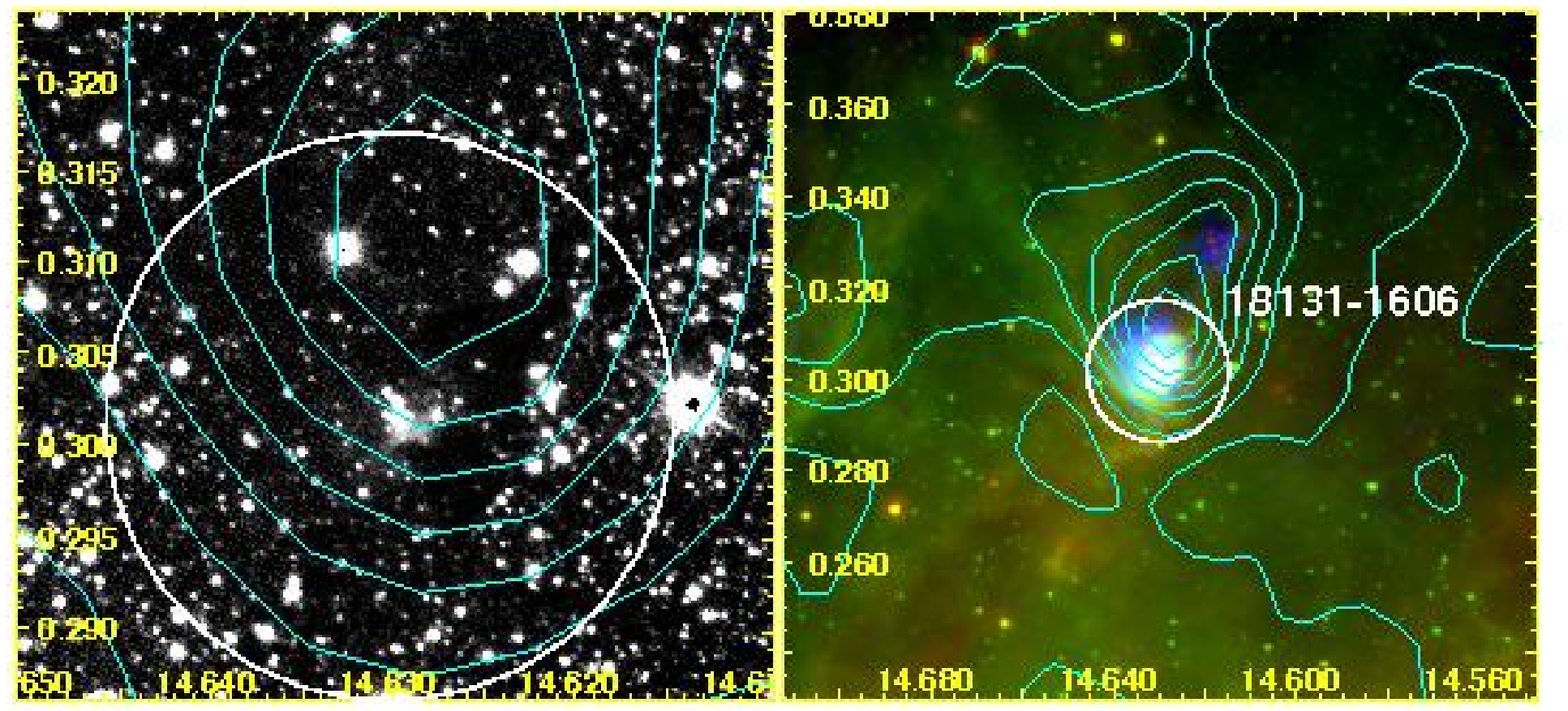}} \newline
\resizebox{3.in}{!}{\includegraphics{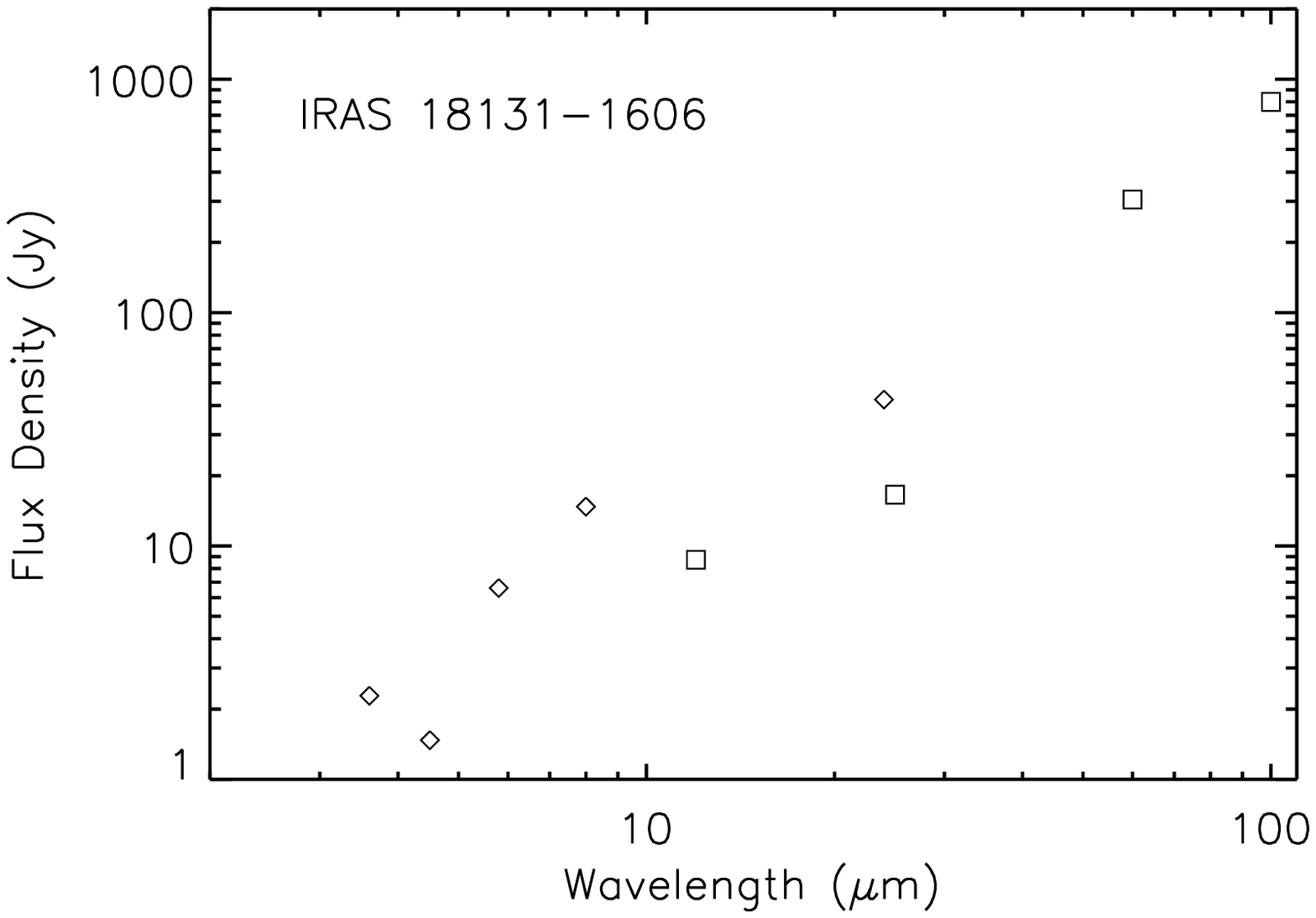}}
\resizebox{3.in}{!}{\includegraphics{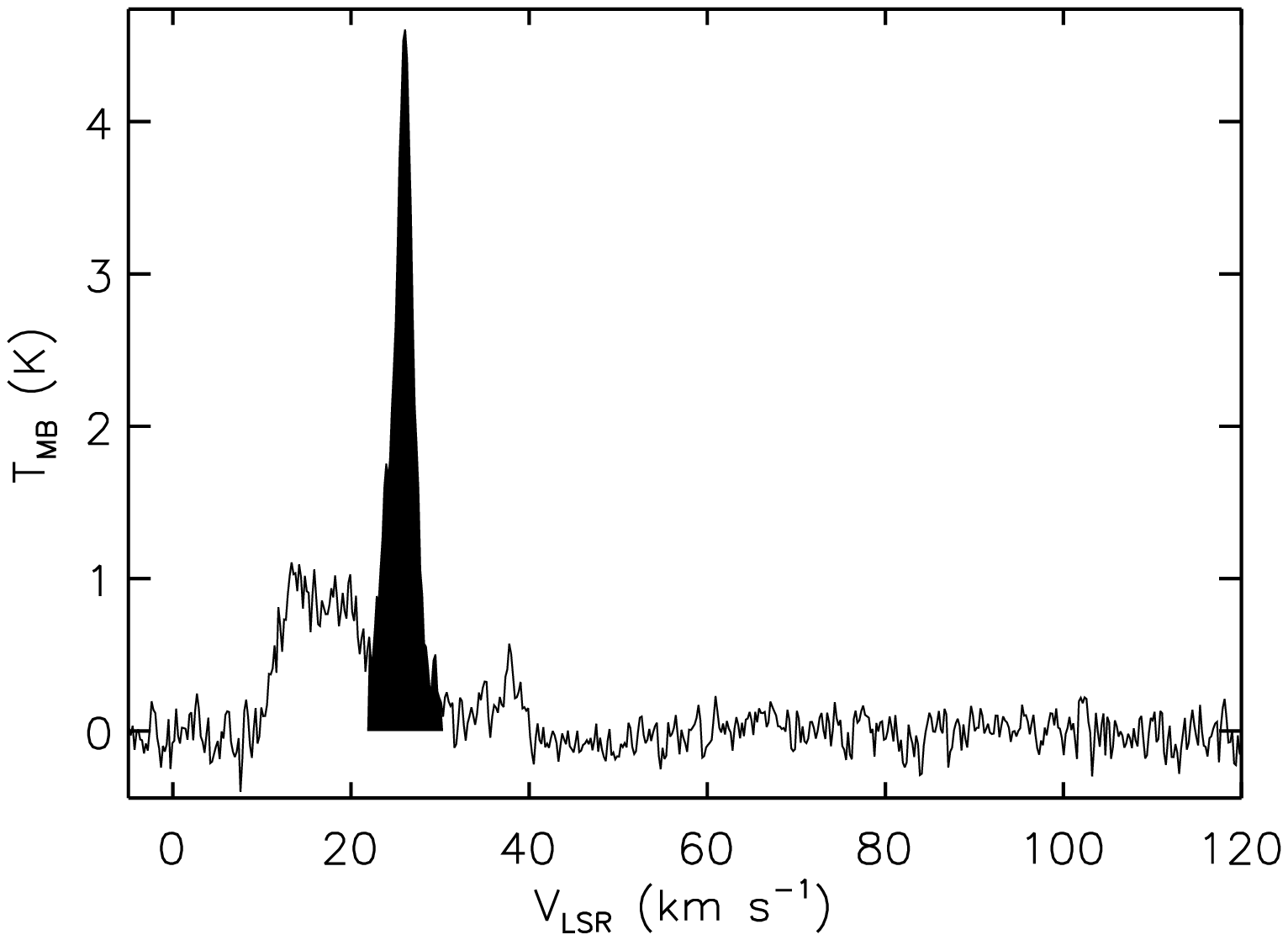}}
\caption{\footnotesize Four views of \emph{IRAS} 18131-1606. (Top left) UKIDSS K-band image. (Top right) RGB composite of \emph{Spitzer} MIPS 24 $\micron$ (red), \emph{Spitzer} IRAC 8.0 $\micron$ (green), and BGPS 1.1 mm continuum (blue) images. The guiding circle shown in the two top panels has the same position and angular size and corresponds to the aperture radius of $54 \arcsec$ in Table~\ref{photometrytable}. The contour levels in both top panels show integrated CO intensity in the zeroth-moment map, in steps of 10\%, from 90\% to 10\% of the peak (in this case $I_{\mathrm{peak}} = 28.5$ K~km~s$^{-1}$, from Table~\ref{moleculartable}). (Bottom left) Spectral energy distribution between 3.6--100 $\micron$. \emph{Spitzer} IRAC and MIPS 24 $\micron$ flux densities are displayed as diamonds. \emph{IRAS} PSC flux densities are displayed as squares. (Bottom right) The spatially averaged GRS 1-D CO spectrum, where the shaded area corresponds to the velocity range used to make the zeroth-moment map ($\Delta V$ from Table~\ref{blobsshells}). \emph{IRAS} 18131-1606 falls within the partially covered GRS area. It is a $1 \arcmin$ blob at 8.0 and 24 $\micron$, with a possible clustering of 2.2 $\micron$ point sources at the center. The \emph{IRAS} source is mentioned in \citet{kwok97} as having PAH features in its low resolution \emph{IRAS} spectrum, which is consistent with our interpretation of the source as an IM SFR. The blob is connected through a 24 $\micron$ arc coming up from the Galactic south, connecting it to G14.6+0.1, a larger \ion{H}{2} region \citep{kuchar97}. The blob is spatially associated with the source G014.634+00.308 from the BGPS source catalog \citep{bgpscatalog}. This BGPS source is clearly seen in blue.\label{18131-1606-figure}}\normalsize
\end{figure}

Figure~\ref{18131-1606-figure} shows a compilation of the basic data for the candidate IM SFR \emph{IRAS} 18131-1606 (higher resolution images will be available on-line through the Astronomical Journal). The top left panel is an UKIRT Infrared Deep Sky Survey \citep[UKIDSS;][]{dr3} K-band image. The top right panel is an RGB composite of \emph{Spitzer} MIPS 24 $\micron$ (red), \emph{Spitzer} IRAC 8.0 $\micron$ (green), and BGPS 1.1 mm continuum (blue) images. \emph{Spitzer} 8.0 $\micron$ shows the presence of polycyclic aromatic hydrocarbons (PAHs) and \emph{Spitzer} 24 $\micron$ shows the presence of hot dust. The BGPS is a 1.1 mm continuum survey of the Galactic Plane with an effective resolution of $33 \arcsec$ and a non-uniform
$1 \sigma$ noise level in the range of 30 to 60 mJy~beam$^{-1}$. The BGPS shows thermal dust emission from cold, dense material. The guiding circle shown has the same position in both panels and is the same angular size (in this case the aperture radius is $54 \arcsec$, from Table~\ref{photometrytable}) as the aperture used to do photometry (Section~\ref{sed}). The contour levels in both top panels show integrated CO intensity in the zeroth-moment map, in steps of 10\%, from 90\% to 10\% of the peak (in this case $I_{\mathrm{peak}} = 28.5$ K~km~s$^{-1}$, from Table~\ref{moleculartable}). The bottom right panel shows the spatially averaged GRS 1-D CO spectrum, where the shaded area corresponds to the velocity range used to make the zeroth-moment map ($\Delta V$ from Table~\ref{blobsshells}). The bottom left panel shows the spectral energy distribution (SED, described further in Section~\ref{sed}) between 3.6--100 $\micron$, where \emph{Spitzer} IRAC and MIPS 24 $\micron$ flux densities are displayed as diamonds, and \emph{IRAS} PSC flux densities are displayed as squares.

\subsection{Distances and Sizes}{\label{near}}
We estimated the LSR velocity ($V_{\mathrm{LSR}}$) of each candidate IM SFR by measuring the peak of the CO feature in a 1-D spectrum produced by averaging the GRS cube of each candidate IM SFR over the area of the mid-IR feature, typically a region covering 4--16 GRS pixels (1 GRS pixel = $22 \arcsec \times 22 \arcsec$). The results are shown in Table~\ref{blobsshells}. Once the velocities were known, we obtained corresponding kinematic distances, near and far. We adopted the Galactic rotation
curve of \citet{clemens85}, assuming $R_0 = 8.5$ kpc, $\Omega_0 = 220$ km s$^{-1}$. The more recent rotation curve of \citet{levine08} yielded very similar distances for the same $R_0$ and $\Omega_0$. Exceptions were when $\ell \gtrsim 45 \degr$ and the velocity $V \lesssim 50$ km s$^{-1}$, then the differences between the distances obtained with the two different rotation curves can be 0.3 kpc or more.

The angular diameters were determined by eye in the 8 and 24 $\micron$ images. Typically the diameter roughly corresponds to where the mid-IR surface brightness falls to $5\% - 25\%$ of the peak surface brightness. However, the varying background levels and irregular shapes make a quantitative angular size estimate imprecise. We estimated that this technique has an uncertainty of $\sim 20\%$. We find that the angular diameter of \emph{all} the blobs and shells range from $0.3 \arcmin$ to $3.0 \arcmin$, with a median angular diameter of $0.7 \arcmin$ and an average of $0.9 \arcmin$.

\begin{figure}
\resizebox{6.in}{!}{\includegraphics{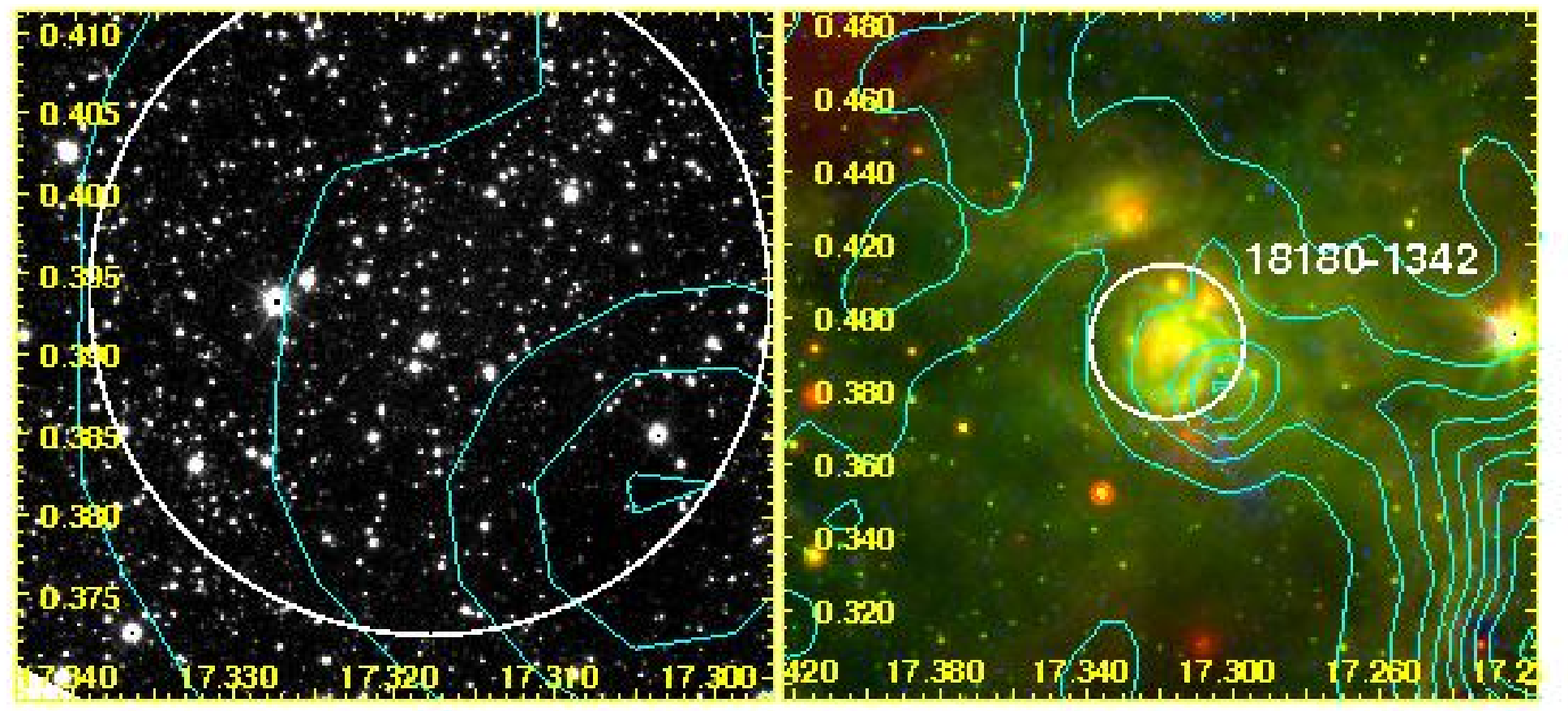}} \newline
\resizebox{3.in}{!}{\includegraphics{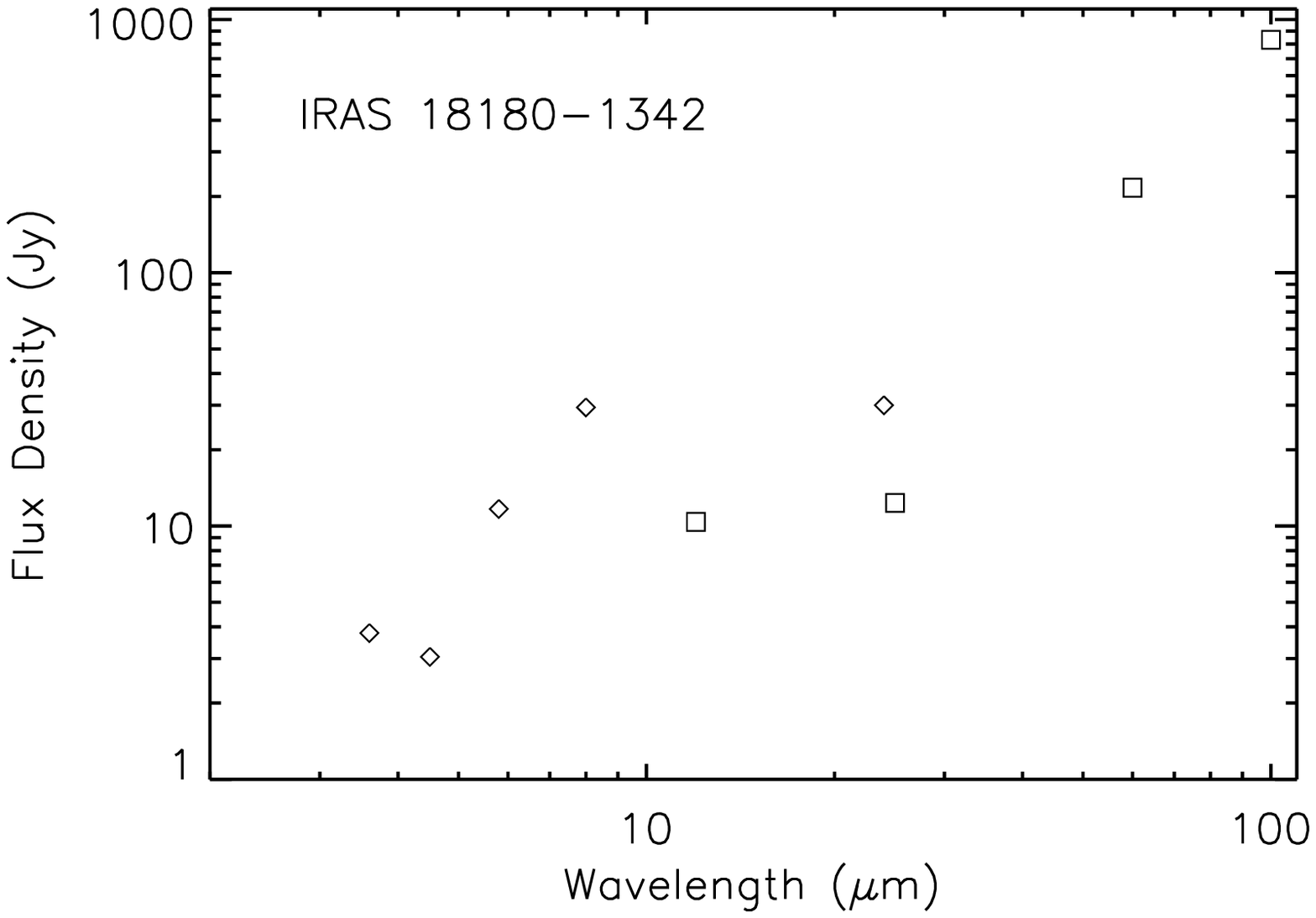}}
\resizebox{3.in}{!}{\includegraphics{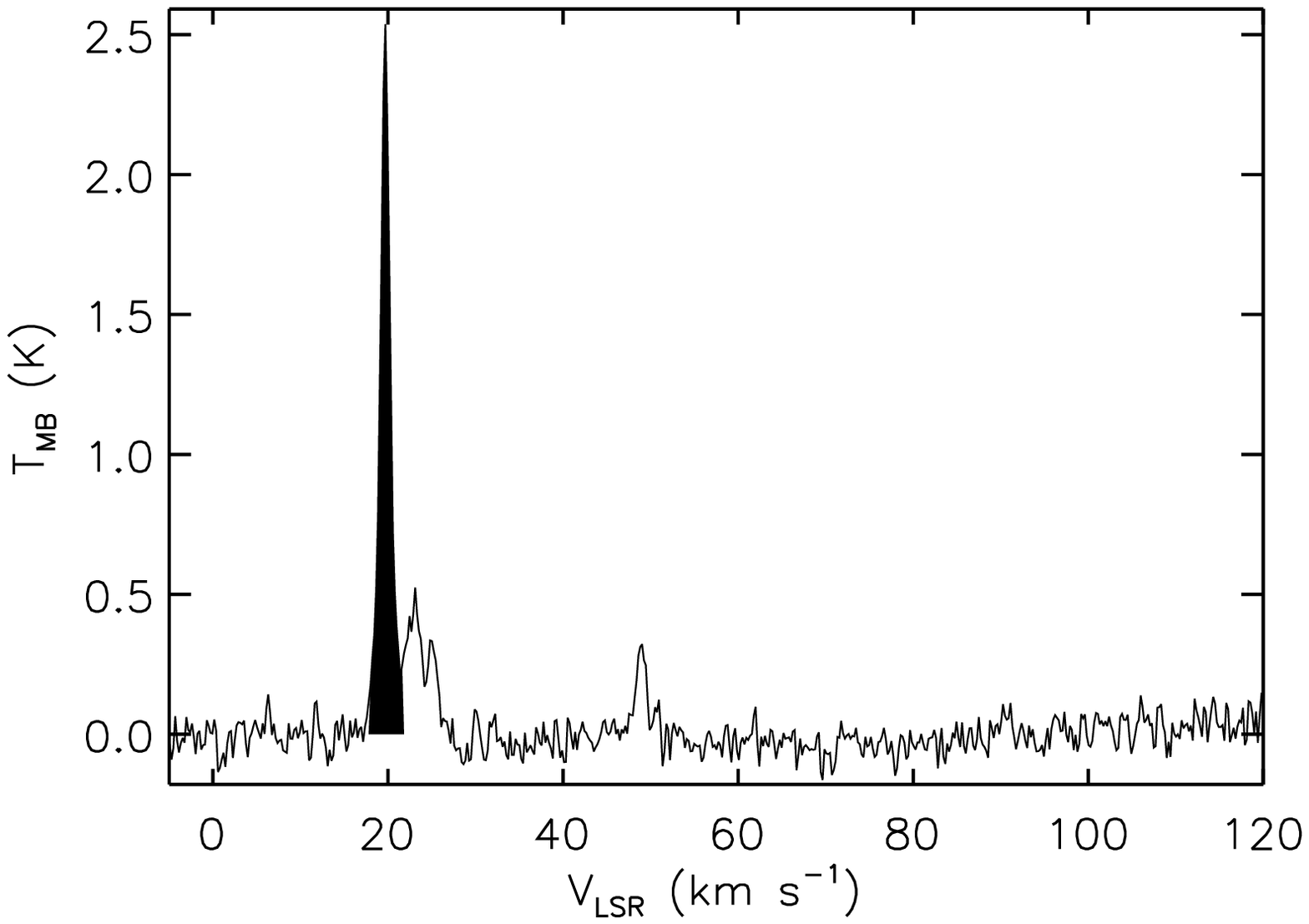}}
\caption{Four views of \emph{IRAS} 18180-1342, like Figure~\ref{18131-1606-figure}. The guiding circle has a radius of $76 \arcsec$. This object also falls within the partially covered GRS area. The blob has an angular diameter of $2 \arcmin$, the largest apparent diameter of the near objects, but given its kinematic 2.0 kpc distance, it has a physical diameter of 1.2 pc.\label{18180-1342-figure}}
\end{figure}

\begin{figure}
\resizebox{6.in}{!}{\includegraphics{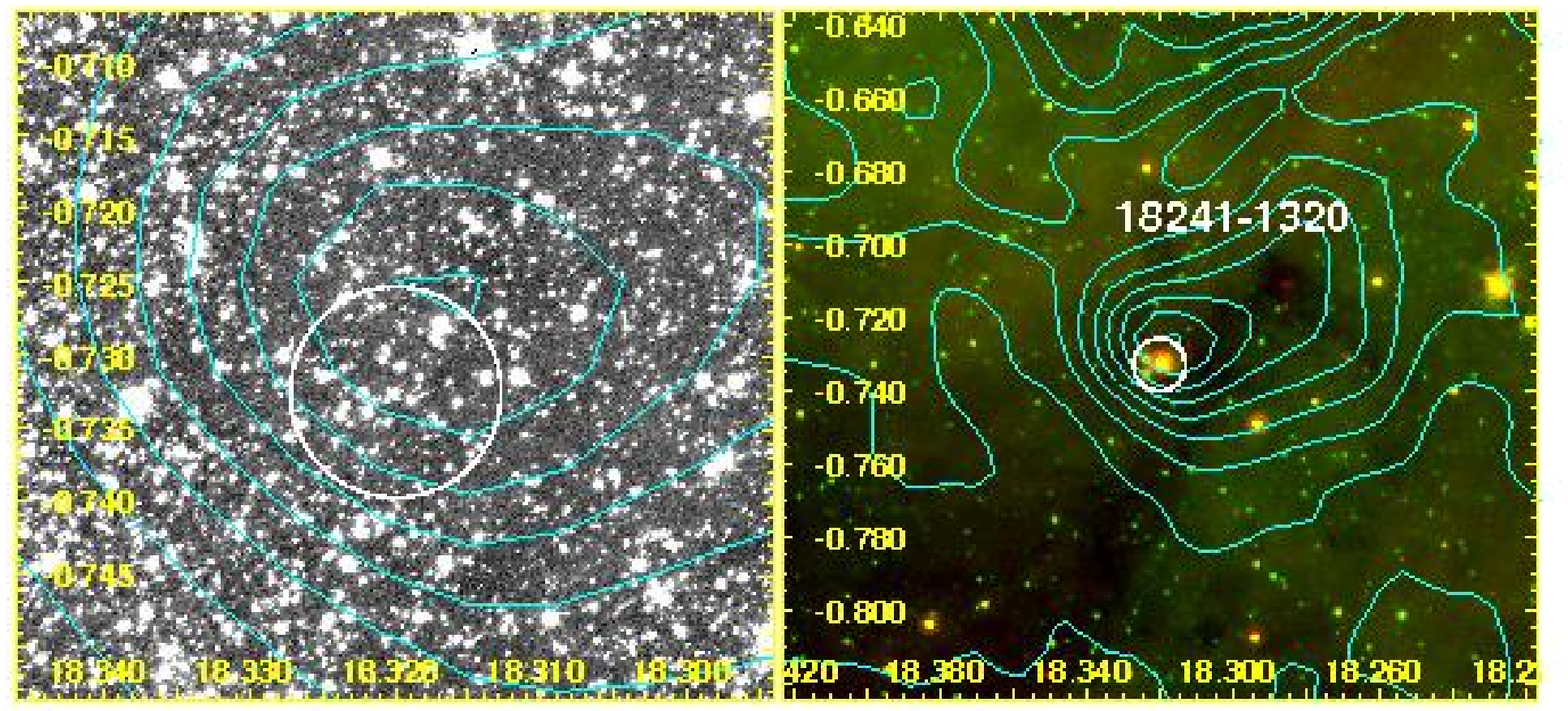}} \newline
\resizebox{3.in}{!}{\includegraphics{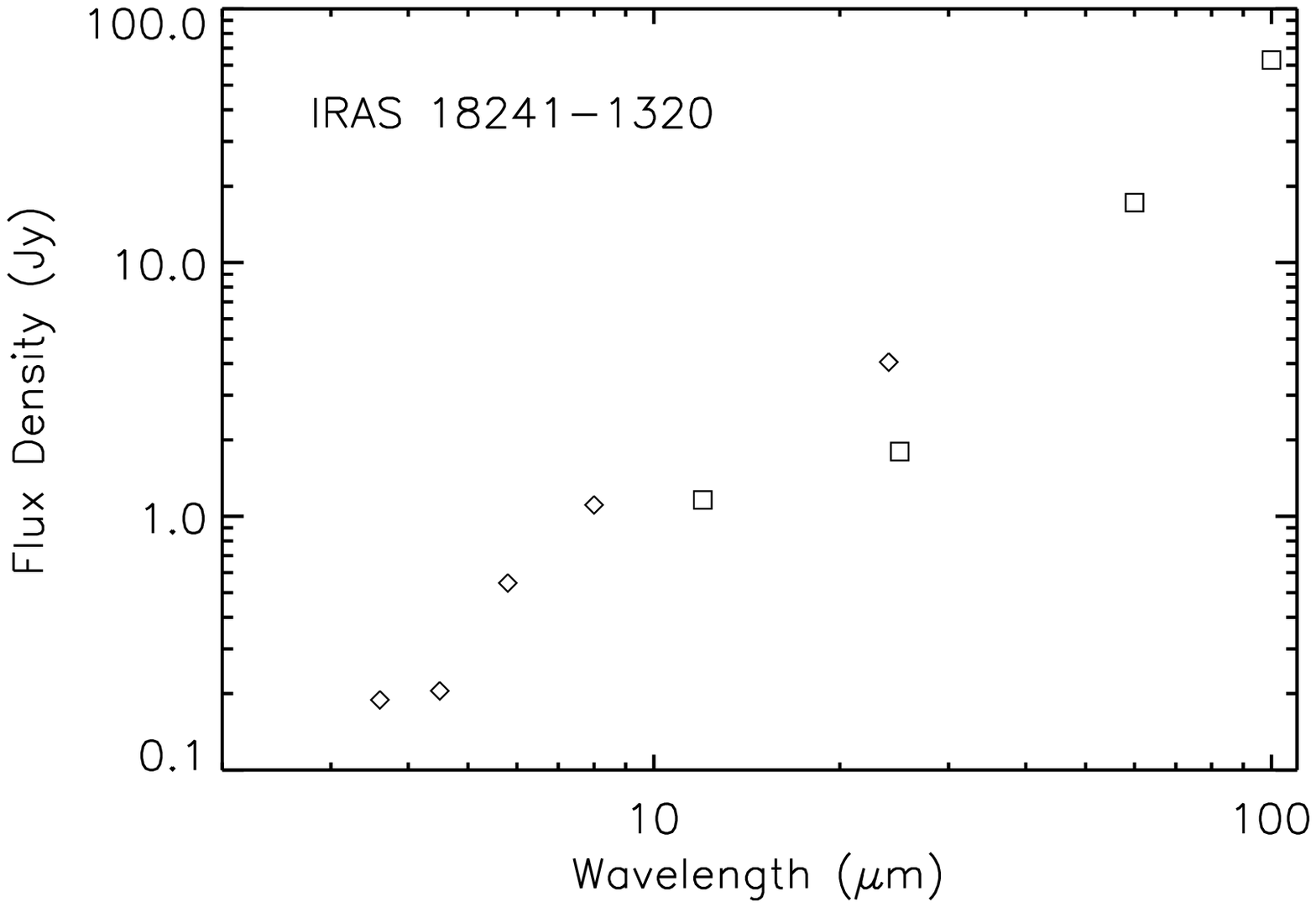}}
\resizebox{3.in}{!}{\includegraphics{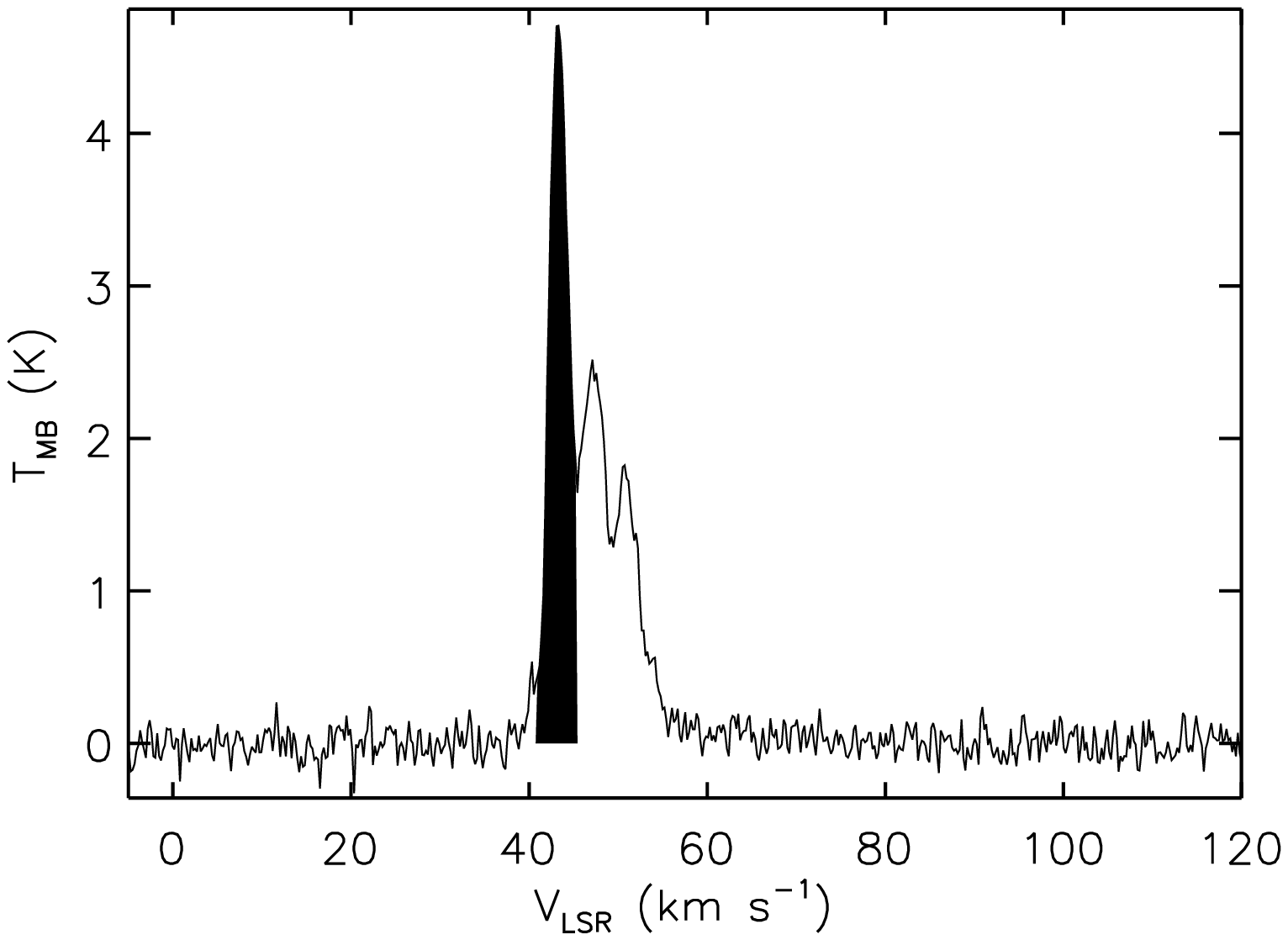}}
\caption{Four views of \emph{IRAS} 18241-1320, like Figure~\ref{18131-1606-figure}. The guiding circle has a radius of $26 \arcsec$. A rather small blob without any obvious clustering of stellar objects at the center of the blob. The blob is very likely, from spatial and velocity considerations, to be associated with the {}``clump'' c7 in the cloud GRSMC G018.39-00.41 \citep{rathborne09}.\label{18241-1320-figure}}
\end{figure}

\begin{figure}
\resizebox{6.in}{!}{\includegraphics{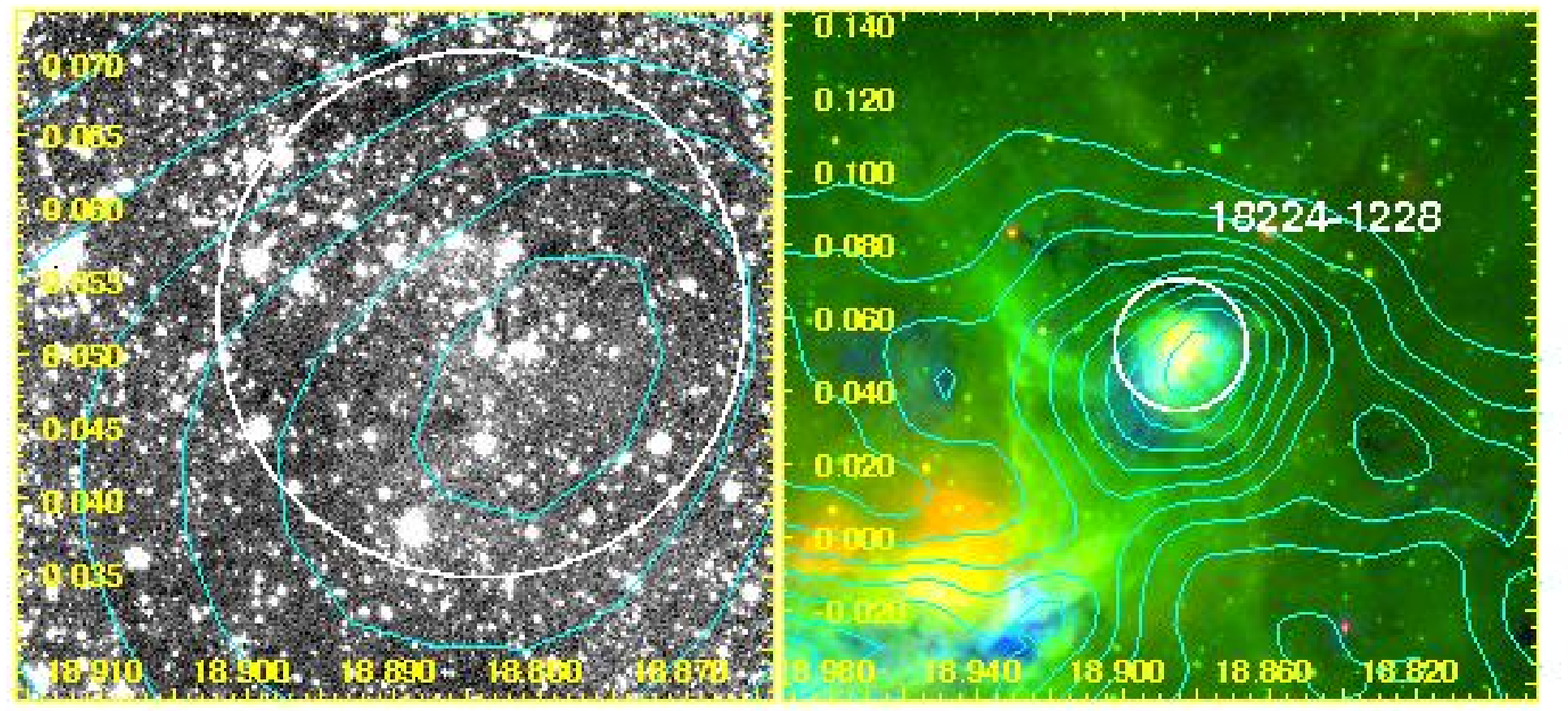}} \newline
\resizebox{3.in}{!}{\includegraphics{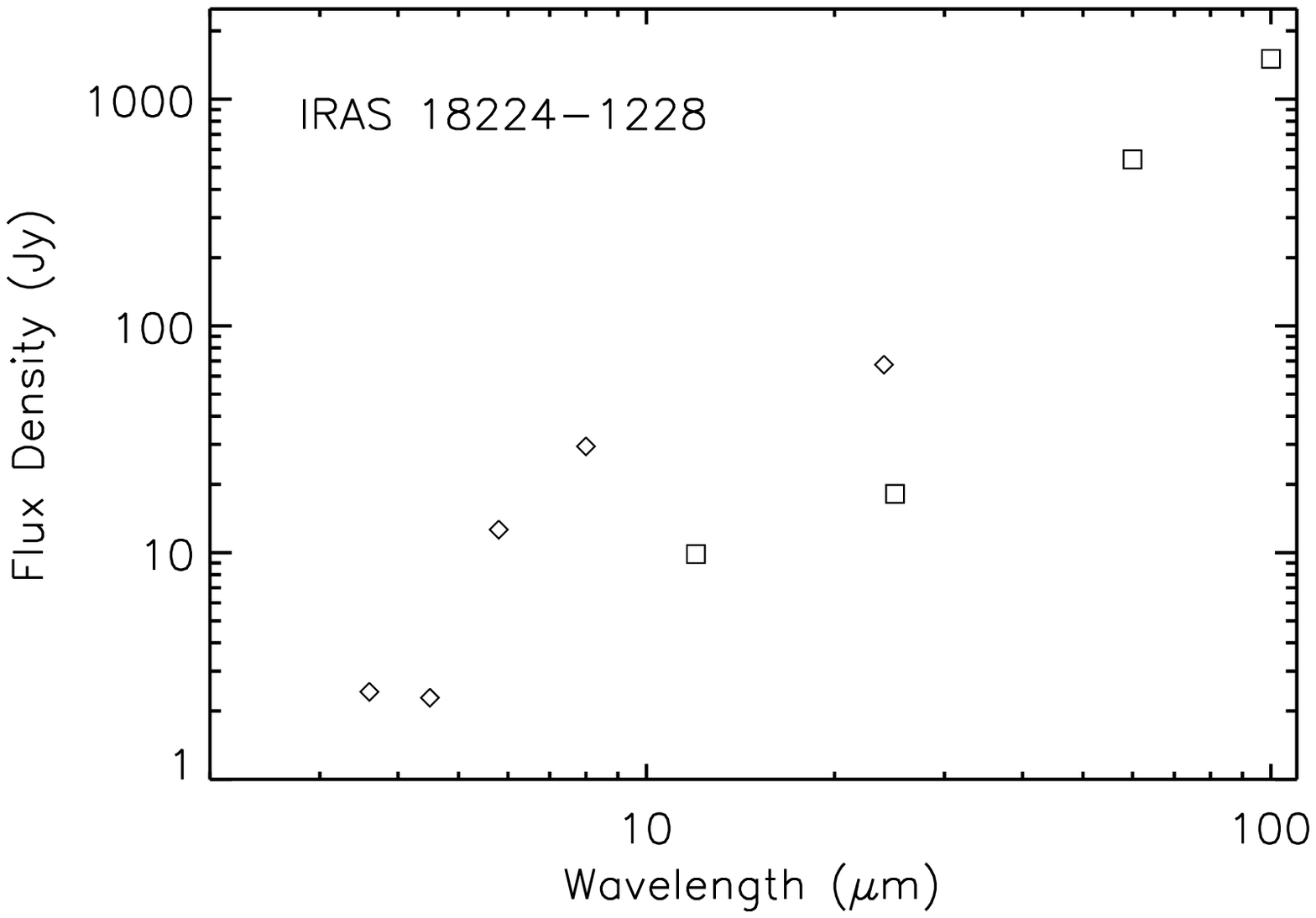}}
\resizebox{3.in}{!}{\includegraphics{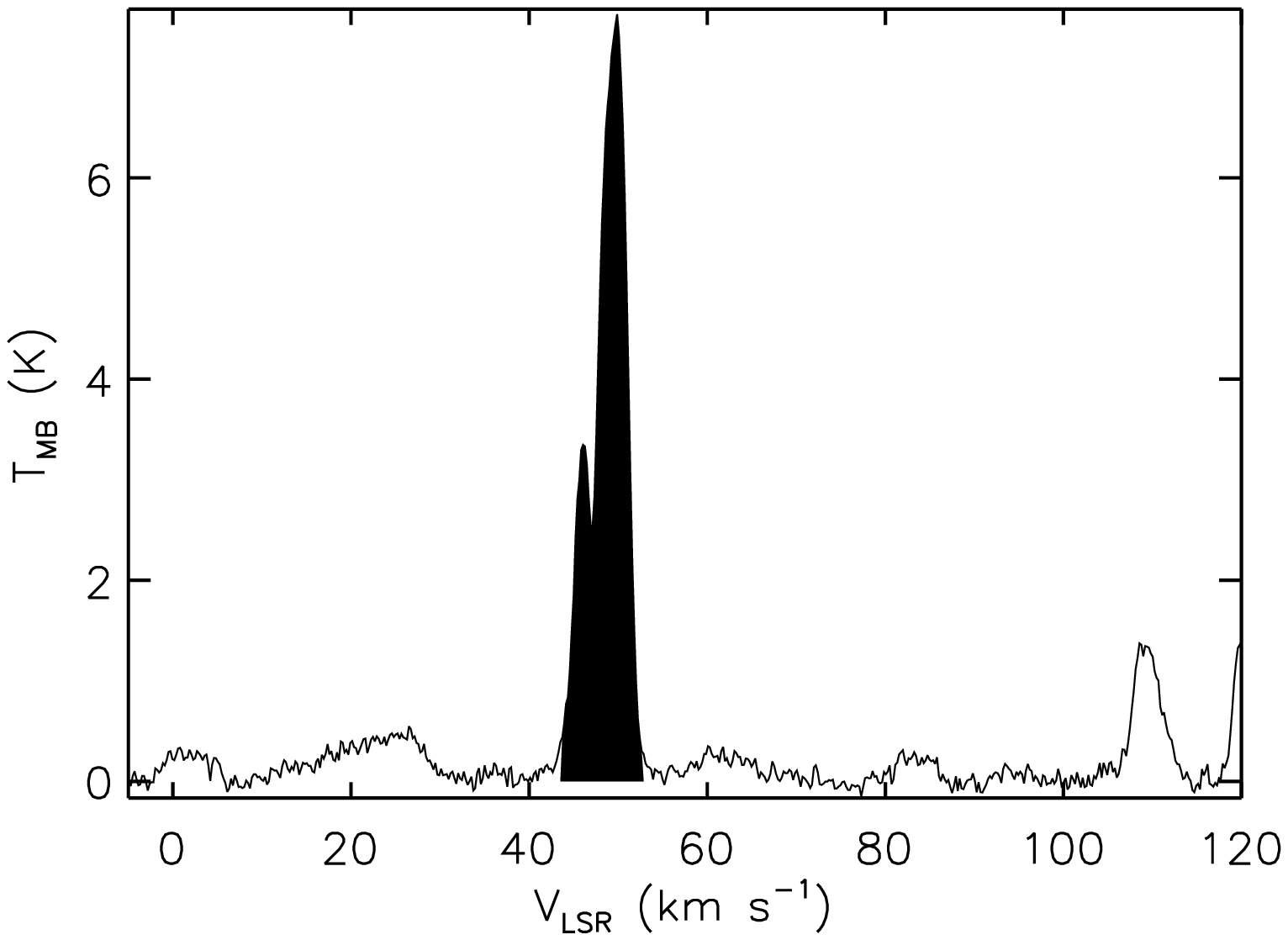}}
\caption{Four views of \emph{IRAS} 18224-1228, like Figure~\ref{18131-1606-figure}. The guiding circle has a radius of $65 \arcsec$. The largest (1.7 pc) and the most luminous ($2.7 \times 10^4 L_{\sun}$) of our near objects sits on the outskirts of the W39 complex. This blob is likely associated with {}``clump'' c1 (another possibility is c2) in the cloud GRSMC G018.89+00.04. Looking at the position of the blob compared to BGPS, an association with BGPS G018.890+00.045 is likely (another possibility is BGPS G018.879+00.053).\label{18224-1228-figure}}
\end{figure}

\begin{figure}
\resizebox{6.in}{!}{\includegraphics{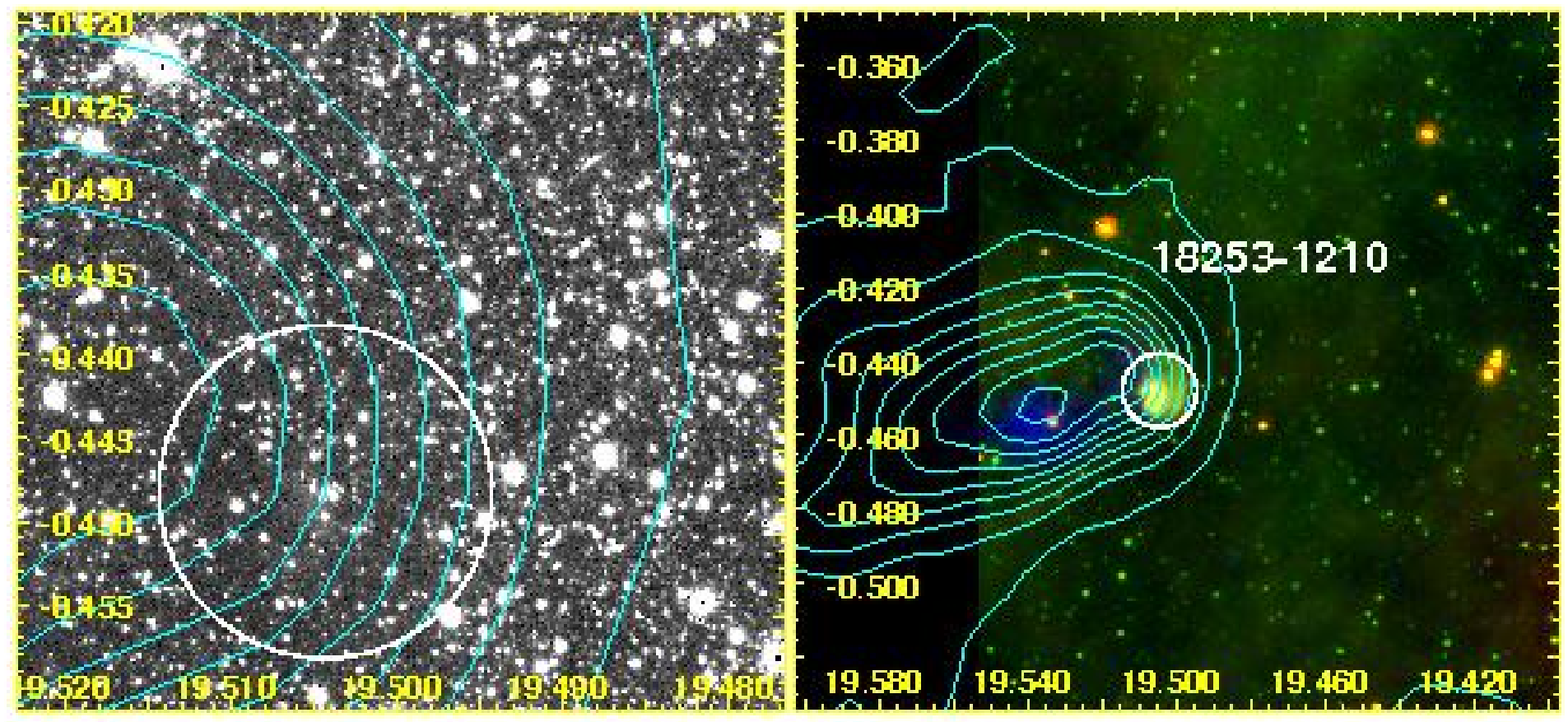}} \newline
\resizebox{3.in}{!}{\includegraphics{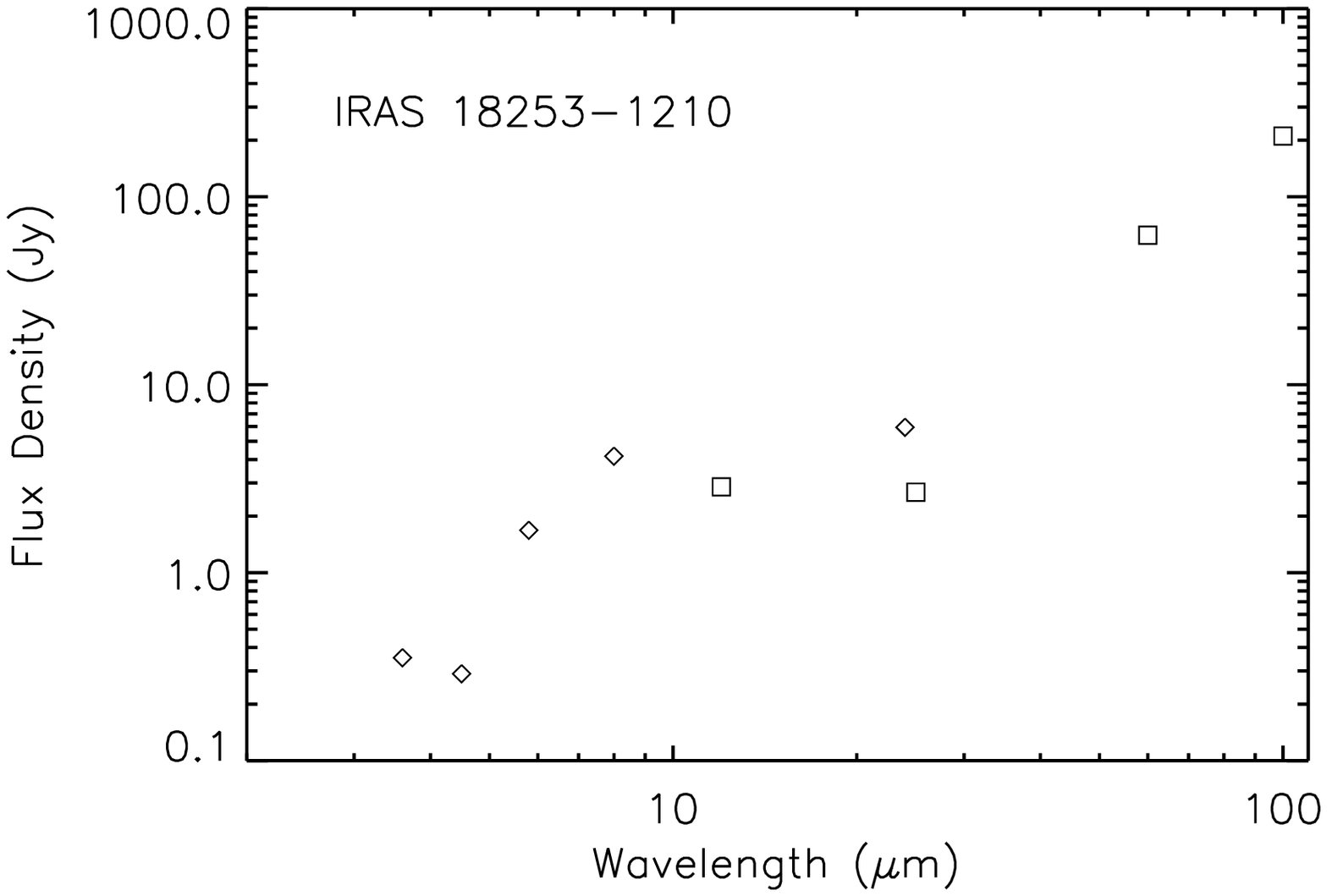}}
\resizebox{3.in}{!}{\includegraphics{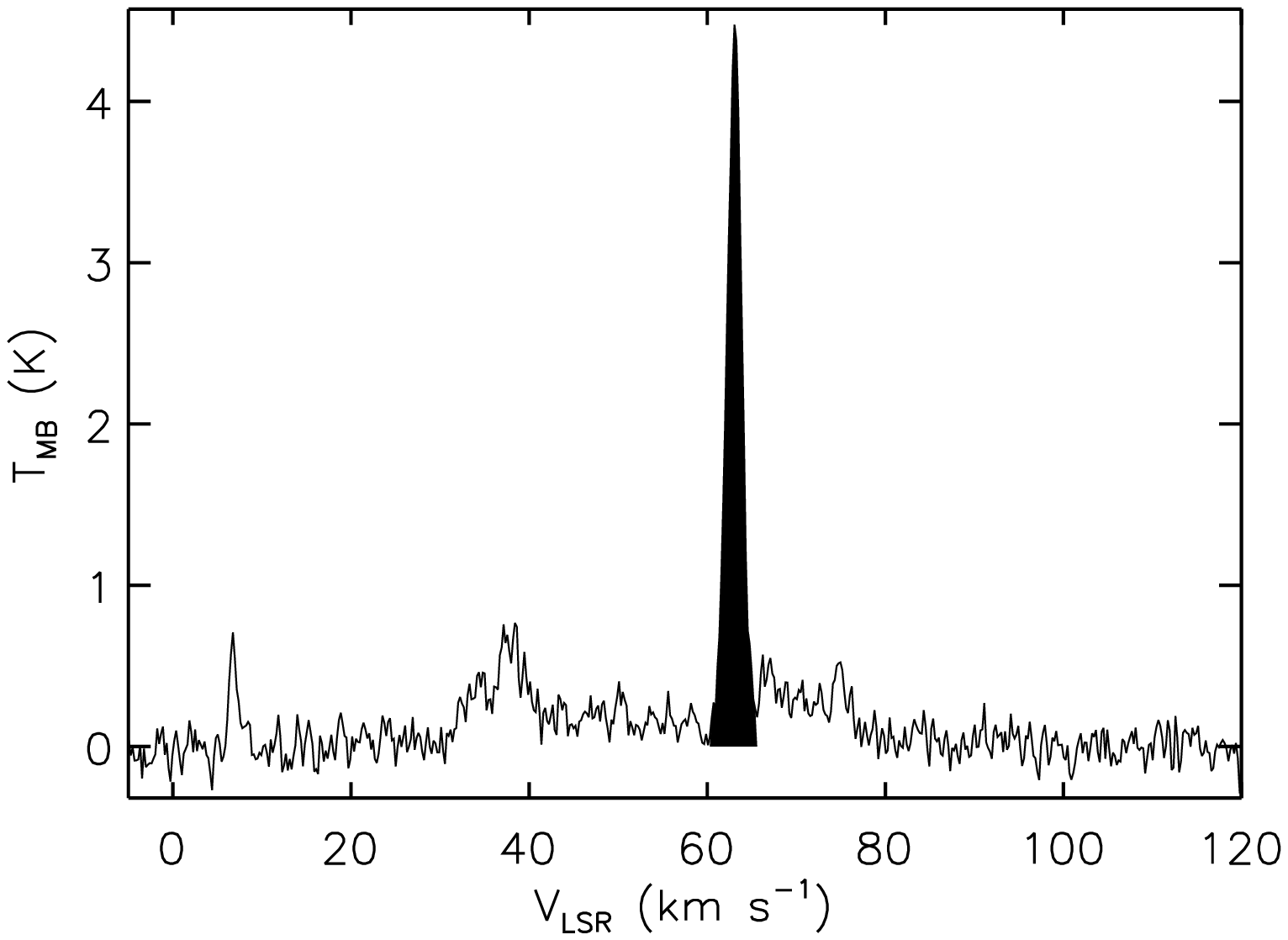}}
\caption{Four views of \emph{IRAS} 18253-1210, like Figure~\ref{18131-1606-figure}. The guiding circle has a radius of $36 \arcsec$. This $1 \arcmin$ blob has a bow-shock like appearance. The molecular material indicated by the contours is associated with {}``clump'' c1 in the cloud GRSMC G019.54-00.46. The \emph{IRAS} blob is spatially associated with BGPS G019.507-00.447, while the position of maximum CO column density is spatially associated with BGPS G019.542-00.457.\label{18253-1210-figure}}
\end{figure}

\begin{figure}
\resizebox{6.in}{!}{\includegraphics{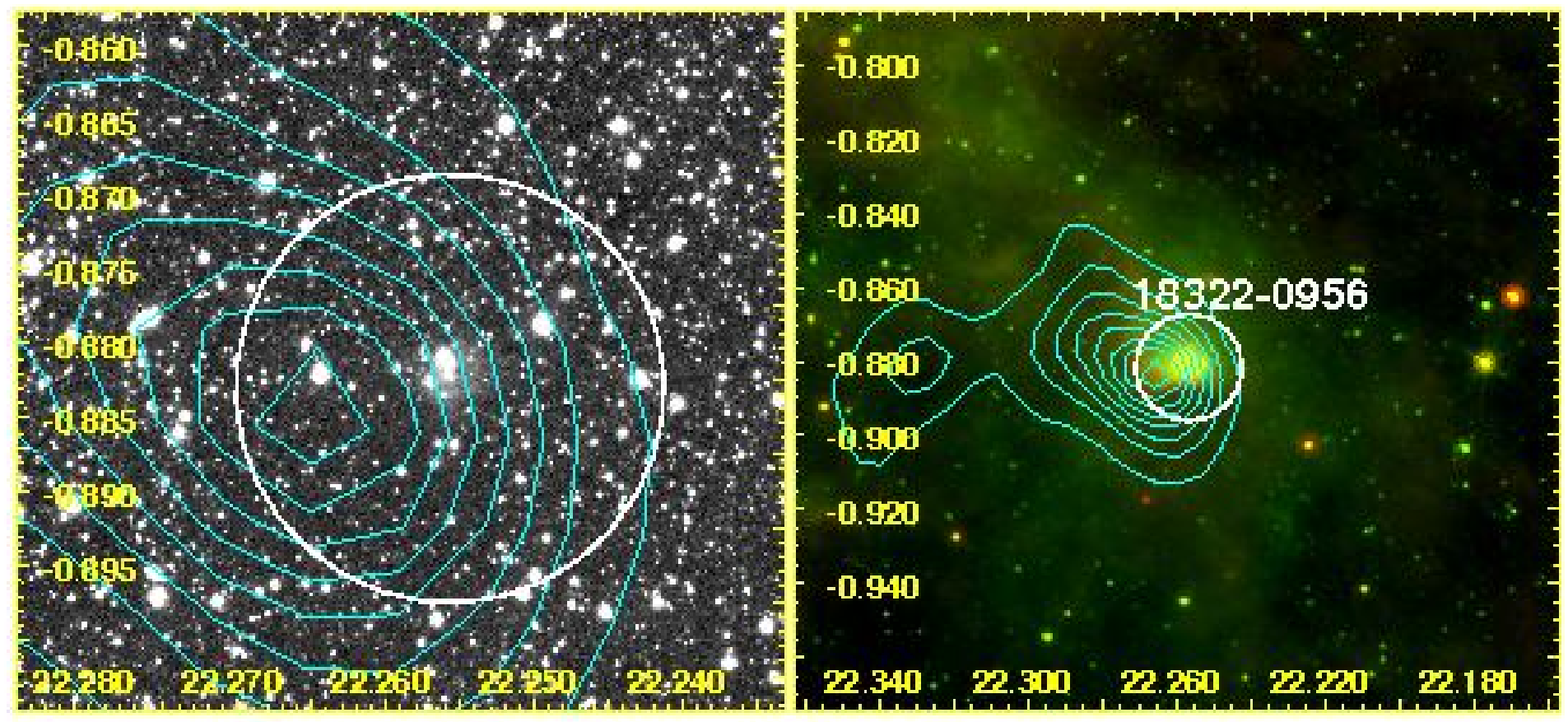}} \newline
\resizebox{3.in}{!}{\includegraphics{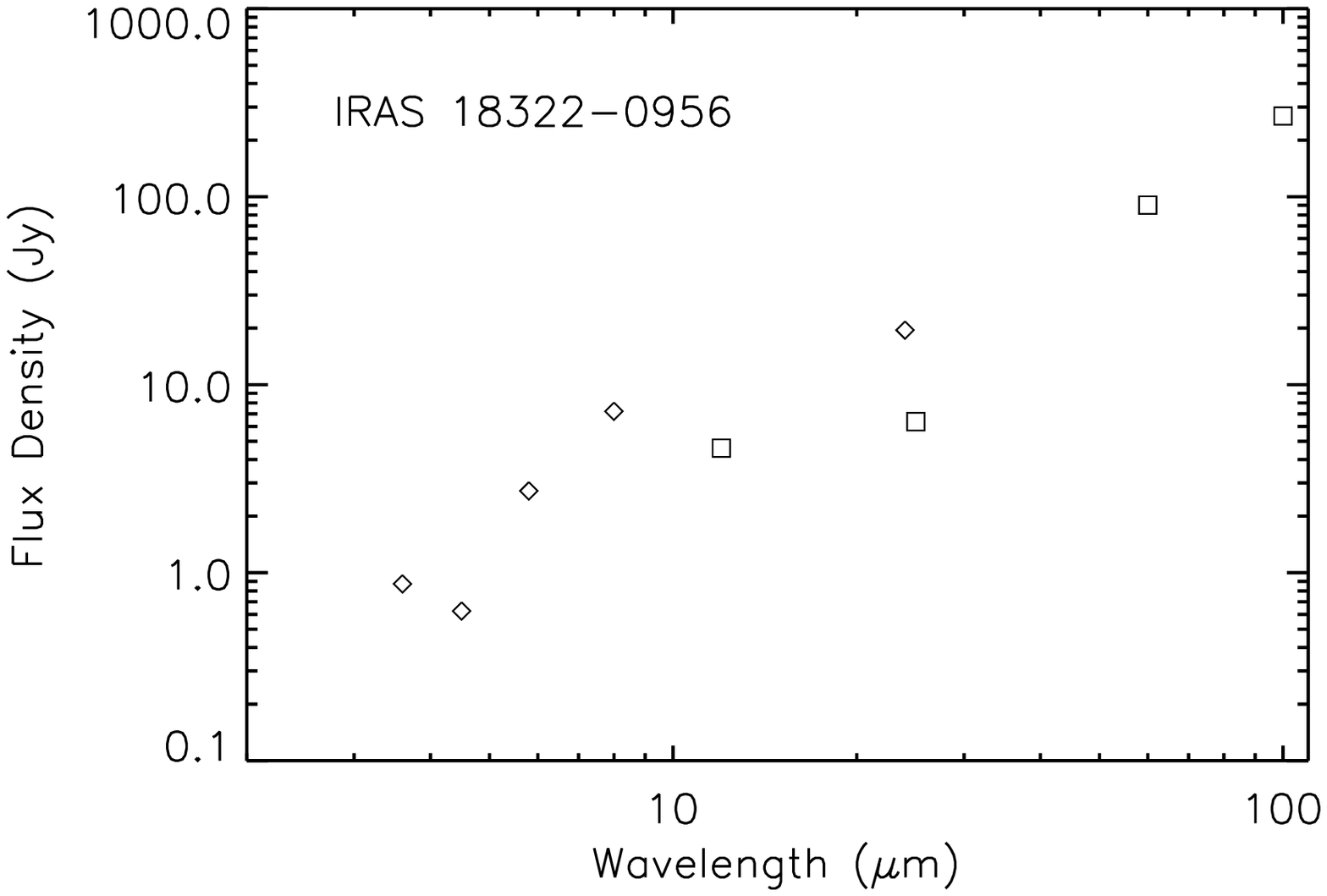}}
\resizebox{3.in}{!}{\includegraphics{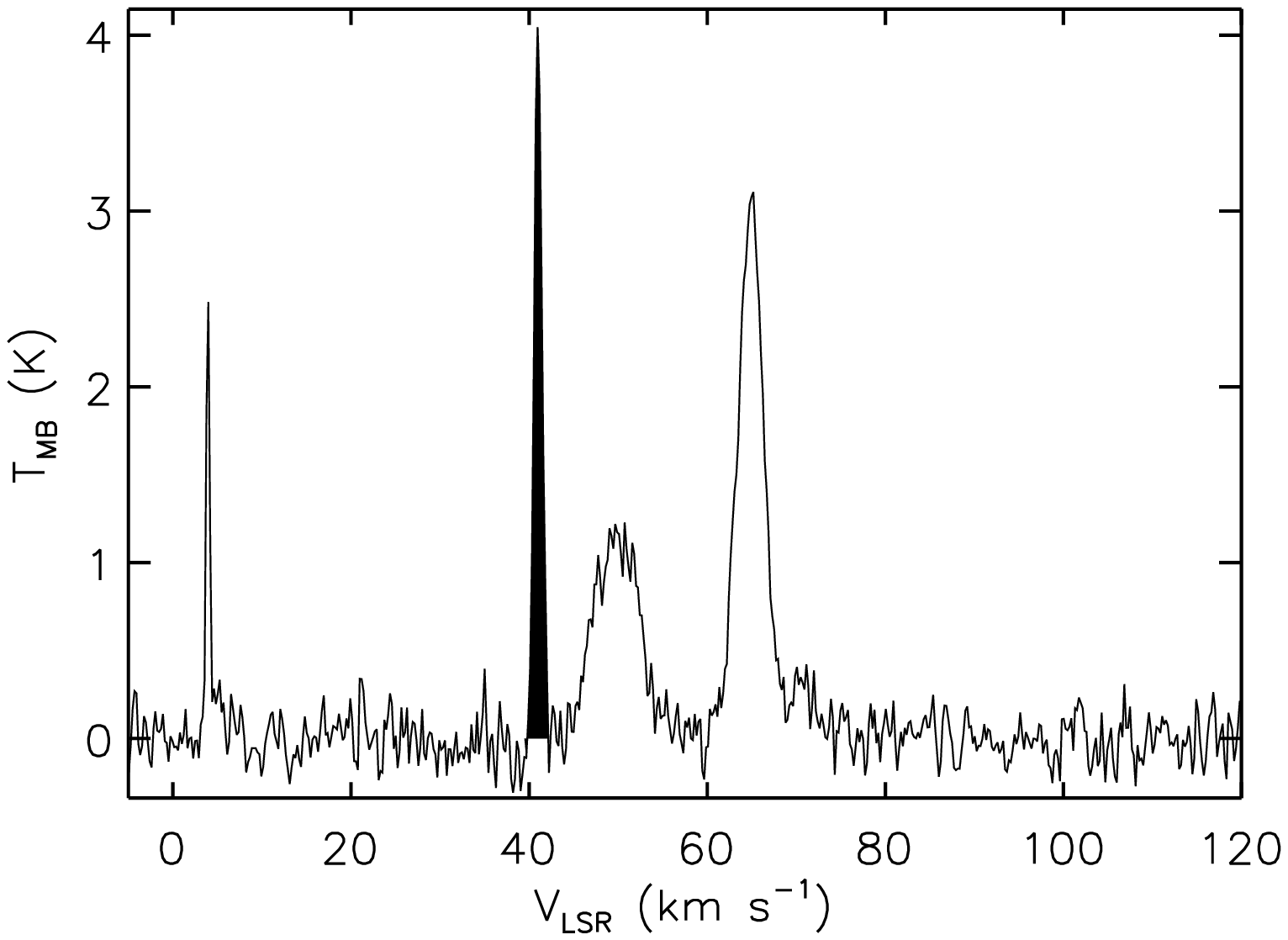}}
\caption{Four views of \emph{IRAS} 18322-0956, like Figure~\ref{18131-1606-figure}. The guiding circle has a radius of $52 \arcsec$. This is an isolated $0.6 \arcmin$ blob, sitting on the edge of molecular material at 41.0 km~s$^{-1}$. We note that it is possible that this source could be at a velocity of $\sim 65$ km~s$^{-1}$, which would increase its physical diameter estimate by a factor of 1.4 and its luminosity estimate by a factor of 2.0.\label{18322-0956-figure}}
\end{figure}

\begin{figure}
\resizebox{6.in}{!}{\includegraphics{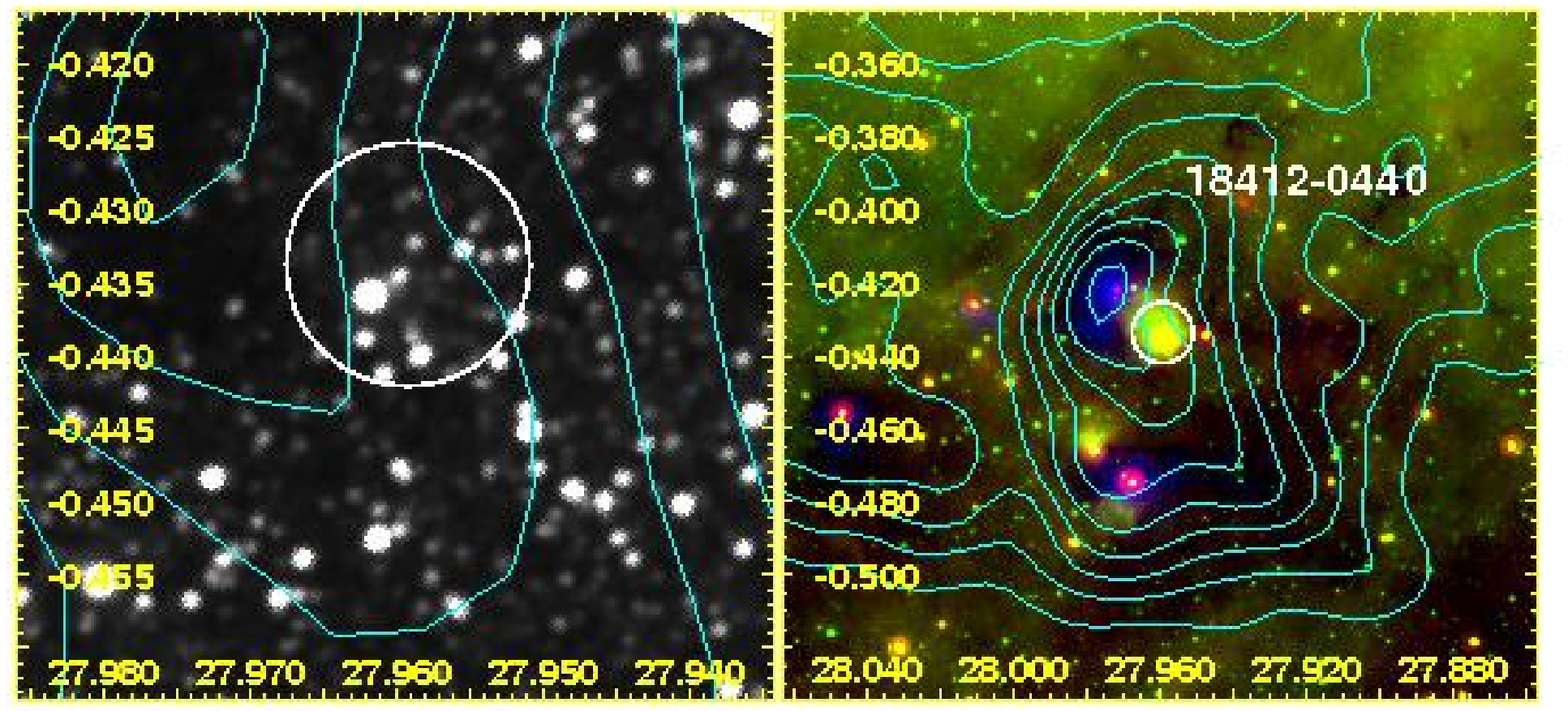}} \newline
\resizebox{3.in}{!}{\includegraphics{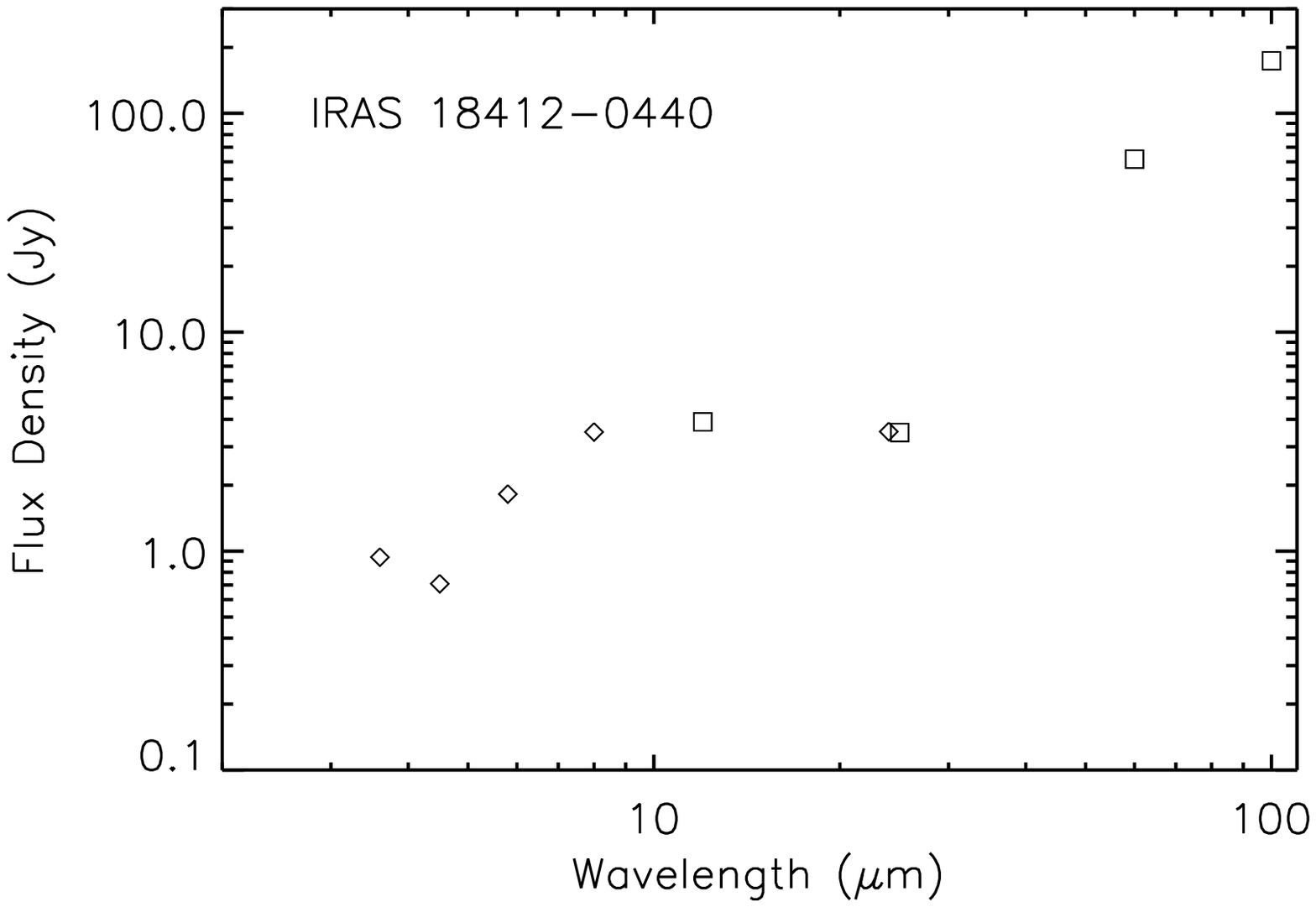}}
\resizebox{3.in}{!}{\includegraphics{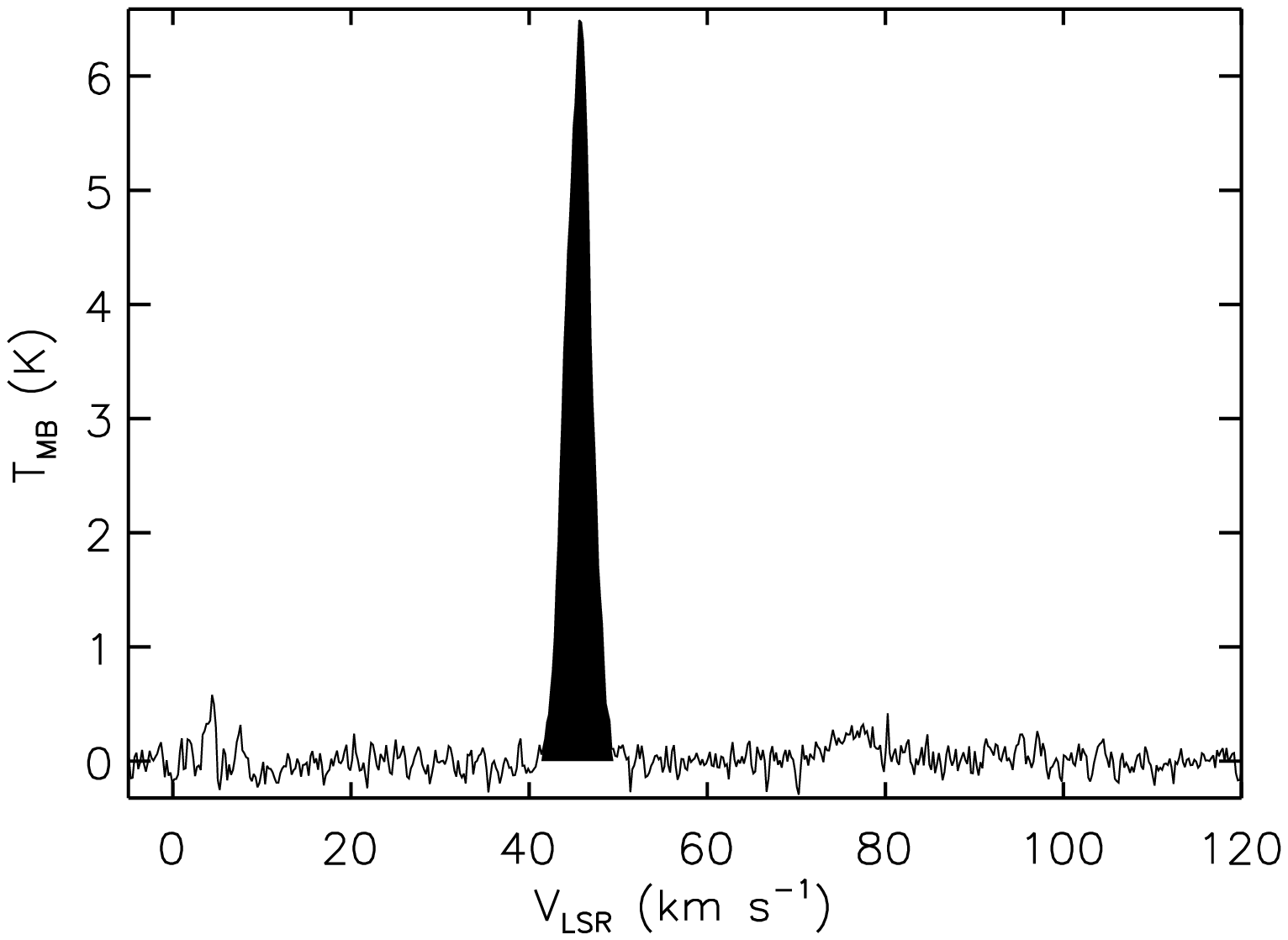}}
\caption{Four views of \emph{IRAS} 18412-0440, like Figure~\ref{18131-1606-figure}. The guiding circle has a radius of $35 \arcsec$. A $0.8 \arcmin$ blob, clearly associated with IRDC G027.94-00.47 from \citet{simon06}, which is seen to the left of the blob. Further associations are the {}``clump'' c2 in GRSMC G027.99-0.46 and BGPS G027.972-00.422.\label{18412-0440-figure}}
\end{figure}

\begin{figure}
\resizebox{6.in}{!}{\includegraphics{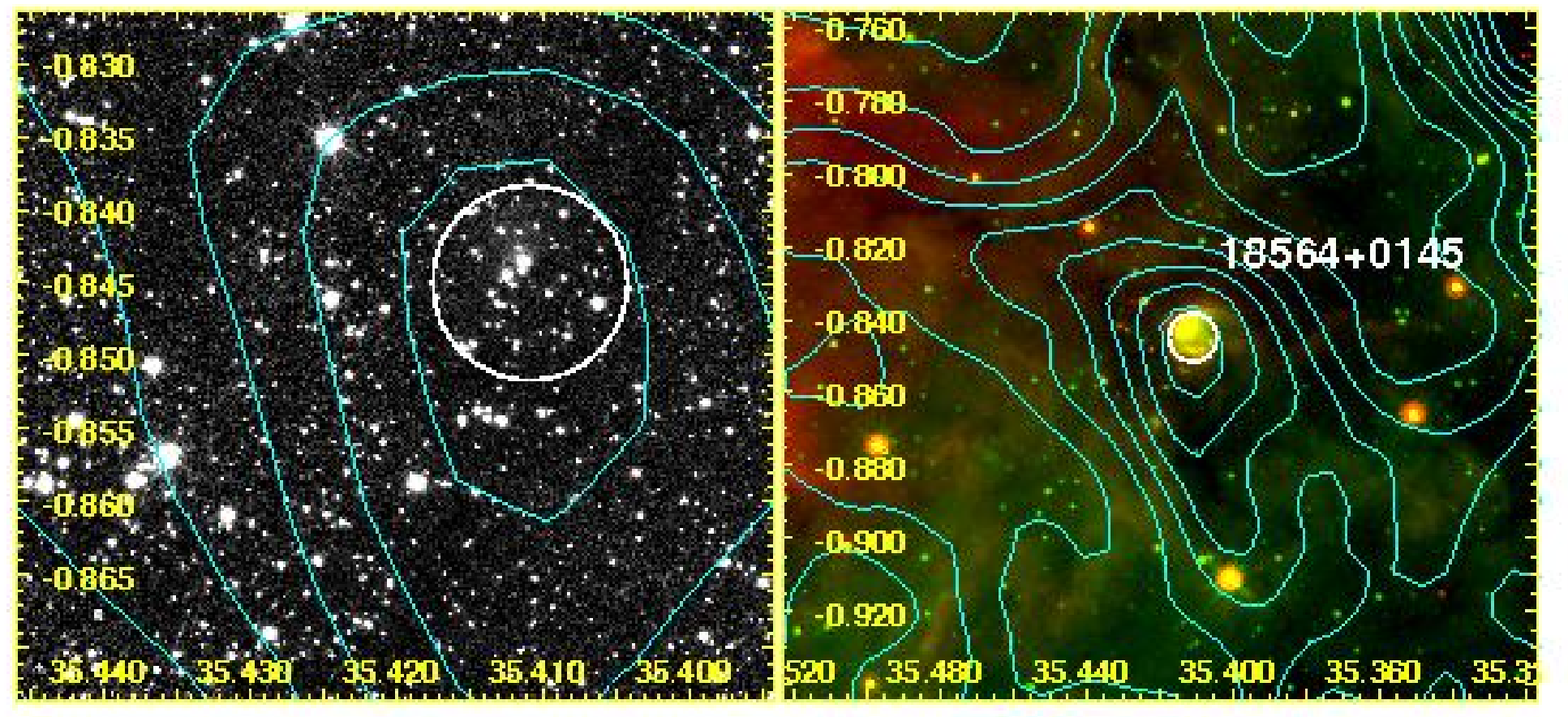}} \newline
\resizebox{3.in}{!}{\includegraphics{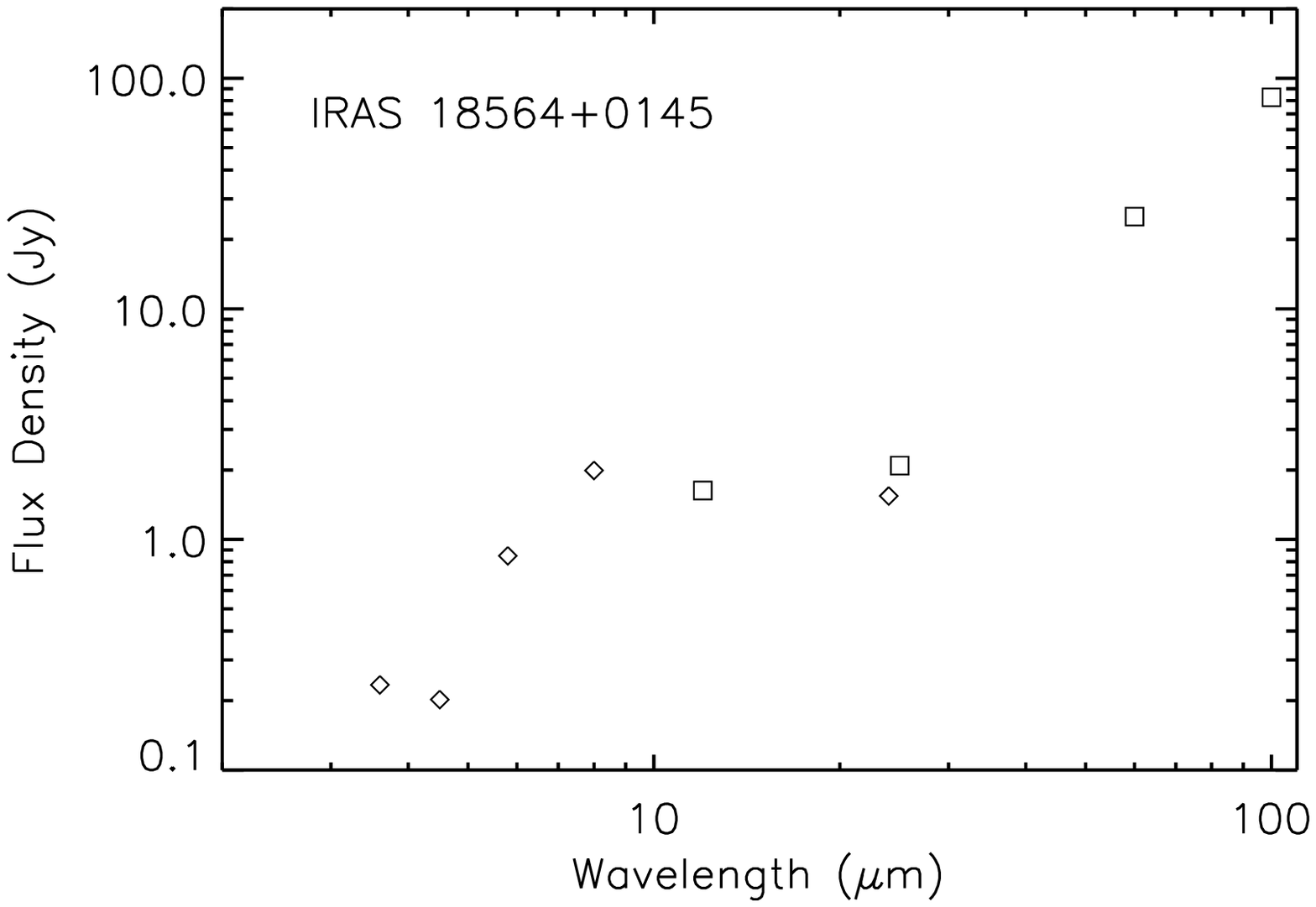}}
\resizebox{3.in}{!}{\includegraphics{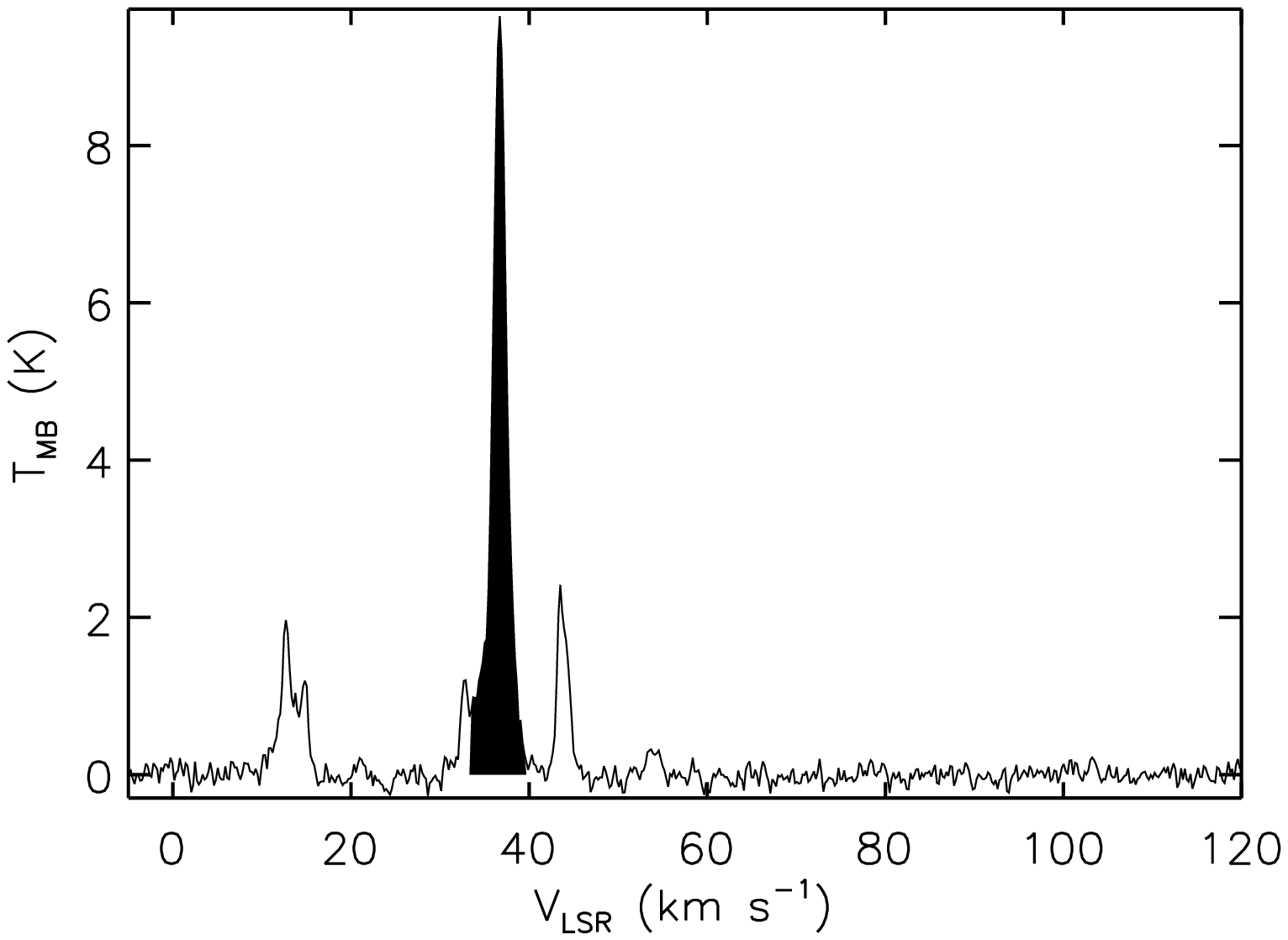}}
\caption{Four views of \emph{IRAS} 18564+0145, like Figure~\ref{18131-1606-figure}. The guiding circle has a radius of $24 \arcsec$. This $0.7 \arcmin$ blob sits on CO that can be seen as dark absorption features in the 8.0 $\micron$ emission. The molecular material is likely referenced as {}``clump'' c1 in GRSMC G035.59-00.91.\label{18564+0145-figure}}
\end{figure}

\begin{figure}
\plottwo{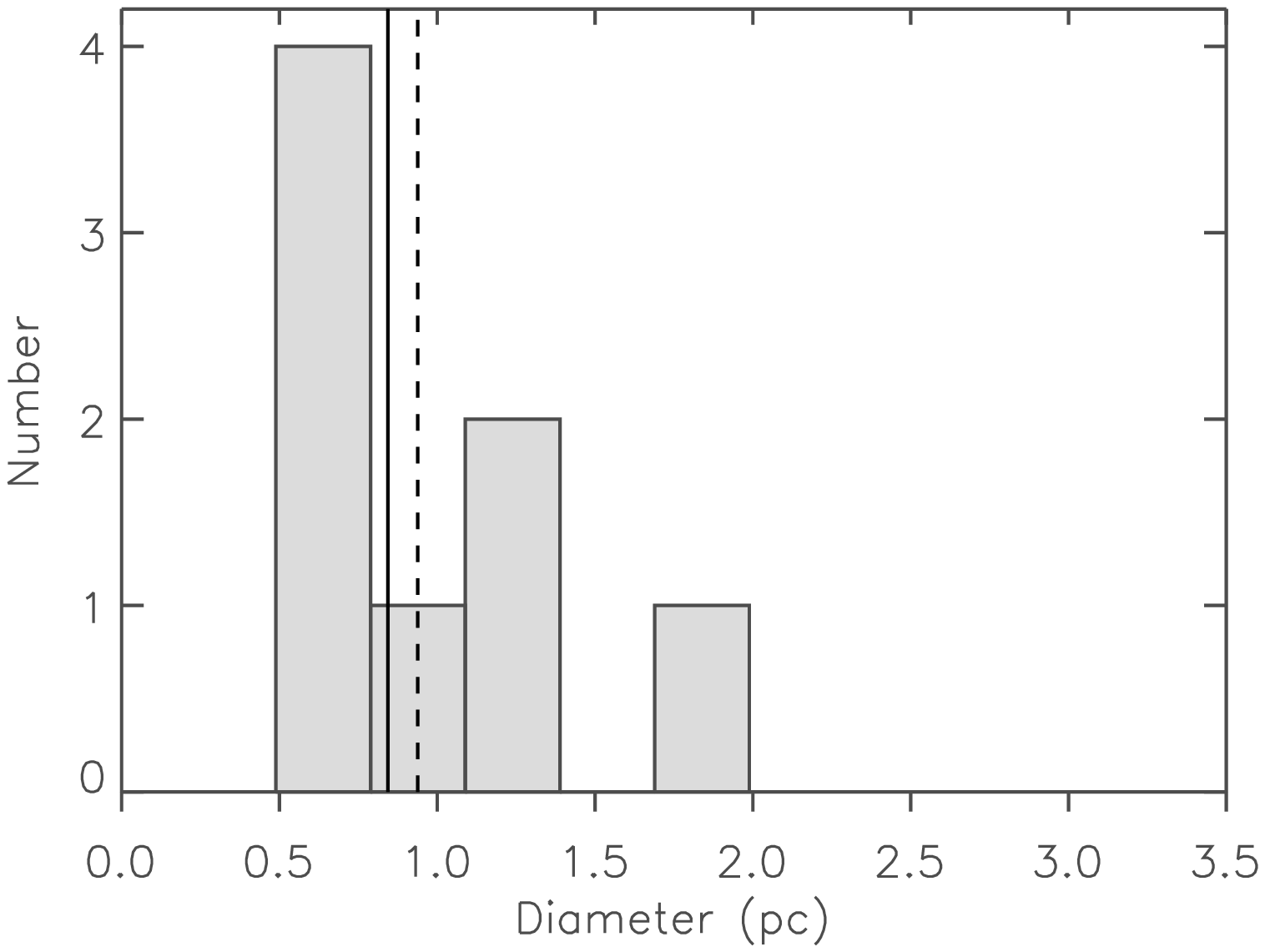}{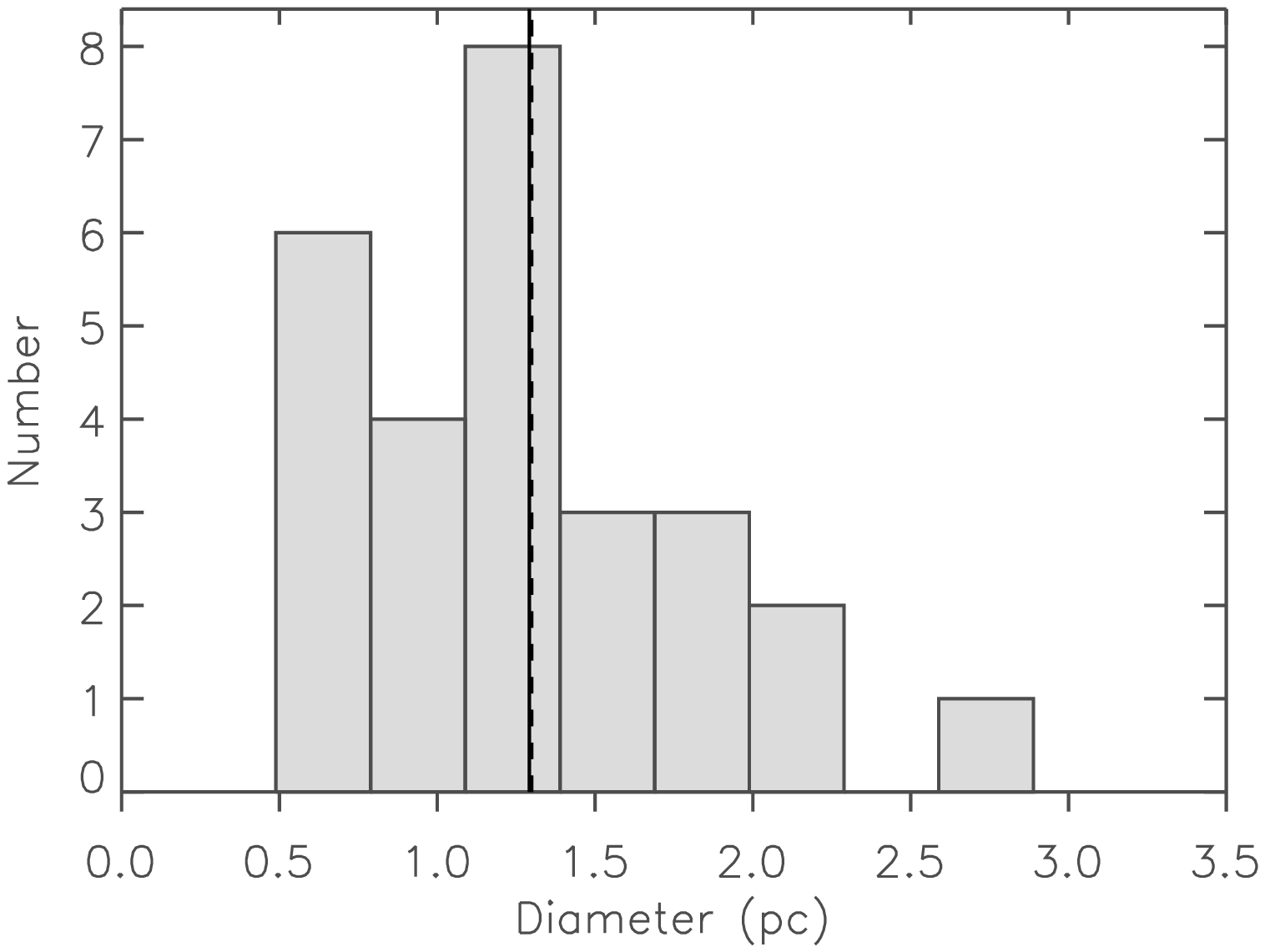}
\caption{(Left) Histogram of diameters for the eight near IM SFRs. The solid line represents the median size (0.8 pc) while the dashed line represents the average size (0.9 pc). (Right) Histogram of diameters of the 27 IM SFRs that have assigned distances. The solid line represents the median size (1.3 pc) while the dashed line represents the average size (1.3 pc). The bin size is 0.3 pc.\label{sizefigure}}
\end{figure}

We can attempt to use the available kinematic distances from the CO associations to convert the angular diameters into (near and far) physical diameters, given in Table~\ref{blobsshells}. We attempted to break the near/far distance degeneracy by considering eight objects among the blobs/shells that are highly likely to be located at the near distance. Seven of them are clearly associated with CO features that morphologically resemble local minima and/or dark lanes in the \emph{Spitzer} 8.0 $\micron$ images. One of them (\emph{IRAS} 18412-0440) is clearly associated with an infrared dark cloud (IRDC) as described in \citet{simon06}. These are likely to be associated with molecular clouds at the near distance, seen in silhouette against the more distant diffuse Galactic emission. The eighth near object (\emph{IRAS} 18180-1342) is on the outskirts of the molecular cloud associated with the Eagle Nebula (M16, NGC 6611). The Eagle Nebula has a photometric distance of $1.8 \pm 0.5$ kpc \citep{bonatto}, putting \emph{IRAS} 18180-1342 at the near distance. Multipanel images of these eight objects are shown as Figures~\ref{18131-1606-figure} to \ref{18564+0145-figure}. Using the angular diameters for these eight near objects, we converted angular sizes into physical diameters. The left panel of Figure~\ref{sizefigure} shows a histogram of the distribution in physical diameter. The diameters range from 0.5 to 1.7 pc with most objects having a diameter less than 1 pc. The median value (solid line) is 0.8 pc and the average value (dashed line) is 0.9 pc. If instead these objects are considered to be at the far distance, the diameters would range from 2.2 to 8.3 with a median value of 3.0 pc and an average value of 3.8 pc. The diameter of $\sim 1$ pc for IM SFRs is consistent with models for PhotoDissociation Regions (PDRs). The \citet{1998ApJ...501..192D} models
predict an equilibrium PDR diameter of $\sim 1$ pc for an IM
star in both a dust-free and a dusty environment. Other PDR models are also consistent with this rough diameter \citep{kerton02}.

Given the small range in physical size among the near distance objects and the consistency with PDR models, we employed a bootstrap {}``Standard Ruler'' method. Using that method, we assigned a more likely distance to objects that cannot be determined to be at the near distance from clouds/dark lanes or association with sources with known distances. We applied the {}``Standard Ruler'' approach to assign objects to the near distance if they have near distance diameters $\geq 0.5$ pc \emph{and} far distance diameters $ > 2.0$ pc. These 14 objects were designated with {}``N'' in Table~\ref{blobsshells}. Objects with near distance diameters $< 0.5$ pc \emph{and} far distance diameters $\leq 2.0$ pc were assigned to be at the far distance. These six objects were designated with {}``F''. Objects at or above tangent point velocities were assigned the tangent point distance and the corresponding diameter. These seven objects were designated with {}``T''. \emph{IRAS} 19012+0505-S was assigned to be near, due to its apparent connection to \emph{IRAS} 19012+0505-N that was assigned the near distance. Fifteen objects remain unassigned. Thirteen of them have both near and far diameters between 0.5 and 2.0 pc, and two objects have near and far diameters that exceed 2.0 pc. These objects were designated {}``U''. The right panel of Figure~\ref{sizefigure} shows the distribution of physical diameters for the 27 IM SFRs that have assigned distances in Table~\ref{blobsshells} (near, far, and tangent). The diameters range from 0.5 to 2.9 pc, with both median (solid line) and average (dashed line) values being 1.3 pc. If we assume that all the unassigned objects are at the near distance, the median diameter drops to 1.0 pc, and the average drops to 1.2 pc. If instead we assume that all the unassigned objects are at the far distance, the median diameter remains at 1.3 pc, and the average increases to 1.4 pc. Regardless of the choice of distances for the remaining unassigned objects, the mean diameter changes by very little.

\section{Spectral Energy Distribution}
\begin{figure}
\plotone{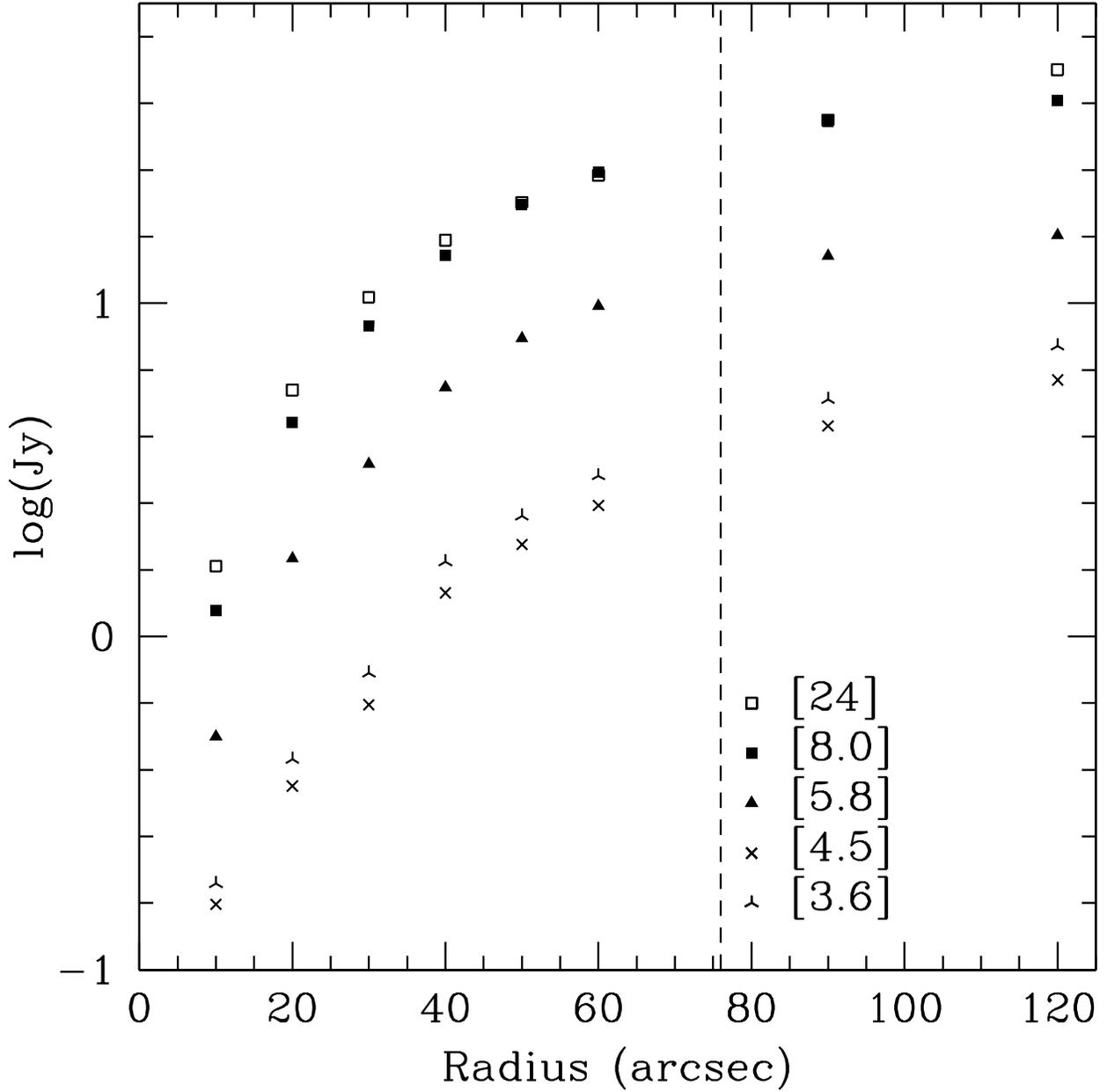}
\caption{Figure showing the integrated flux density of \emph{IRAS} 18180-1342 as a function of aperture radius at each of the four IRAC and the MIPS 24 $\micron$ bandpasses. The radial profiles are similar in each bandpass. The 8.0 and 24 $\micron$ flux densities asymptotically approach $\sim 42$ and $\sim 50$ Jy at a radius of $120 \arcsec$. The chosen aperture radius of $76 \arcsec$ is indicated by the dashed line and encompasses about half of this total.\label{fluxdistr-18180-1342}}
\end{figure}

{\label{sed}}
We performed aperture photometry using circular apertures and
background annuli, defined by eye, on the Spitzer pipeline mosaics based on the morphology of each object in the 8.0 and 24 $\micron$ images. Typical aperture radii range between $14 \arcsec$ and $162 \arcsec$ and are listed in Table~\ref{photometrytable}. The irregular shapes and complex backgrounds near most targets, particularly at 8.0 and 24 $\micron$, in addition to bright stars at the shorter bands, do not easily lend themselves to more quantitative approaches that could uniformly be applied to all bandpasses. Nevertheless, the relatively isolated nature of these objects makes it possible to obtain a meaningful measure of their total flux densities using basic aperture photometry. Figure~\ref{fluxdistr-18180-1342} shows the integrated flux density of one object, \emph{IRAS} 18180-1342, as a function of aperture size, at each of the four IRAC and the MIPS 24 $\micron$ bandpasses. Note that the radial profiles are similar in each bandpass. The 8.0 and 24 $\micron$ flux densities asymptotically approach $\sim 42$ and $\sim 50$ Jy at a radius of $120 \arcsec$. The chosen aperture radius of $76 \arcsec$ encompasses about half of this total. Typically, the aperture defined by eye encompasses about half of the asymptotic flux density, while an extended envelope accounts for the other half. In some cases emission from a more extended envelope, and possibly other unrelated emission along the line of sight, exceeds half of the asymptotic flux density. In an effort to measure only the flux density from the immediate vicinity of the IM SFRs, we adopted the by-eye apertures and note that these produce flux densities that are typically a factor of two (0.3 dex) lower than the asymptotic flux densities. Table~\ref{photometrytable} lists the adopted aperture radii in arcsec and resultant flux densities ($F_{\lambda}$) for each object. Aperture
corrections were applied as given by \emph{Spitzer} SSC webpage handbooks. Specifically, the IRAC aperture correction\footnote{http://ssc.spitzer.caltech.edu/irac/iracinstrumenthandbook/} is of the analytical form $a \cdot e^{-b} + c$ where $b$ depends on aperture and $a$ and $c$ are constants that depend on bandpass. For the 24 $\micron$ MIPS bandpass we estimated the aperture correction\footnote{http://ssc.spitzer.caltech.edu/mips/mipsinstrumenthandbook/} based on aperture size and it is typically 1.06. Uncertainties are dominated
more by systematic effects stemming from aperture definition than statistical
fluctuations, so we conservatively estimated relative systematic uncertainties to be
10\% in each band, and added that in quadrature with the uncertainty in the flux calibration at each bandpass. We used IRAC flux calibration uncertainties of $2\%$ \citep{iraccalibration2005PASP..117..978R} and a MIPS 24 $\micron$ flux calibration uncertainty of $4\%$ \citep{mipascalibration2007PASP..119..994E}. Saturation was not an issue for these sources, and total uncertainties are typically $\sim 10\% $ for \emph{Spitzer} photometry. Since MIPS 70 $\micron$ and 160 $\micron$ mosaics were unavailable at the time of this project, we used the flux densities from the \emph{IRAS} PSC for the 60 and 100 $\micron$ bands ($\sim 15\%$ flux density uncertainties). All flux densities are listed in Table~\ref{photometrytable}. We then plot the final SEDs, from 3.6 to 100 $\micron$, see for example Figure~\ref{18131-1606-figure}, bottom left.

In the vast majority of cases, the flux density in the shorter wavelength IRAC bands is $\sim 1$ Jy and decreases from 3.6 to 4.5 $\micron$. The SED then rises to a few Jy at 5.8 $\micron$ and continues to rise at 8.0 $\micron$. This SED shape is a signature of PAH dominated objects, owing to the lack of PAH lines within the 4.5 $\micron$ band. All SEDs continue to rise through 24 $\micron$, indicating the presence of hot dust. They are still rising in the \emph{IRAS} 60 and 100 $\micron$ bands, typically reaching $F_{\nu}(100) \approx  200$ Jy. Typically the \emph{IRAS} PSC $F_{\nu}(12)$ and $F_{\nu}(25)$ are lower than $F_{8.0}$ and $F_{24}$ found using aperture photometry. This effect is expected for extended sources since the \emph{IRAS} PSC filter surpresses extended emission, thus underestimating the true flux density. This effect is most dramatic at 12 and 25 $\micron$ \citep{fich1996ApJ...472..624F}.

\subsection{Infrared Luminosities}{\label{subsec:irlum}}
From the SEDs and distance estimates, the infrared luminosity ($L_{\mathrm{IR}}$) can be estimated. We estimated the flux by integrating under the curve. For the contribution from the \emph{IRAS} PSC flux densities, the method of \citet{emerson} is used to obtain the contribution to the fluxes. We obtain the luminosities (near and far) by multiplying the fluxes by $4 \pi d^2$, where $d$ is the distance (near and far). These infrared luminosities are listed in Table~\ref{photometrytable}. These tabulated luminosities are lower limits since they do not consider contributions from wavelengths longer than 135 $\micron$ (the edge of the \emph{IRAS} 100 $\micron$ bandpass). Comparison of \emph{IRAS}-derived luminosities with SEDs from the Diffuse Infrared Background Experiment (DIRBE) and the Balloon-borne Large Aperture Submillimeter Telescope (BLAST) suggest that we are missing $\sim 10 \%$ of the total luminosity \citep{kerton2000PhDT.........6K,chapin2008ApJ...681..428C}

\begin{figure}
\plottwo{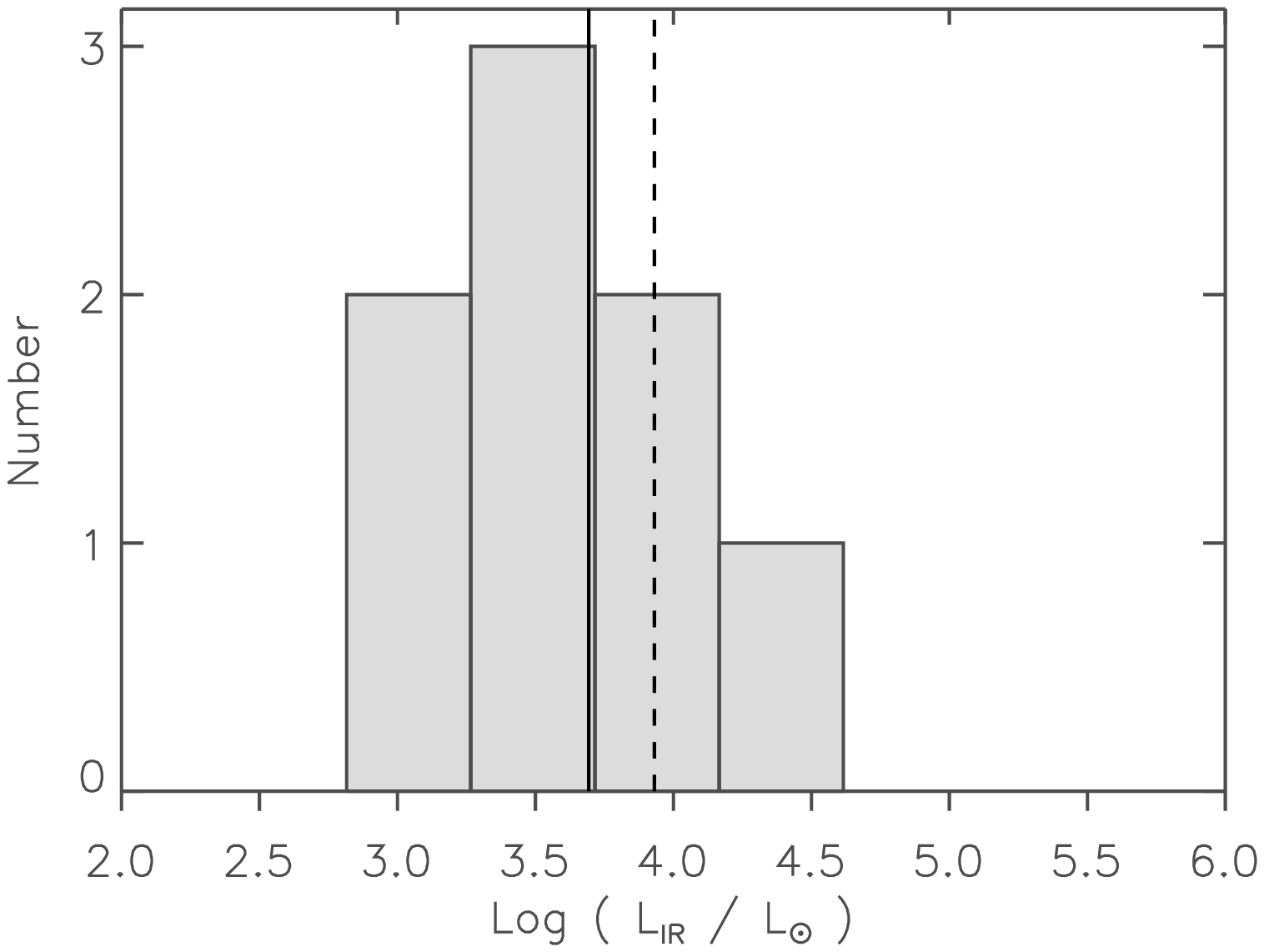}{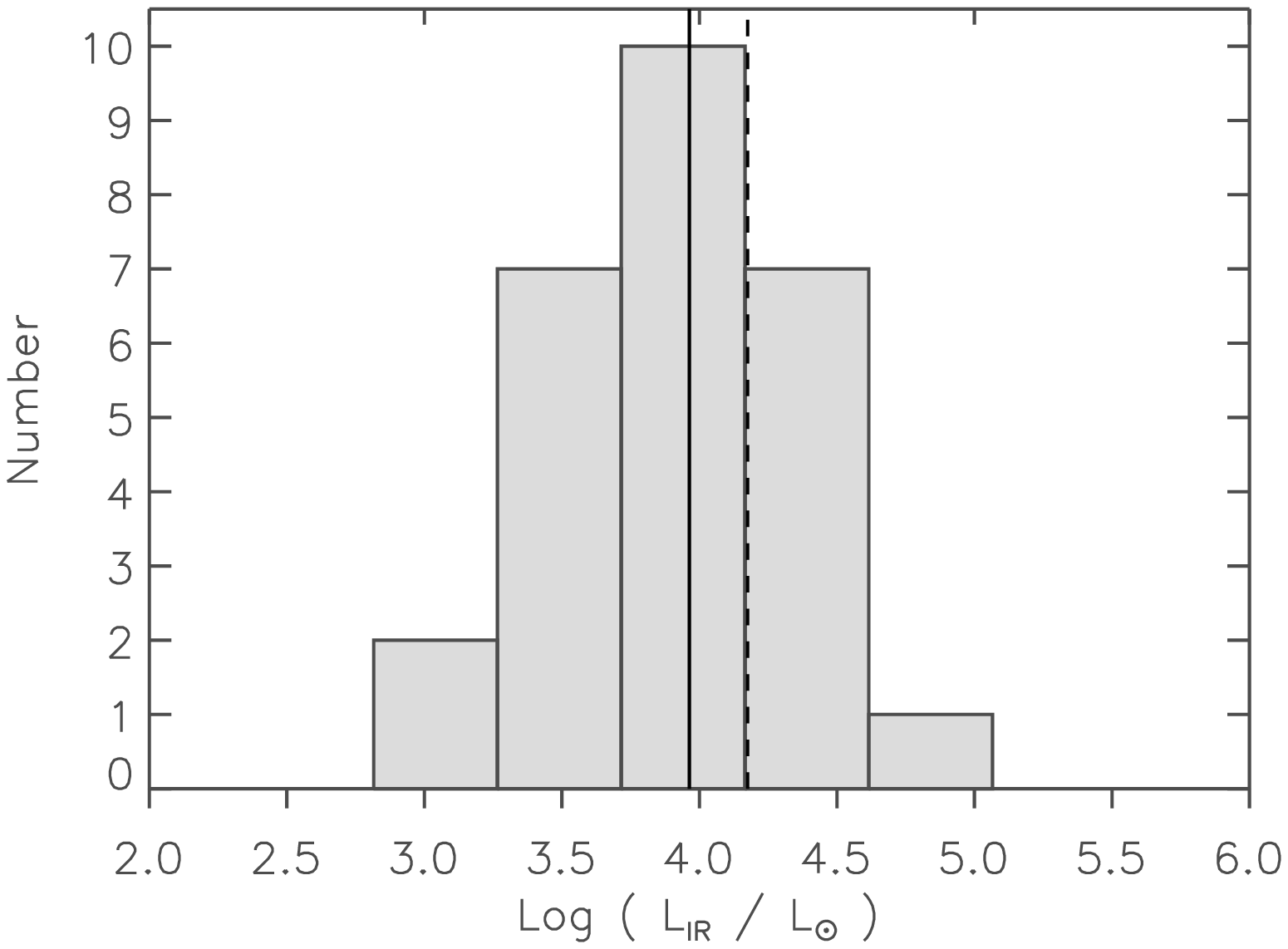}
\caption{(Left) Histogram of $\log ( L_{\mathrm{IR}} / L_{\sun} )$ for the eight near IM SFRs. The solid line represents $\log (\mathrm{median}) = 3.7$ while the dashed line represents $\log (\mathrm{average}) = 3.9$. (Right) Histogram of $\log ( L_{\mathrm{IR}} / L_{\sun} )$ for the 27 IM SFRs that have assigned distances. The solid line represents $\log (\mathrm{median}) = 4.0$ while the dashed line represents $\log (\mathrm{average}) = 4.2$. The bin size is 0.45 dex.\label{lumfigure}}
\end{figure}

The distribution of $L_{\mathrm{IR}}$ for the near eight objects is shown as a histogram in Figure~\ref{lumfigure} (left). The luminosities range from $7 \times 10^2$ to $3.7 \times 10^4$ L$_{\mathrm{\sun}}$. Most are less than $10^4$ solar luminosities, which is reflected in a median value of $4.9 \times 10^3$ L$_{\mathrm{\sun}}$ (solid line) and an average value of $8.4 \times 10^3$ L$_{\mathrm{\sun}}$ (dashed line). If instead these objects were considered to be at the far distance, the luminosities would range from $1.4 \times 10^4$ to $3.3 \times 10^5$ L$_{\mathrm{\sun}}$, with a median value of $7.0 \times 10^4$ L$_{\mathrm{\sun}}$ and an average value of $1.2 \times 10^5$ L$_{\mathrm{\sun}}$.

\begin{figure}
\plotone{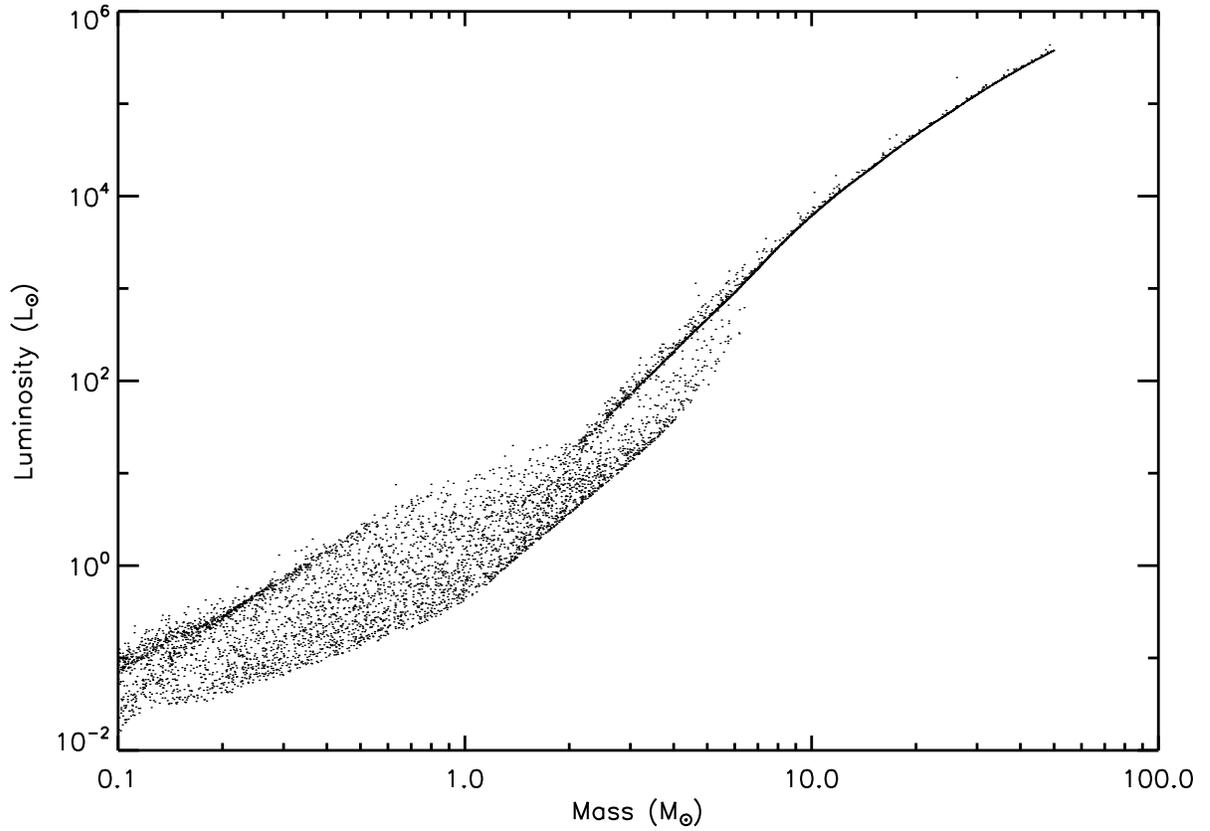}
\caption{\citet{Robitaille} model YSO luminosities of a range of Stage II objects as a function of stellar mass. The $L_{\mathrm{IR}}$ range found in Section~\ref{subsec:irlum}, $10^3$ to $10^4$ L$_{\mathrm{\sun}}$, corresponds roughly to $\sim 4 - 10$ M$_{\sun}$ YSOs, i.e., intermediate-mass.\label{robitaille-figure}}
\end{figure}

The right panel of Figure~\ref{lumfigure} shows a histogram of the distribution of $L_{\mathrm{IR}}$ for the 27 objects that have assigned distances. The luminosities range from $7 \times 10^2$ to $1.1 \times 10^5$ L$_{\mathrm{\sun}}$. The median value is $9.2 \times 10^3$ L$_{\mathrm{\sun}}$ (solid line) and the average value is $1.5 \times 10^4$ L$_{\mathrm{\sun}}$ (dashed line). We find that this luminosity range is consistent with recent models of YSO SEDs \citep{Robitaille}. Figure~\ref{robitaille-figure} shows the model YSO luminosities of a range of objects that have optically thick disks (Stage II) as a function of stellar mass. Evidently, $10^3$ to $10^4$ L$_{\mathrm{\sun}}$ corresponds roughly to $\sim 4 - 10$ M$_{\sun}$ YSOs, i.e., intermediate-mass. In Section~\ref{uchiiIR}, we compare this range in luminosity to luminosities of UC\ion{H}{2} regions, showing that IM SFRs and UC\ion{H}{2} regions are quite distinct sets.

\section{Molecular Material and Masses}{\label{masssection}}
In this section, we investigate the molecular material that is spatially associated with the IM SFRs, using the corresponding zeroth-moment maps. Usually, the location of the maximum value in the zeroth-moment map (i.e., the peak integrated intensity) for the local CO {}``clump'' lies outside of the mid-IR blob/shell seen in \emph{Spitzer} mosaics. Table~\ref{moleculartable} lists the \emph{IRAS} name for sources having a velocity association from Table~\ref{blobsshells}, the peak integrated intensity ($I_{\mathrm{peak}}$), the corresponding peak H$_2$ column density ($N(\mathrm{H_2})_{\mathrm{peak}}$) and peak mass column density ($N(M)_{\mathrm{peak}}$), the solid angle ($\Omega$) occupied by the CO clump (used to calculate the mass), the flag from Table~\ref{blobsshells} indicating distance assumption, the near and far distance clump masses ($M_{\mathrm{LTE}}$), the parent GRS Molecular Cloud (GRSMC), associated GRSMC-clump, the mass of the GRSMC ($M_{\mathrm{Cloud}}$), and a note on the distance determination of the GRSMC.

\subsection{Column densities and masses from CO/H$_2$}{\label{coh2masses}}
To estimate column densities and molecular masses from GRS CO observations, we assumed an optically thin, LTE \citep[CO is almost always close to LTE, see][]{tools} situation for the $^{13}$CO $J = 1 \rightarrow 0$ rotational line. Furthermore, we assumed the excitation temperature is approximately equal the main beam temperature; typically $\sim 10$ K for dense cold molecular material. The derived $^{13}$CO column densities are only weakly sensitive to the exact choice of
the excitation temperature, changing it to 20 K increases the derived mass by 40\% \citep{simon01}. Under these assumptions, the expression for the column density of $^{13}$CO is \citep[eq. 14.40]{tools},
\begin{equation}{\label{eq:13co}}
N({^{13}\mathrm{CO}}) = 7.3 \times 10^{14} \int T_{\mathrm{MB}} \mathrm{d} V ~ \mathrm{cm}^{-2}
\end{equation}
Using conversion factors from \citet{simon01} and references therein (R($^{12}$CO/$^{13}$CO$)=45$ and R($^{12}$CO/H$_2) = 8 \times 10^{-5}$) we calculated the peak H$_2$ column density ($N(\mathrm{H_2})_{\mathrm{peak}}$) in cm$^{-2}$. We got the peak mass column density by multiplying $N(\mathrm{H_2})_{\mathrm{peak}}$ with the H$_2$ molecular mass and a factor of 1.36 to account for elements heavier than hydrogen \citep{simon01}. Summing over the area of the CO clump, we obtained the (near and far distance) clump mass, given in Table~\ref{moleculartable}.

The uncertainty in the mass calculation is dominated by the background\footnote{Using the Dominion Radio Astrophysical Observatory (DRAO) software \textsc{imview}, a polygon incorporating the local CO feature seen in the zeroth-moment maps is drawn by eye. This polygon encloses the area given in Table~\ref{moleculartable}. Typically, the polygon is drawn between $20\%$ to $40\%$ of $I_{\mathrm{peak}}$. The polygon's perimeter defines the background level, and a plane is fitted to the perimeter in order estimate the background across the CO feature. The sum of $N(M)$ above the fitted background plane inside the polygon is used to calculate $M_{\mathrm{LTE}}$.} determination, and is estimated to be $\sim 10 \%$. More importantly, subthermal excitation of higher rotational levels may result in an overestimate of the column density. This overestimate would be of the same order as the underestimate due to moderate opacity effects. This, taken together with the above assumptions, make column densities and derived masses lower limits, that are estimated to be accurate within a factor of a few \citep{simon01}.

\begin{figure}
\plottwo{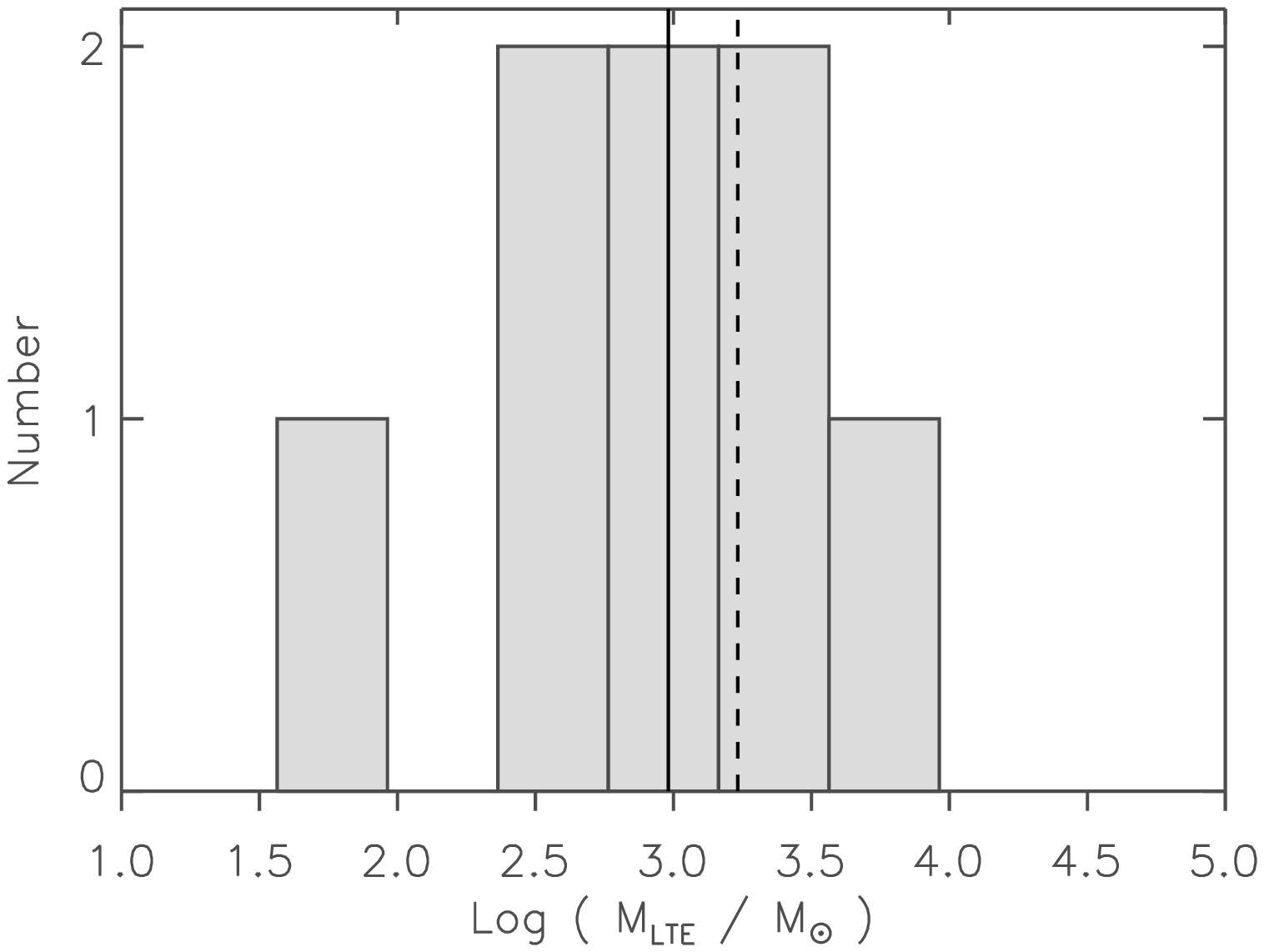}{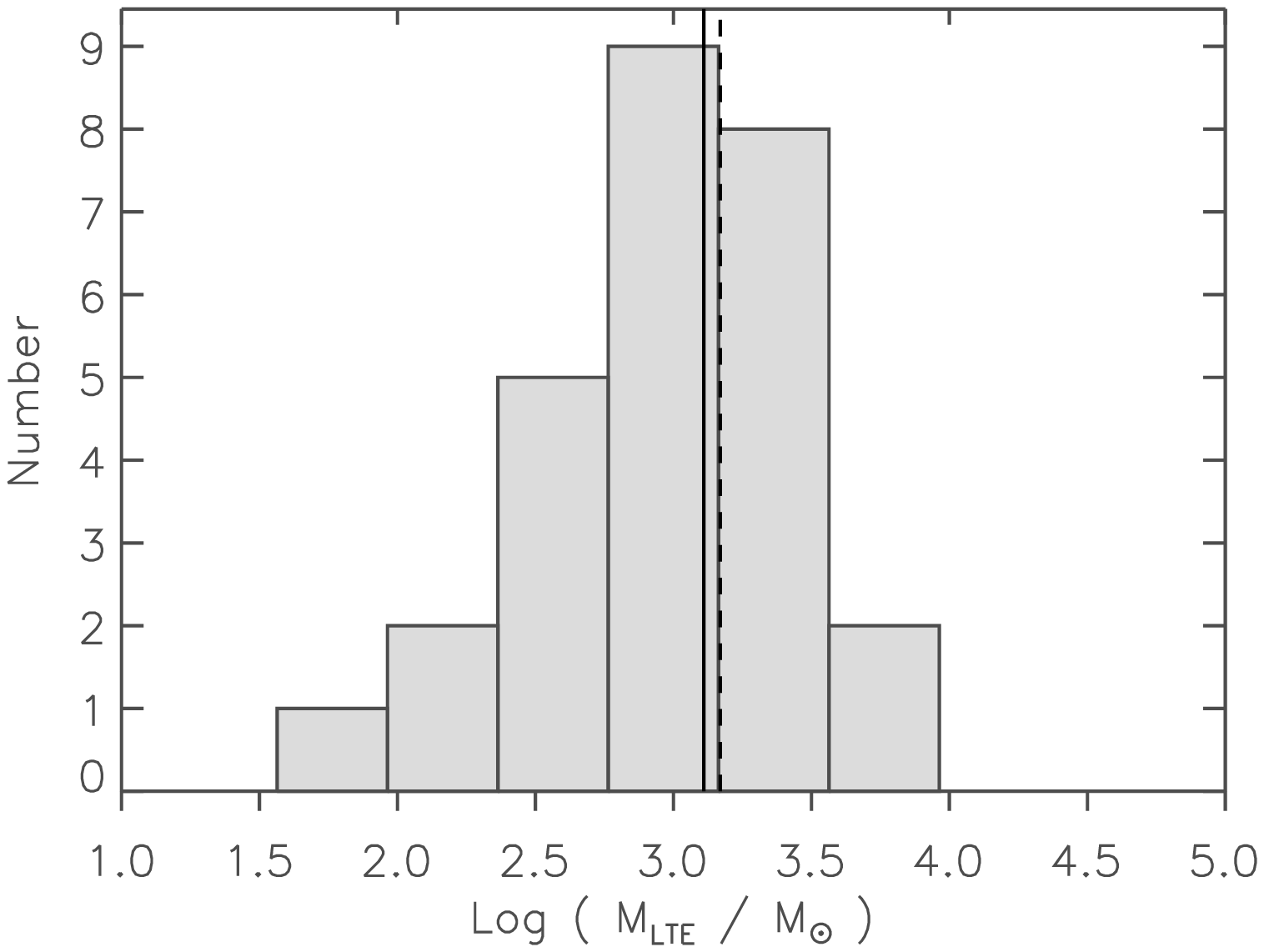}
\caption{(Left) Histogram of $\log ( M_{\mathrm{LTE}} / M_{\sun} )$ for the local molecular material associated with the eight near IM SFRs. The solid line represents $\log (\mathrm{median}) = 3.0$ while the dashed line represents $\log (\mathrm{average}) = 3.2$. (Right) Histogram of $\log ( M_{\mathrm{LTE}} / M_{\sun} )$ for the local molecular material associated with the 27 IM SFRs that have assigned distances. The solid line represents $\log (\mathrm{median}) = 3.1$ while the dashed line represents $\log (\mathrm{average}) = 3.2$. The bin size is 0.4 dex.\label{molfigure}}
\end{figure}

The median and average peak mass column density obtained this way is 0.03 g~cm$^{-2}$ with a maximum of 0.08 g~cm$^{-2}$. The mass of the local molecular material associated with the eight near sources, $M_{\mathrm{LTE}}$, range from 40 M$_{\sun}$ to 2800 M$_{\sun}$, and Figure~\ref{molfigure} (left) shows a histogram of this distribution. The median of the masses is 900 M$_{\sun}$ (solid line) and the average is 1700 M$_{\sun}$ (dashed line). If instead these objects were considered to be at the far distance, the masses would range from $1.8 \times 10^3$ M$_{\sun}$ to $5.6 \times 10^5$ M$_{\sun}$, with a median value of $1.4 \times 10^4$ M$_{\sun}$ and an average value of $1.9 \times 10^4$ M$_{\sun}$. The right panel of Figure~\ref{molfigure} shows a histogram of the distribution of $M_{\mathrm{LTE}}$ for the molecular material associated with the 27 objects that have assigned distances. The masses range from 40 M$_{\sun}$ to 6100 M$_{\sun}$, with a median of 1300 M$_{\sun}$ (solid line) and an average of 1500 M$_{\sun}$ (dashed line).

\subsection{Column densities and masses from BGPS}{\label{bgpsmasses}}
Out of the 50 objects, the CO material is spatially associated with a millimeter-wave continuum source from the BGPS catalog \citep{bgpscatalog} in 28 objects, 10 objects have no BGPS source associated, and 12 objects are not covered by the BGPS. Typically for an association, the BGPS continuum source is located on or very near the location of $I_{\mathrm{peak}}$. For some objects there are several BGPS sources within the CO feature, but Table~\ref{bgpstable} only lists the source closest to $I_{\mathrm{peak}}$. The peak brightness within the BGPS source, $F_{\mathrm{peak}}$,  is related to the peak column density, $N(\mathrm{H_2})_{\mathrm{peak}}$, through the conversion factor of $N(\mathrm{H_2})/F = 6.77 \times 10^{19}$ cm$^{-2}$~(mJy beam$^{-1})^{-1}$ \citep[Table A.1 in][]{kauffmann2008A&A...487..993K}. From the the peak H$_2$ column density traced by millimeter continuum emission we obtain the peak mass column density as in Section~\ref{coh2masses}. We note that column densities derived in this manner are lower limits for unresolved sources in the BGPS. The peak mass column densities traced by the BGPS are typically $\sim 0.1$ g~cm$^{-2}$, with a minimum of 0.05 g~cm$^{-2}$ and a maximum of 0.5 g~cm$^{-2}$. These peak mass column densities are higher than the corresponding peak mass column densities determined using the GRS. This is likely an effect of $^{13}$CO becoming optically thick in the denser regions.

\begin{figure}
\plotone{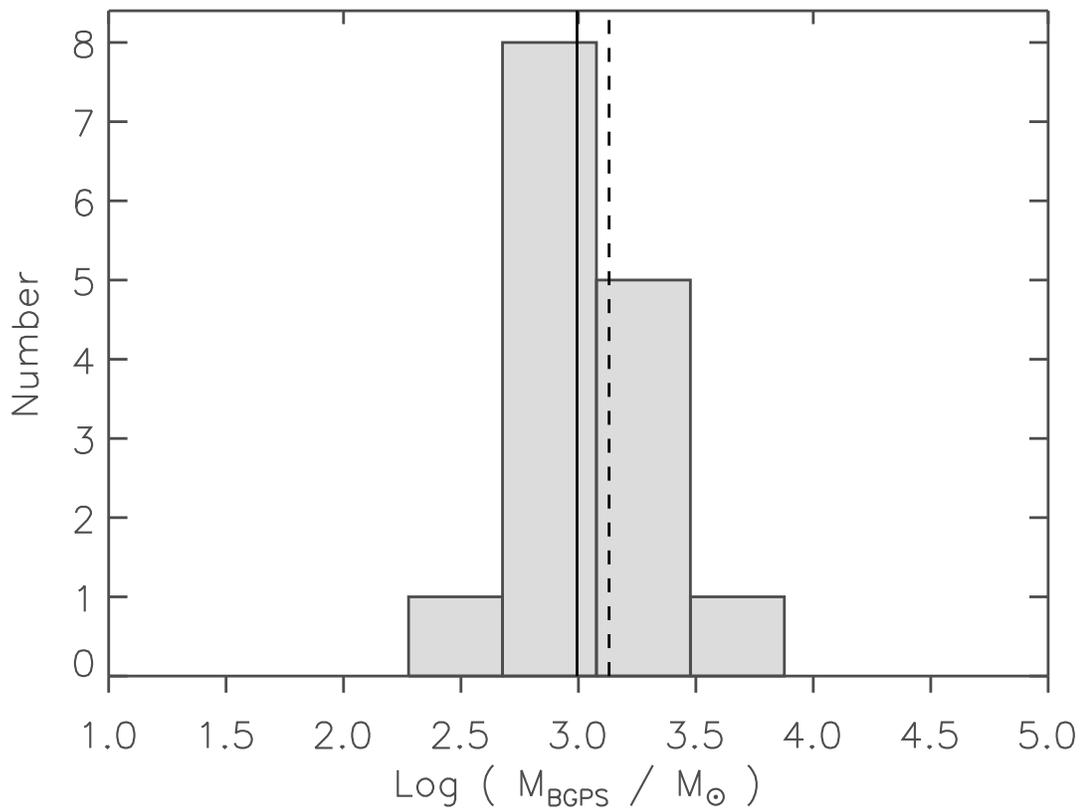}
\caption{Histogram of $\log ( M_{\mathrm{BGPS}} / M_{\sun} )$ for the 15 objects with assigned distances that are associated with BGPS sources. The solid line represents $\log (\mathrm{median}) = 3.0$ while the dashed line represents $\log (\mathrm{average}) = 3.1$. The object with the highest $M_{\mathrm{BGPS}}$ in the figure is \emph{IRAS} 18502-0018, an object dominated by an \ion{H}{2} region (see Section~\ref{discussion}). The bin size is 0.4 dex.\label{bgpsmassfigure}}
\end{figure}

Out of the 28 BGPS sources, 22 have velocities from CO associations. 15 of them have distance determinations while seven do not. Using the integrated flux density $S$ from the BGPS catalog together with our distance estimates, the source masses can be estimated using
\citep[eq. 10]{bgpscatalog},
\begin{equation}{\label{eq:bgpsmass}}
M = 13.1 \mathrm{M}_{\sun} \left( \frac{d}{1 ~\mathrm{kpc}} \right)^2 \left( \frac{S}{1~\mathrm{Jy}} \right) \left[ \frac{\exp(13.0~\mathrm{K}/T)-1}{\exp(13.0/20)-1} \right]
\end{equation}
where we use $T=10$ K like before. The results are given in Table~\ref{bgpstable}. Figure~\ref{bgpsmassfigure} shows a histogram of the distribution of $M_{\mathrm{BGPS}}$ for the 15 objects with distance determinations. The masses range from 200 M$_{\sun}$ to 6700 M$_{\sun}$, with a median of 1000 M$_{\sun}$ (solid line) and an average of 1400 M$_{\sun}$ (dashed line). This result largely agrees with CO-based mass estimates.

\subsection{Molecular Clouds}{\label{molecularclouds}}
To investigate the possibility that IM SFRs are only found in low mass molecular clouds, we use the catalog of GRS molecular clouds \citep[GRSMCs;][]{rathborne09} to identify the large scale molecular clouds containing the IM SFRs. The identification was made by comparing the locations and velocities of so called {}``clumps'' in \citet{rathborne09} to our identifications of molecular material. An identification was made if the clump has a tabulated velocity within $\sim 1$ km~s$^{-1}$ of the velocity listed in Table~\ref{blobsshells} and has a position centered on or very near the location of $I_{\mathrm{peak}}$. In 28 out of 42 cases we were able to associate a clump with a particular molecular cloud (see Table~\ref{moleculartable}).

\begin{figure}
\plotone{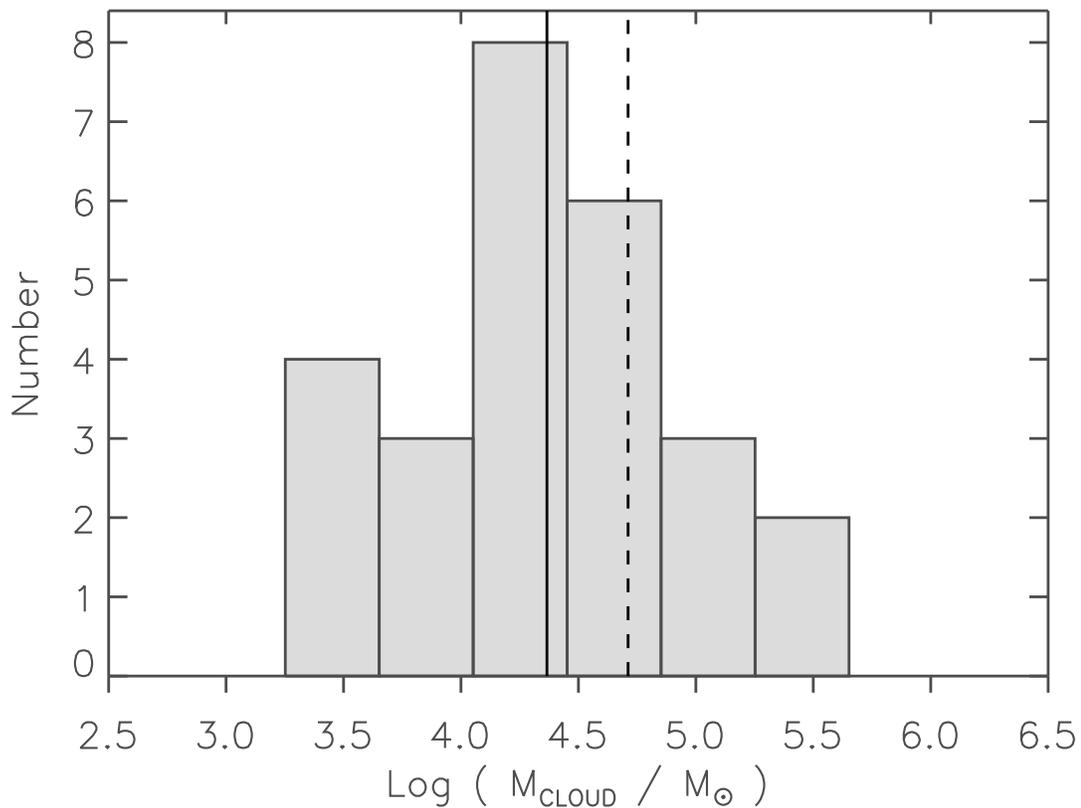}
\caption{Histogram of $\log ( M_{\mathrm{Cloud}} / M_{\sun} )$ for the 26 GRSMCs associated with the  IM SFRs. The four GRSMCs listed as (!) in Table~\ref{moleculartable} have been scaled to the near distance using $d_{near}$ from Table~\ref{blobsshells}. The solid line represents $\log (\mathrm{median}) = 4.4$ while the dashed line represents $\log (\mathrm{average}) = 4.7$. The bin size is 0.4 dex.\label{grsmcmassfigure}}
\end{figure}

\citet{roman-duval} lists distances to 26 of the 28 GRSMCs we found to be associated with IM SFRs. We note that none of these GRSMCs have strong 21 cm continuum sources associated with them, therefore the kinematic distance ambiguity was resolved by the presence or absence of \ion{H}{1} self-absorption (HISA). Only four of the 26 GRSMC distance determinations disagree with our IM SFR standard ruler method determinations, they are marked as (!) in Table~\ref{moleculartable}. We use the $^{13}$CO luminosity ($L(^{13}$CO)) of the cloud from \citet{roman-duval} and the same conversion factors used in Section~\ref{coh2masses} to compute the cloud mass. The results of the calculations are listed in Table~\ref{moleculartable}. Figure~\ref{grsmcmassfigure} shows a histogram of the 26 GRSMC masses where the four GRSMCs listed as (!) in Table~\ref{moleculartable} have been recalculated using the near distance $d_{near}$ from Table~\ref{blobsshells}. The GRSMC masses range from 1800 M$_{\sun}$ to $2.3 \times 10^5$ M$_{\sun}$, with a median of $2.3 \times 10^4$ M$_{\sun}$ (solid line) and an average of $5.1 \times 10^4$ M$_{\sun}$ (dashed line). These fall toward the lower end of the range in mass expected of Giant Molecular Clouds (GMCs; $10^4 - 10^6$ M$_{\sun}$).

\section{Comparison to UC\ion{H}{2} regions}{\label{uchiisection}}
UC\ion{H}{2} regions are manifestations of newly formed massive stars that are still embedded in their natal molecular clouds. They are small ($< 0.1$ pc), dense ($>10^4$ cm$^{-3}$), bright photoionized nebulae. At \emph{IRAS} 100 $\micron$ they are among the brightest objects in the Galaxy because of the warm dust envelope converting the entire stellar luminosity into FIR radiation. We compare these relatively well-studied objects to our IM SFRs and show that UC\ion{H}{2} regions are quite distinct from IM SFRs in luminosity, in column density, and in the mass of associated material traced with millimeter-wave continuum emission.

\subsection{UC\ion{H}{2} region luminosities}
\begin{figure}
\plotone{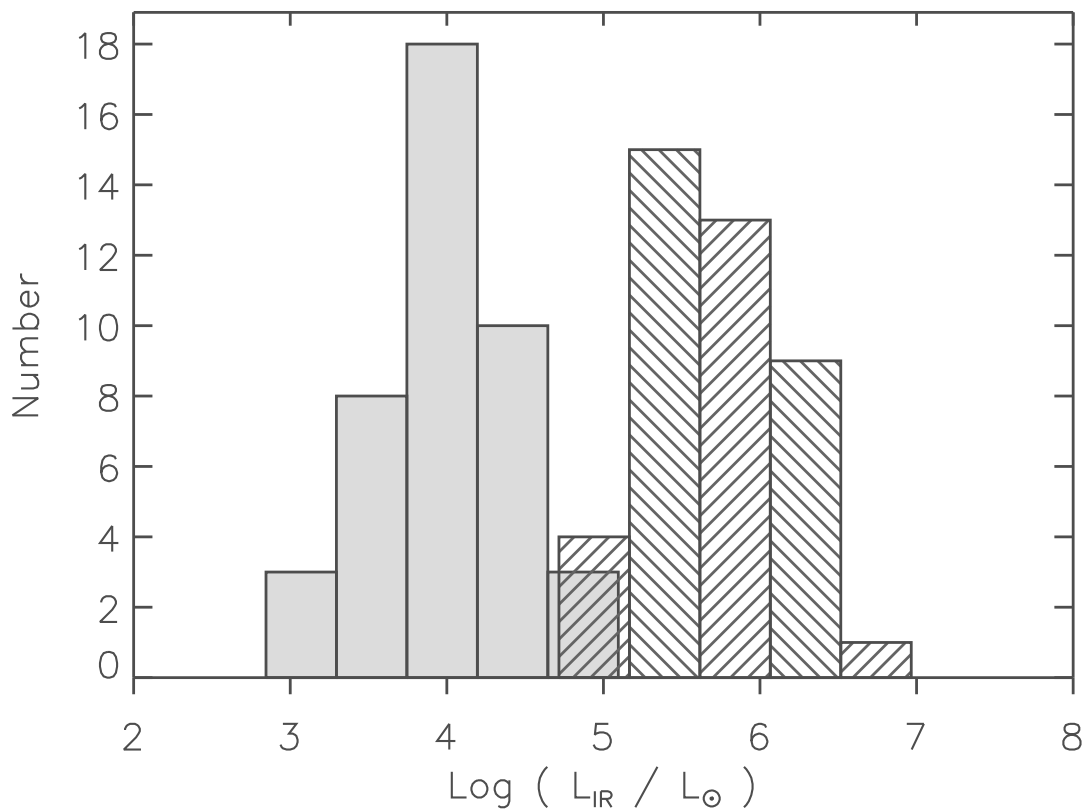}
\caption{Histogram of $\log ( L_{\mathrm{IR}} / L_{\sun} )$ for 42 IM SFR luminosities (shaded, with unassigned objects assumed to be at the far distance) and for 42 UC\ion{H}{2} region luminosities (barred) from \citet{uchii1989ApJS...69..831W}. The luminosity distribution is bi-modal, with UC\ion{H}{2} regions and IM SFRs forming two distinct populations separated by roughly an order of magnitude. The IM SFR overlap consists of unassigned objects and \emph{IRAS} 18502-0018, an object dominated by an \ion{H}{2} region (see Section~\ref{discussion}). The bin size is 0.45 dex.\label{uchiilumfigure}}
\end{figure}

{\label{uchiiIR}}
Figure~\ref{uchiilumfigure} shows a histogram of the infrared luminosities ($L_{\mathrm{IR}}$) of the IM SFRs (shaded), with all the objects labeled Unassigned ({}``U'' in Table~\ref{blobsshells}) assumed to be at the far distance (as a most extreme case scenario). The barred histogram shows $L_{\mathrm{IR}}$ for 42 of the UC\ion{H}{2} regions in Table 18 of \citet{uchii1989ApJS...69..831W}. In both cases contributions to $L_{\mathrm{IR}}$ from wavelengths longer than the edge of the \emph{IRAS} 100 $\micron$ bandpass were not considered, and so in both cases $L_{\mathrm{IR}}$ are lower limits to the bolometric luminosities. The luminosity distribution is bi-modal, with UC\ion{H}{2} regions and IM SFRs forming two distinct populations separated by roughly an order of magnitude. In fact, the overlapping region at the higher end of the IM SFR luminosity distribution (shaded) consists of objects with unassigned distances (which were all assumed at the far distance for the figure). If they instead were assumed to be at near distance only one object would overlap with the lower end of the UC\ion{H}{2} region distribution, and that object turns out to be dominated by an \ion{H}{2} region (\emph{IRAS} 18502-0018, see Section~\ref{discussion}). Comparing the median (or average) shows that UC\ion{H}{2} regions are typically 40--80 times more luminous than IM SFRs.

\subsection{UC\ion{H}{2} regions in the BGPS}
\begin{figure}
\plotone{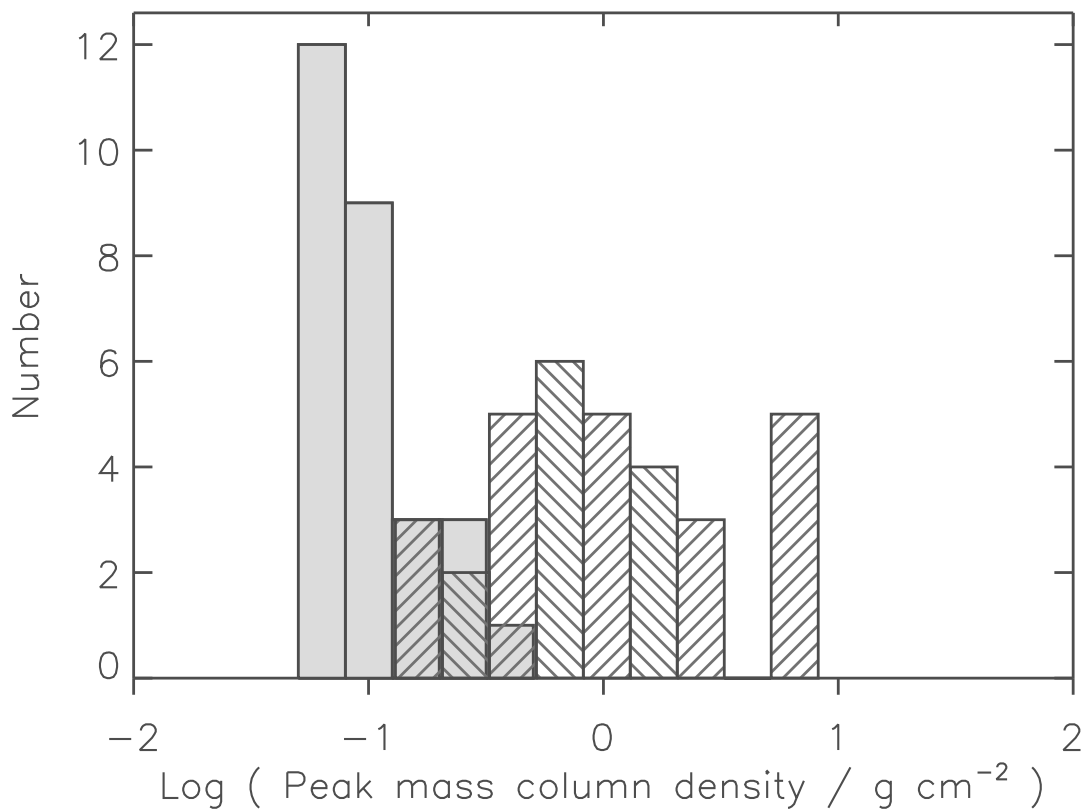}
\caption{Histogram of the peak mass column density, $\log ( N(M)_{\mathrm{peak}} / \mathrm{g~cm}^{-2})$ for the material associated with 28 IM SFRs, traced by millimeter-wave continuum (shaded) and for the material associated with 33 UC\ion{H}{2} regions (barred). The mass column density distribution is roughly bi-modal, with UC\ion{H}{2} regions and IM SFRs forming two populations separated by roughly an order of magnitude. The bin size is 0.2 dex.\label{uchiibgpscolumndensityfigure}}
\end{figure}

{\label{uchiibgps}}
We examined the positions of UC\ion{H}{2} regions with adopted distances \citep[Table 4 in][]{uchii1989ApJS...69..831W} by eye for association with a millimeter-wave continuum source from the BGPS catalog \citep{bgpscatalog}. For 33 UC\ion{H}{2} regions, an associated BGPS source is found. Using the same technique used for IM SFRs in Section~\ref{bgpsmasses} we calculate peak mass column densities toward these UC\ion{H}{2} regions. Figure~\ref{uchiibgpscolumndensityfigure} shows peak mass column densities for IM SFRs (shaded) and UC\ion{H}{2} regions (barred). The peak mass column density distribution is roughly bi-modal, with UC\ion{H}{2} regions and IM SFRs forming two populations separated by roughly an order of magnitude. Comparing the median (or average) shows that peak mass column densities for the material associated with UC\ion{H}{2} regions are typically $\sim10$ times greater than for IM SFRs.

\begin{figure}
\plotone{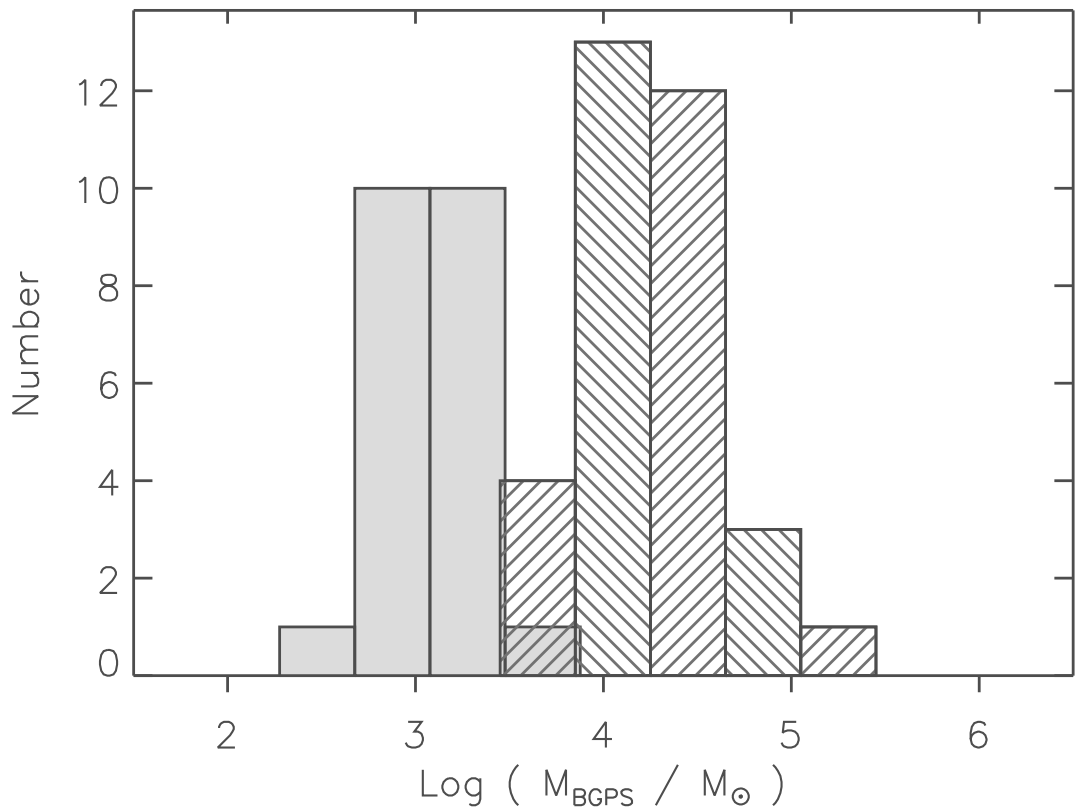}
\caption{Histogram of $\log ( M_{\mathrm{BGPS}} / M_{\sun} )$ for the material associated with 22 IM SFRs, traced by millimeter-wave continuum (shaded, with unassigned objects assumed to be at the far distance) and for the material associated with 33 UC\ion{H}{2} regions (barred). The mass distribution is bi-modal, with UC\ion{H}{2} regions and IM SFRs forming two distinct populations separated by roughly an order of magnitude. The IM SFR overlap consists of \emph{IRAS} 18502-0018, an object dominated by an \ion{H}{2} region (see Section~\ref{discussion}). The bin size is 0.4 dex.\label{uchiibgpsmassfigure}}
\end{figure}

We use equation~\ref{eq:bgpsmass} with the same assumptions as in Section~\ref{bgpsmasses} together with the adopted distances from Table 4 in \citet{uchii1989ApJS...69..831W} to compute the masses of the cold, dense material associated with the UC\ion{H}{2} regions. The masses of the material associated with UC\ion{H}{2} regions is shown as the barred histogram in Figure~\ref{uchiibgpsmassfigure} while the shaded histogram is the mass of the material closely associated with the IM SFRs (from Table~\ref{bgpstable}, with unassigned objects being assumed to be at the far distance). As with the distributions of luminosities and peak mass column densities, the masses of the closely associated material form a bi-modal distribution, with UC\ion{H}{2} regions and IM SFRs forming two populations separated by roughly an order of magnitude. The one overlapping shaded object is again \emph{IRAS} 18502-0018. Comparing the median (or average) of the two populations shows material associated with UC\ion{H}{2} regions is typically $\sim20$ times more massive than the material associated with IM SFRs.

\section{Discussion}{\label{discussion}}
Radio continuum images from the NRAO/VLA Sky Survey \citep[NVSS;][]{condonnvss1998AJ....115.1693C} at 1.42 GHz show that all but one of the investigated blobs/shells lack 1.42 GHz continuum emission from \ion{H}{2}. NVSS has a $45 \arcsec$ FWHM resolution and rms brightness uncertainties of about $\sim0.45$ mJy~beam$^{-1} \approx 0.14$ K (Stokes I). The completeness limit is about 2.5 mJy. Radio continuum measurements of \ion{H}{2} regions can provide a lower limit on the total number of ionizing photons ($N_L$),
\begin{equation}{\label{eq:N_L}}
N_L \geq 7.5 \times 10^{43} S_{\nu} d^2 \nu^{0.1} T_e^{-0.45} ~\textrm{s}^{-1}\\
\end{equation}
where $\nu$ is the frequency in GHz, $S_{\nu}$ is the flux density measured at frequency $\nu$ in mJy, $d$ is the distance to the source in kpc, and $T_e$ is the electron temperature in units of $10^4$ K \citep{rud96}. This assumes an optically thin, spherical, constant-density \ion{H}{2} region. Given the adopted distances in Table~\ref{blobsshells}, a $3 \sigma$ NVSS detection would correspond to about $\log(N_L)\sim 45 $ (with a maximum of 46.1). For comparison, a B0.5 V star powered \ion{H}{2} region has $\log(N_L) = 47.77$ \citep{Schaerer1997A&A...322..598S} and would easily be detected at a $3 \sigma$ level in NVSS if $d \lesssim 75$ kpc.

Highly embedded massive stars can potentially be radio quiet due to the absorption of ultra-violet photons by surrounding dust. We think it is very unlikely that we are probing massive stars at such a very early stage because our sample's range in luminosity, total molecular mass and mass column density are all much lower than those associated with regions of known massive star formation. We conclude that the lack of radio continuum emission in these sources is consistent with our interpretation of them as IM SFRs. To confirm the IM nature of these sources, direct IR spectroscopical observations of the YSO content in these sources are being planned.

The only apparent exception, \emph{IRAS} 18502-0018, actually consists of a $0.4 \arcmin$ blob located on the edge of a dusty $2.5 \arcmin$ diameter \ion{H}{2} region. Applying the standard ruler technique to the $0.4 \arcmin$ blob results in a distance of 11.1 kpc, which agrees with the distance found for the associated GRSMC \citep{roman-duval}. The 1.42 GHz continuum emission associated with the \ion{H}{2} region has a flux density of $S_{\nu} = 126.7 \pm 5.1$ mJy in the NVSS catalog. This corresponds to $\log(N_L) = 48.1$ or a single B0 V star \citep{Schaerer1997A&A...322..598S}. We interpret this region as a possible IM SFR sitting on the edge of a dusty shell surrounding a small \ion{H}{2} region.

We found that the large-scale molecular environment associated with IM SFRs is similar to that associated with high-mass star formation; IM SFRs are found in molecular clumps of $\sim 10^3$ M$_{\sun}$ sitting within Giant Molecular Clouds (GMCs) with a range of masses between $10^4 - 10^5$ M$_{\sun}$. The implication of this result is that massive star formation, while possible in GMCs in this mass range, is not inevitable. IM SFRs are also associated with lower mass $\sim 10^2$ M$_{\sun}$ clumps, much lower than those associated with high mass star formation \citep{zinn07}. In contrast, no IM SFRs are found in GMCs with masses $\gtrsim 10^6$ M$_{\sun}$, suggesting that for the most massive GMCs, high-mass stars will always form.

\begin{figure}
\plottwo{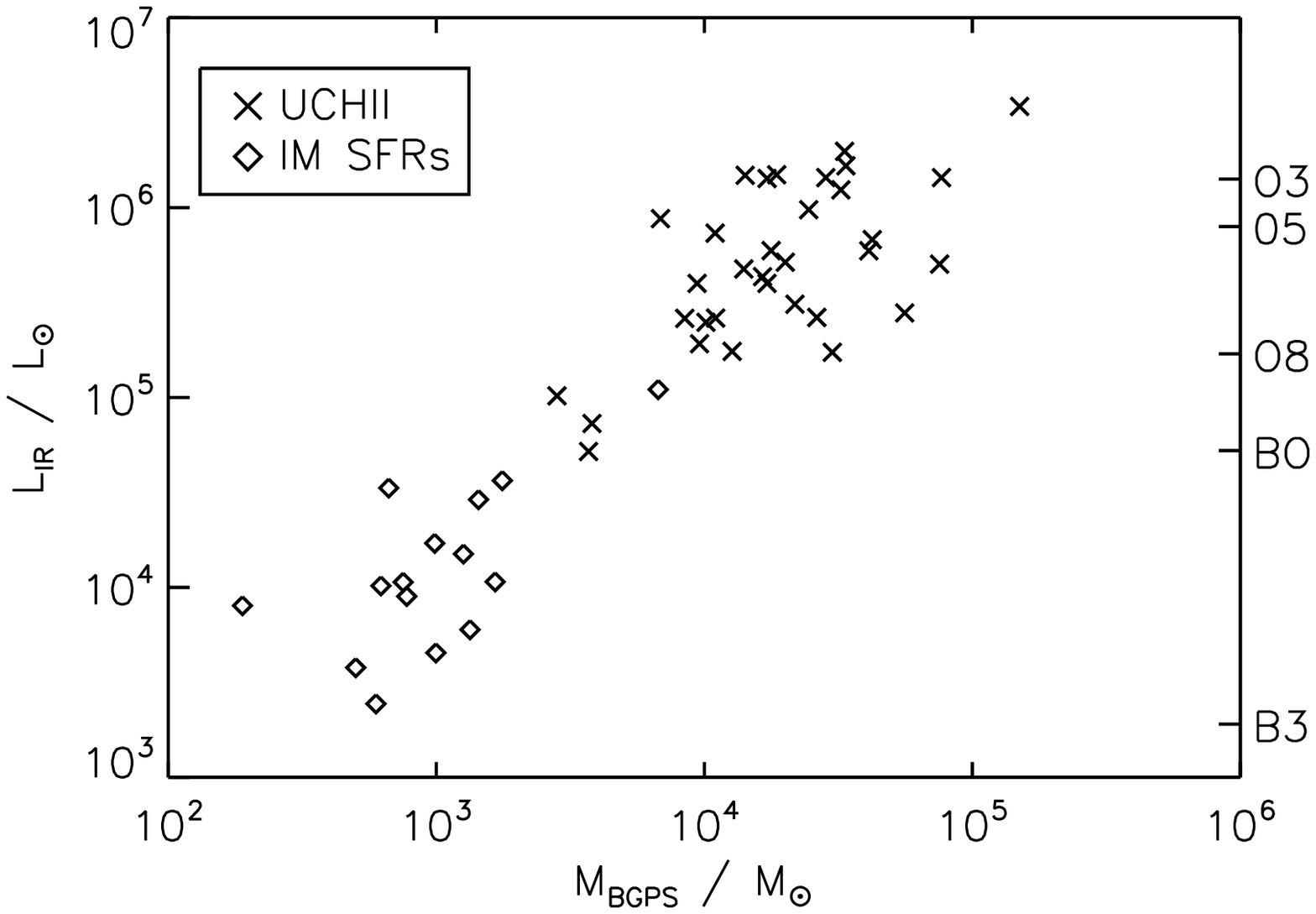}{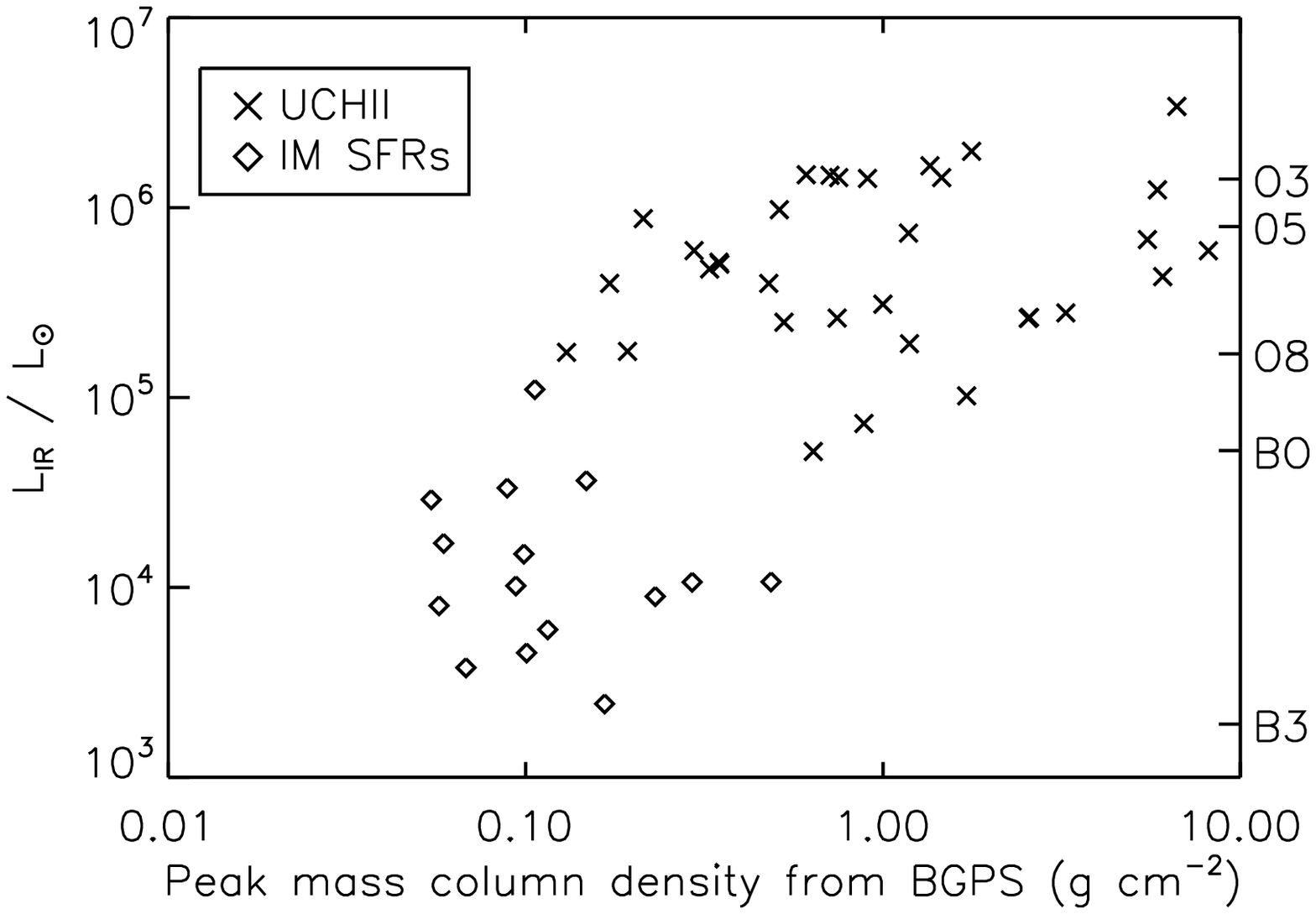}
\caption{(Left) Luminosity versus associated mass. (Right) Luminosity versus the peak mass column density, traced by millimeter-wave continuum. IM SFRs (15 objects) are diamonds. UC\ion{H}{2} regions (33 objects) are crosses. The scale on the right is spectral type of a single class V star.\label{luminosityvscolumndensityfigure}}
\end{figure}

The left hand panel of Figure~\ref{luminosityvscolumndensityfigure} shows a strong correlation between luminosity and associated clump mass for both IM SFRs and UC\ion{H}{2} regions. The data are fit by $L \propto M^{1.12 \pm 0.08}$ which agrees with the mass-luminosity relationship found by \citet{chini1987A&A...181..378C} that was derived using only a sample of compact \ion{H}{2} regions. This relationship clearly holds over three orders of magnitude, a result only hinted at in \citet{chini1987A&A...181..378C}.

In the right panel of Figure~\ref{luminosityvscolumndensityfigure} we plot luminosity versus peak mass column density for our sample of IM SFRs and UC\ion{H}{2} regions. \citet{2008Natur.451.1082K} provide a theoretical expectation that mass column densities $\gtrsim 1$ g~cm$^{-2}$ are needed for massive star formation. We observe the maximum mass column density for IM SFRs is 0.5 g~cm$^{-2}$ and both the median and the average are $\sim 0.1$ g~cm$^{-2}$, all well below the proposed threshold. A simple interpretation of the data shown in the right panel of Figure~\ref{luminosityvscolumndensityfigure} is that if the mass column density is above 1 g~cm$^{-2}$, only massive stars will form, below 1 g~cm$^{-2}$ you can form intermediate through massive stars, and at or below 0.1 g~cm$^{-2}$, no massive stars will form.

As this paper went to press, we became aware of the methanol maser
survey of \citet{pandianII-2007ApJ...669..435P} who report methanol masers near two of the
candidate IM SFRs studied here: 19012+0505-N (as 38.92-0.36) and 19049+0712 (as 41.23-0.20). These two IM SFRs, along with 18224-1228, were also detected in previous methanol maser surveys \citep{szymczak-2000A&AS..143..269S,pestalozzi-2005A&A...432..737P}. Methanol masers are believed to occur during the early
phases of massive star formation.  Association with some of the IM SFRs
suggests either that some of our sample objects are more massive (and
possibly more distant) than inferred here, or that intermediate-mass
star-forming regions are also capable of producing maser activity.
Interestingly, these three sources having methanol maser associations
(18224-1228, 19012+0505N, and 19049+0712) are flagged in Table~\ref{moleculartable} as
having previous distance determinations \citep[based on the absence of HISA;][]{roman-duval,pandianIII-2009ApJ...706.1609P} that
would place them at the far kinematic distance, resulting in larger
luminosities and molecular masses that would make them similar to some
of the least luminous UC\ion{H}{2} regions in our comparison
sample.

\section{Conclusions}{\label{conclusions}}
We have identified a sample of 50 IM SFRs in the inner Galaxy. The IM SFRs share some basic properties. They have typical luminosities of $\sim 10^4$ L$_{\sun}$ and the PDRs that demarcate the IM SFRs have typical diameters of $\sim 1$ pc. All but one lack radio continuum emission and some show small stellar clusters visible in the NIR. These properties are consistent with these objects being regions containing stars in the $2-8$ M$_{\sun}$ range with their associated clusters of low-mass stars. The most massive stars in these regions are likely to be precursors to Herbig AeBe stars.

IM SFRs do show some similarities with high-mass star-forming regions. On parsec-scales IM SFRs are typically associated with molecular clumps of mass $\sim 10^3$ M$_{\sun}$ and on larger scales they are found within GMCs with masses between $10^4-10^5$ M$_{\sun}$. We conclude that massive star formation in GMCs in that mass range is not inevitable.

IM SFRs are also distinct from regions of more massive star formation, forming in clumps with masses as low as $\sim 10^2$ M$_{\sun}$ and not being found in the most massive GMCs ($\gtrsim 10^6$ M$_{\sun}$). IM SFRs typically have an order of magnitude less luminosity than UC\ion{H}{2} regions.

Our sample of IM SFRs, combined with UC\ion{H}{2} region data, shows that a strong correlation between luminosity and associated molecular mass in star-forming regions applies over three orders of magnitude encompassing both IM SFRs and UC\ion{H}{2} regions.

The molecular material associated with IM SFRs typically have an order of magnitude smaller peak mass column density and clump mass compared to UC\ion{H}{2} regions. The lack of IM SFRs found with associated peak mass column densities $> 0.5$ g~cm$^{-2}$ supports the idea that there is a threshold in mass column density above which only massive star formation occurs.

\begin{acknowledgements}
This work was funded by the \emph{Spitzer Space Telescope} archival research grant 50623 and NASA ADP grant NNX10AD55G. This work is based [in part] on archival data obtained with the Spitzer Space Telescope, which is operated by the Jet Propulsion Laboratory, California Institute of Technology under a contract with NASA. Support for this work was provided by NASA. The authors thank Jonathan Tan, Charles Telesco, and Peter Barnes for
comments on an early version of the manuscript. H. A. K. thanks the Department of Astronomy at the University of Florida for
hospitality during a sabbatical semester over which this work was
completed. B. U. acknowledges support from a NASA Graduate Student Researchers Program fellowship, grant NNX06AI28H. This work makes use of molecular line data from the Boston University-FCRAO Galactic Ring Survey (GRS). The GRS is a joint project of Boston University and Five College Radio Astronomy Observatory, funded by the National Science Foundation under grants AST-9800334, AST-0098562, AST-0100793, AST-0228993, \& AST-0507657. This research has made use of the NASA/IPAC Infrared Science Archive, which is
operated by the Jet Propulsion Laboratory, California
Institute of Technology, under contract with the National
Aeronautics and Space Administration. The UKIDSS project is defined in \citet{ukidss}. We have used data from the DR3 data release, which is described in detail in \citet{dr3}. The BGPS project is supported by the National Science Foundation through NSF grant AST-0708403.
\end{acknowledgements}

\clearpage

\begin{deluxetable}{ccc}
\tablecolumns{3}
\tabletypesize{\footnotesize}
\tablewidth{0pc}
\tablecaption{Number of potential sources.\label{numbers}}
\tablehead{
\colhead{Region} & \colhead{Number of sources}
}
\startdata
All sky&
984\\
&
&
\\
Galactic plane&
360\\
($|b|<1 \degr$)&
&
\\
GLIMPSE I&
178\\
$(10 \degr <| \ell |<65 \degr$, $|b|<1 \degr)$&
&
\\
GRS&
70\\
$(18 \degr < \ell <55 \fdg 7$, $|b|<1 \degr)$&
&
\\
\enddata
\tablecomments{Number of \emph{IRAS} point sources for various regions using the color criteria from \citet{kerton02}: $-0.18 < \log(F_{\nu}(25)/F_{\nu}(12)) < 0.40$, $0.96 < \log(F_{\nu}(60)/F_{\nu}(25)) < 1.56$ and $0.29 < \log(F_{\nu}(100)/F_{\nu}(60)) < 0.69$.}
\end{deluxetable}

\clearpage

\begin{deluxetable}{crrc}
\tablecolumns{4}
\tabletypesize{\footnotesize}
\tablewidth{0pc}
\tablecaption{Filamentary and star-like objects.\label{rejects}}
\tablehead{
\colhead{\emph{IRAS} name} & \colhead{$\ell$} & \colhead{$b$} & \colhead{Class}\\
\colhead{} & \colhead{($\degr$)} & \colhead{($\degr$)} & \colhead{}
}
\startdata
18163-1603 & 15.03450 & $-0.35301$ & F \\
18169-1338 & 17.23340 & $0.68040$ & F \\
18227-1218 & 19.07937 & $0.06120$ & F \\
18257-1226 & 19.30243 & $-0.65138$ & S \\
18279-1039 & 21.12765 & $-0.28842$ & F \\
18317-0926 & 22.63746 & $-0.54983$ & S \\
18328-0900 & 23.14221 & $-0.58687$ & S \\
18318-0741 & 24.20073 & $0.25168$ & S \\
18416-0518 & 27.43590 & $-0.81472$ & S \\
18465-0018 & 32.44363 & $0.39727$ & S \\
18559+0308 & 36.58437 & $-0.09614$ & S \\
18571+0326 & 36.99180 & $-0.22897$ & F \\
18550+0358 & 37.22322 & $0.47883$ & F \\
18588+0350 & 37.54749 & $-0.40816$ & S \\
19040+0616 & 40.30611 & $-0.43082$ & F \\
19079+0919 & 43.45117 & $0.11756$ & F \\
19137+1013 & 44.92225 & $-0.71903$ & F \\
19150+1116 & 45.98566 & $-0.50465$ & S \\
19152+1209 & 46.80001 & $-0.12841$ & F \\
19273+1637 & 52.11558 & $-0.57587$ & S \\
19300+1852 & 54.38650 & $-0.03721$ & F \\
\enddata
\tablecomments{Filamentary and star-like objects. Listed are the \emph{IRAS} PSC names, the Galactic longitudes ($\ell$), Galactic latitudes ($b$), and the classifications. Filamentary objects are denoted with a {}``F'', and star-like objects with a {}``S''.}
\end{deluxetable}

\clearpage

\begin{deluxetable}{crrrccccccccc}
\tablecolumns{13}
\tabletypesize{\scriptsize}
\tablewidth{0pc}
\rotate
\tablecaption{IM SFRs - Properties of blobs and shells.\label{blobsshells}}
\tablehead{
\colhead{\emph{IRAS} name} & \colhead{$\ell$} & \colhead{$b$} & \colhead{$\theta$} & \colhead{$V_{\mathrm{LSR}}$} & \colhead{$ \Delta V$} & \colhead{$d_{\mathrm{near}}$} & \colhead{$d_{\mathrm{far}}$} & \colhead{Adopted} & \colhead{Flag} & \colhead{Diameter} & \colhead{$L_{\mathrm{IR}}$} & \colhead{$M_{\mathrm{LTE}}$}\\
\colhead{} & \colhead{($\degr$)} & \colhead{($\degr$)} & \colhead{($\arcmin$)} & \colhead{(km~s$^{-1}$)} & \colhead{(km~s$^{-1}$)} & \colhead{(kpc)} & \colhead{(kpc)} & \colhead{(kpc)} & \colhead{} & \colhead{(pc)} & \colhead{($10^3$ L$_{\sun}$)} & \colhead{(M$_{\sun}$)}
}
\startdata
18131-1606 & 14.63111 & $0.30075$ & 1.0 & $26.1$ & 8.2 & 2.9 & 13.6 & 2.9 & N & 0.8 & 10.7 & 500 \\
18180-1342 & 17.31025 & $0.40066$ & 2.0 & $19.7$ & 3.7 & 2.0 & 14.2 & 2.0 & N & 1.2 & 4.9 & 40 \\
18241-1320 & 18.32426 & $-0.73709$ & 0.6 & $43.3$ & 4.4 & 3.7 & 12.5 & 3.7 & N & 0.6 & 1.2 & 800 \\
18224-1228 & 18.88269 & $0.05205$ & 1.5 & $49.9$ & 9.3 & 4.0 & 12.1 & 4.0 & N & 1.7 & 36.5 & 6100 \\
18253-1210 & 19.50327 & $-0.44381$ & 1.0 & $63.1$ & 5.0 & 4.6 & 11.4 & 4.6 & N & 1.3 & 6.0 & 2800 \\
18322-0956 & 22.25764 & $-0.87999$ & 0.6 & $41.0$ & 2.6 & 3.2 & 12.5 & 3.2 & N & 0.6 & 4.6 & 200 \\
18308-0741 & 24.09699 & $0.45326$ & 1.8 & $94.7$ & 8.2 & 5.9 & 9.6 & \nodata & U & 3.1/5.0 & 27.7/73.3 & 4300/11500 \\
18370-0607 & 26.19714 & $-0.18673$ & 0.6 & $108.3$ & 4.5 & 7.6 & 7.6 & 7.6 & T & 1.3 & 15.0 & 1700 \\
18367-0529 & 26.71852 & $0.17412$ & 0.5 & \nodata & \nodata & \nodata & \nodata & \nodata & A & \nodata & \nodata & \nodata \\
18362-0517 & 26.83021 & $0.38216$ & 2.2 & \nodata & \nodata & \nodata & \nodata & \nodata & A & \nodata & \nodata & \nodata \\
18412-0440 & 27.96505 & $-0.44254$ & 0.8 & $45.6$ & 7.8 & 3.1 & 11.9 & 3.1 & N & 0.7 & 2.4 & 2300 \\
18411-0312 & 29.24142 & $0.25972$ & 0.5 & \nodata & \nodata & \nodata & \nodata & \nodata & A & \nodata & \nodata & \nodata \\
18433-0327 & 29.28703 & $-0.33671$ & 0.5 & \nodata & \nodata & \nodata & \nodata & \nodata & A & \nodata & \nodata & \nodata \\
18441-0118 & 31.27801 & $0.48597$ & 1.5 & $89.2$ & 14.4 & 5.8 & 8.8 & \nodata & U & 2.5/3.8 & 5.7/13.2 & 2100/4900 \\
18463-0052 & 31.91297 & $0.19629$ & 1.0 & $107.9$ & 7.3 & 7.2 & 7.2 & 7.2 & T & 2.1 & 29.0 & 1600 \\
18502-0018 & 32.86081 & $-0.41053$ & 0.4 & $48.0$ & 9.2 & 3.2 & 11.1 & 11.1 & F & 1.3 & 110.3 & 2600 \\
18504+0025 & 33.53779 & $-0.12907$ & 0.5 & \nodata & \nodata & \nodata & \nodata & \nodata & A & \nodata & \nodata & \nodata \\
18537+0145 & 35.09790 & $-0.24120$ & 0.5 & $51.6$ & 7.6 & 3.4 & 10.6 & \nodata & U & 0.5/1.5 & 4.4/42.9 & 100/900 \\
18527+0203 & 35.26396 & $0.12033$ & 1.0 & \nodata & \nodata & \nodata & \nodata & \nodata & A & \nodata & \nodata & \nodata \\
18564+0145 & 35.41394 & $-0.84999$ & 0.7 & $36.7$ & 6.2 & 2.4 & 11.4 & 2.4 & N & 0.5 & 0.7 & 1000 \\
19014+0451 & 38.75274 & $-0.52447$ & 0.5 & $66.2$ & 7.8 & 4.5 & 8.7 & \nodata & U & 0.7/1.3 & 4.9/18.5 & 200/800 \\
19012+0505-N & 38.92659 & $-0.37007$ & 1.8 & $39.8$ & 6.7 & 2.6 & 10.6 & 2.6 & N & 1.4 & 10.7 & 700 \\
19012+0505-S & 38.92659 & $-0.37007$ & 3.0 & $41.3$ & 6.7 & 2.7 & 10.5 & 2.7 & N & 2.4 & 9.0 & 1400 \\
19023+0545 & 39.64763 & $-0.29262$ & 0.6 & $25.8$ & 4.8 & 1.8 & 11.3 & 11.3 & F & 2.0 & 26.6 & 200 \\
19023+0601 & 39.87238 & $-0.17333$ & 1.3 & $57.3$ & 12.4 & 3.8 & 9.2 & 3.8 & N & 1.4 & 8.0 & 200 \\
19027+0656 & 40.74324 & $0.15801$ & 0.5 & $16.9$ & 8.8 & 1.3 & 11.6 & 11.6 & F & 1.7 & 17.1 & 3100 \\
19060+0657 & 41.14428 & $-0.56201$ & 2.2 & \nodata & \nodata & \nodata & \nodata & \nodata & A & \nodata & \nodata & \nodata \\
19072+0649 & 41.14722 & $-0.88475$ & 0.5 & $66.5$ & 6.0 & 4.9 & 7.9 & \nodata & U & 0.7/1.1 & 2.1/5.5 & 700/1700 \\
19049+0712 & 41.22959 & $-0.19367$ & 1.0 & $59.4$ & 6.3 & 4.1 & 8.7 & 4.1 & N & 1.2 & 10.2 & 400 \\
19062+0758 & 42.05222 & $-0.11977$ & 1.2 & $70.3$ & 6.9 & 5.6 & 7.1 & 5.6 & N & 2.0 & 3.8 & 1600 \\
19105+0852 & 43.35147 & $-0.66408$ & 0.6 & $58.0$ & 6.3 & 4.2 & 8.2 & \nodata & U & 0.7/1.4 & 3.3/12.5 & 500/1800 \\
19056+0947 & 43.59957 & $0.85377$ & 0.8 & \nodata & \nodata & \nodata & \nodata & \nodata & No CO & \nodata & \nodata & \nodata \\
19139+1045 & 45.40416 & $-0.51347$ & 1.1 & $60.9$ & 11.1 & 4.9 & 7.0 & 4.9 & N & 1.6 & 9.2 & 300 \\
19138+1055 & 45.53808 & $-0.39856$ & 0.7 & $57.7$ & 13.3 & 4.5 & 7.4 & \nodata & U & 0.9/1.5 & 3.0/8.0 & 1000/2600 \\
19157+1319 & 47.87449 & $0.31098$ & 0.4 & $63.5$ & 5.5 & 5.7 & 5.7 & 5.7 & T & 0.7 & 4.5 & 1400 \\
19207+1329 & 48.60178 & $-0.66833$ & 0.5 & $52.6$ & 11.8 & 4.4 & 6.9 & \nodata & U & 0.6/1.0 & 2.9/7.2 & 500/1200 \\
19207+1348 & 48.87743 & $-0.51817$ & 0.6 & $57.3$ & 5.0 & 5.6 & 5.6 & 5.6 & T & 1.0 & 33.4 & 4300 \\
19156+1441 & 49.07458 & $0.98102$ & 0.4 & $53.1$ & 4.7 & 4.5 & 6.6 & \nodata & U & 0.5/0.8 & 0.8/1.7 & 300/600 \\
19193+1443 & 49.53260 & $0.20918$ & 0.5 & $63.1$ & 5.5 & 5.5 & 5.5 & 5.5 & T & 0.8 & 4.2 & 600 \\
19233+1413 & 49.54016 & $-0.88136$ & 0.5 & $43.3$ & 8.0 & 3.3 & 7.7 & \nodata & U & 0.5/1.1 & 2.9/16.1 & 600/3200 \\
19205+1447 & 49.72280 & $-0.01595$ & 0.6 & $48.2$ & 10.4 & 3.9 & 7.1 & \nodata & U & 0.7/1.2 & 13.4/44.5 & 2000/6700 \\
19221+1456 & 50.03463 & $-0.28204$ & 1.8 & $61.8$ & 10.4 & 5.5 & 5.5 & 5.5 & T & 2.9 & 16.3 & 1300 \\
19214+1556 & 50.84320 & $0.33493$ & 0.7 & $40.5$ & 8.4 & 3.2 & 7.6 & \nodata & U & 0.7/1.5 & 2.7/15.3 & 1000/5400 \\
19255+1531 & 50.93431 & $-0.72378$ & 0.6 & $38.6$ & 5.1 & 3.0 & 7.7 & \nodata & U & 0.5/1.3 & 2.2/14.3 & 300/1800 \\
19256+1705 & 52.33394 & $0.00598$ & 0.5 & $12.0$ & 5.4 & 1.1 & 9.3 & 9.3 & F & 1.4 & 9.7 & 600 \\
19248+1730 & 52.60285 & $0.37402$ & 0.7 & $41.2$ & 10.1 & 3.5 & 6.8 & \nodata & U & 0.7/1.4 & 1.9/7.0 & 100/500 \\
19247+1829 & 53.44999 & $0.87096$ & 0.4 & $-2.6$ & 4.4 & 0.3 & 9.9 & 9.9 & F & 1.2 & 6.9 & 1300 \\
19260+1821 & 53.48407 & $0.52532$ & 0.4 & $49.9$ & 6.7 & 5.1 & 5.1 & 5.1 & T & 0.6 & 2.1 & 2000 \\
19266+1926 & 54.51342 & $0.92616$ & 0.6 & $32.4$ & 7.1 & 2.7 & 7.1 & \nodata & U & 0.5/1.2 & 2.5/17.5 & 400/2500 \\
19330+1956 & 55.66724 & $-0.13004$ & 0.5 & $31.6$ & 5.6 & 2.8 & 6.9 & 6.9 & F & 1.0 & 10.2 & 900 \\
\enddata
\tablecomments{Properties of the objects labeled as blobs or shells. $\theta$ is the angular diameter. $V_{\mathrm{LSR}}$ is the associated CO velocity from the GRS, and $ \Delta V$ is the width of the associated CO feature. $d_{\mathrm{near}}$ and $d_{\mathrm{far}}$ are the near and far distances. Adopted indicates the value for the distance used in subsequent analysis. The flags are: {}``N'' for near, {}``F'' for far, {}``T'' for tangent point, {}``U'' for unassigned, and {}``A'' for ambiguous CO association. The diameter is calculated using the adopted distance. $L_{\mathrm{IR}}$ is taken from Table~\ref{photometrytable}, and $M_{\mathrm{LTE}}$ from Table~\ref{moleculartable}.}
\end{deluxetable}

\begin{deluxetable}{ccrrrrrrrrrrrrcrr}
\tablecolumns{17}
\tabletypesize{\scriptsize}
\tablewidth{0pc}
\rotate
\tablecaption{IM SFRs - Photometry and luminosity.\label{photometrytable}}
\tablehead{
\colhead{\emph{IRAS} name} & \colhead{Aperture} & \colhead{F$_{3.6}$} & \colhead{$\sigma_{3.6}$} & \colhead{F$_{4.5}$} & \colhead{$\sigma_{4.5}$} & \colhead{F$_{5.8}$} & \colhead{$\sigma_{5.8}$} & \colhead{F$_{8.0}$} & \colhead{$\sigma_{8.0}$} & \colhead{F$_{24}$} & \colhead{$\sigma_{24}$} & \colhead{F$_{\nu}(60)$} & \colhead{$F_{\nu}(100)$} & \colhead{Flag} & \colhead{$L_{\mathrm{near}}$} & \colhead{$L_{\mathrm{far}}$} \\
\colhead{} & \colhead{($\arcsec$)} & \colhead{(Jy)} & \colhead{(Jy)} & \colhead{(Jy)} & \colhead{(Jy)} & \colhead{(Jy)} & \colhead{(Jy)} & \colhead{(Jy)} & \colhead{(Jy)} & \colhead{(Jy)} & \colhead{(Jy)} & \colhead{(Jy)} & \colhead{(Jy)} & \colhead{} & \colhead{($10^3$ L$_{\sun}$)} & \colhead{($10^3$ L$_{\sun}$)}
}
\startdata
18131-1606 & 54 & 2.28 & 0.23 & 1.47 & 0.15 & 6.60 & 0.68 & 14.74 & 1.51 & 42.39 & 5.67 & 305.0 & 799.3 & N & 10.7 & 234.3 \\
18180-1342 & 76 & 3.78 & 0.54 & 3.05 & 0.31 & 11.68 & 1.20 & 29.40 & 3.01 & 29.98 & 4.01 & 216.6 & 830.6 & N & 4.9 & 248.5 \\
18241-1320 & 26 & 0.19 & 0.02 & 0.21 & 0.02 & 0.55 & 0.06 & 1.11 & 0.11 & 4.05 & 0.37 & 17.2 & 62.8 & N & 1.2 & 14.3 \\
18224-1228 & 65 & 2.43 & 0.25 & 2.29 & 0.24 & 12.63 & 1.29 & 29.43 & 3.02 & 67.49 & 6.92 & 542.6 & 1506.0 & N & 36.5 & 334.2 \\
18253-1210 & 36 & 0.35 & 0.04 & 0.29 & 0.03 & 1.68 & 0.17 & 4.17 & 0.43 & 5.93 & 0.61 & 62.4 & 210.2 & N & 6.0 & 36.8 \\
18322-0956 & 52 & 0.87 & 0.09 & 0.63 & 0.06 & 2.72 & 0.28 & 7.22 & 0.74 & 19.52 & 2.00 & 90.2 & 268.7 & N & 4.6 & 69.5 \\
18308-0741 & 72 & 2.13 & 0.22 & 1.57 & 0.16 & 4.58 & 0.47 & 9.19 & 0.94 & 27.19 & 2.79 & 166.6 & 560.2 & U & 27.7 & 73.3 \\
18370-0607 & 43 & 0.70 & 0.07 & 0.49 & 0.05 & 1.02 & 0.10 & 1.91 & 0.20 & 5.55 & 0.57 & 72.4 & 170.3 & T & 15.0 & 15.0 \\
18367-0529 & 44 & 0.45 & 0.05 & 0.36 & 0.04 & 1.53 & 0.16 & 3.67 & 0.38 & 12.02 & 1.23 & 147.2 & 445.1 & A & \nodata & \nodata \\
18362-0517 & 91 & 3.75 & 0.38 & 2.73 & 0.28 & 5.72 & 0.59 & 11.76 & 1.21 & 18.13 & 1.86 & 130.5 & 399.1 & A & \nodata & \nodata \\
18412-0440 & 35 & 0.94 & 0.10 & 0.71 & 0.07 & 1.82 & 0.19 & 3.50 & 0.36 & 3.52 & 0.36 & 61.7 & 173.8 & N & 2.4 & 35.9 \\
18411-0312 & 42 & 0.45 & 0.05 & 0.40 & 0.04 & 1.79 & 0.18 & 4.26 & 0.44 & 16.43 & 1.68 & 122.0 & 382.8 & A & \nodata & \nodata \\
18433-0327 & 19 & 0.11 & 0.01 & 0.11 & 0.01 & 0.77 & 0.08 & 1.75 & 0.18 & 4.11 & 0.42 & 75.4 & 215.2 & A & \nodata & \nodata \\
18441-0118 & 64 & 1.08 & 0.11 & 0.79 & 0.08 & 0.93 & 0.10 & 0.97 & 0.10 & 2.79 & 0.29 & 32.0 & 156.5 & U & 5.7 & 13.2 \\
18463-0052 & 101 & 6.62 & 0.68 & 6.01 & 0.62 & 16.26 & 1.67 & 24.46 & 2.51 & 81.06 & 8.31 & 248.9 & 531.5 & T & 29.0 & 29.0 \\
18502-0018 & 146 & 9.33 & 0.96 & 6.54 & 0.67 & 10.19 & 1.04 & 18.64 & 1.91 & 42.04 & 4.31 & 118.3 & 497.6 & F & 9.2 & 110.3 \\
18504+0025 & 34 & 0.28 & 0.03 & 0.21 & 0.02 & 1.27 & 0.13 & 3.06 & 0.31 & 2.22 & 0.23 & 98.1 & 304.9 & A & \nodata & \nodata \\
18537+0145 & 30 & 0.50 & 0.05 & 0.28 & 0.03 & 1.28 & 0.13 & 2.92 & 0.30 & 11.03 & 1.13 & 101.9 & 253.1 & U & 4.4 & 42.9 \\
18527+0203 & 40 & 0.25 & 0.03 & 0.28 & 0.03 & 0.96 & 0.10 & 2.09 & 0.21 & 5.60 & 0.57 & 57.4 & 257.2 & A & \nodata & \nodata \\
18564+0145 & 24 & 0.23 & 0.02 & 0.20 & 0.02 & 0.85 & 0.09 & 1.99 & 0.20 & 1.54 & 0.16 & 25.1 & 82.7 & N & 0.7 & 14.8 \\
19014+0451 & 31 & 0.81 & 0.08 & 0.36 & 0.04 & 1.23 & 0.13 & 2.63 & 0.27 & 3.50 & 0.36 & 61.3 & 178.1 & U & 4.9 & 18.5 \\
19012+0505-N & 113 & 4.95 & 0.51 & 3.89 & 0.40 & 20.49 & 2.10 & 49.67 & 5.09 & 93.25 & 9.56 & 497.2 & 1156.0 & N & 10.7 & 177.8 \\
19012+0505-S & 122 & 4.79 & 0.49 & 3.59 & 0.37 & 20.36 & 2.09 & 37.42 & 3.83 & 25.64 & 2.63 & 497.2 & 1156.0 & N & 9.0 & 135.6 \\
19023+0545 & 18 & 0.05 & 0.01 & 0.05 & 0.01 & 0.42 & 0.04 & 0.98 & 0.10 & 3.61 & 0.37 & 46.0 & 197.0 & F & 0.7 & 26.6 \\
19023+0601 & 103 & 1.99 & 0.20 & 1.41 & 0.14 & 1.46 & 0.15 & 4.88 & 0.50 & 3.57 & 0.37 & 128.2 & 488.1 & N & 8.0 & 47.0 \\
19027+0656 & 32 & 0.19 & 0.02 & 0.12 & 0.01 & 0.64 & 0.07 & 0.62 & 0.07 & 1.87 & 0.19 & 34.2 & 98.3 & F & 0.2 & 17.1 \\
19060+0657 & 48 & 0.38 & 0.04 & 0.30 & 0.03 & 1.46 & 0.15 & 4.88 & 0.50 & 3.57 & 0.37 & 59.3 & 150.4 & A & \nodata & \nodata \\
19072+0649 & 40 & 0.19 & 0.02 & 0.14 & 0.01 & 0.62 & 0.06 & 0.94 & 0.10 & 0.88 & 0.09 & 25.5 & 56.4 & U & 2.1 & 5.5 \\
19049+0712 & 162 & 4.95 & 0.51 & 3.97 & 0.41 & 10.09 & 1.03 & 10.07 & 1.03 & 28.95 & 2.97 & 110.0 & 271.4 & N & 10.2 & 45.9 \\
19062+0758 & 35 & 0.91 & 0.09 & 0.59 & 0.06 & 1.12 & 0.12 & 2.04 & 0.21 & 2.75 & 0.28 & 22.9 & 80.6 & N & 3.8 & 6.1 \\
19105+0852 & 25 & 0.14 & 0.02 & 0.16 & 0.02 & 0.81 & 0.08 & 1.72 & 0.18 & 2.29 & 0.24 & 42.1 & 162.9 & U & 3.3 & 12.5 \\
19056+0947 & 22 & 0.25 & 0.03 & 0.15 & 0.02 & 0.34 & 0.04 & 0.71 & 0.07 & 1.83 & 0.19 & 23.9 & 54.8 & No CO & \nodata & \nodata \\
19139+1045 & 61 & 0.89 & 0.09 & 0.62 & 0.06 & 2.93 & 0.30 & 6.59 & 0.68 & 7.81 & 0.80 & 81.5 & 270.5 & N & 9.2 & 18.8 \\
19138+1055 & 19 & 0.07 & 0.01 & 0.06 & 0.01 & 0.56 & 0.06 & 1.38 & 0.14 & 2.59 & 0.27 & 38.6 & 106.6 & U & 3.0 & 8.0 \\
19157+1319 & 38 & 0.24 & 0.02 & 0.19 & 0.02 & 0.73 & 0.07 & 1.69 & 0.17 & 6.80 & 0.70 & 32.1 & 76.6 & T & 4.5 & 4.5 \\
19207+1329 & 19 & 0.08 & 0.01 & 0.07 & 0.01 & 0.46 & 0.05 & 1.20 & 0.12 & 2.78 & 0.29 & 35.2 & 127.7 & U & 2.9 & 7.2 \\
19207+1348 & 68 & 1.11 & 0.11 & 0.89 & 0.09 & 5.17 & 0.53 & 13.69 & 1.40 & 36.48 & 3.74 & 214.9 & 789.6 & T & 33.4 & 33.4 \\
19156+1441 & 14 & 0.05 & 0.01 & 0.04 & 0.00 & 0.22 & 0.02 & 0.56 & 0.06 & 0.77 & 0.08 & 7.8 & 33.8 & U & 0.8 & 1.7 \\
19193+1443 & 19 & 0.07 & 0.01 & 0.06 & 0.01 & 0.44 & 0.05 & 0.86 & 0.09 & 1.19 & 0.12 & 35.0 & 123.3 & T & 4.2 & 4.2 \\
19233+1413 & 38 & 0.26 & 0.03 & 0.46 & 0.05 & 1.51 & 0.16 & 2.85 & 0.29 & 7.29 & 0.75 & 68.1 & 179.7 & U & 2.9 & 16.1 \\
19205+1447 & 54 & 0.88 & 0.09 & 0.69 & 0.07 & 4.24 & 0.44 & 10.21 & 1.05 & 35.26 & 3.61 & 216.3 & 523.3 & U & 13.4 & 44.5 \\
19221+1456 & 73 & 1.61 & 0.17 & 1.26 & 0.13 & 4.39 & 0.45 & 10.44 & 1.07 & 10.45 & 1.07 & 109.9 & 376.5 & T & 16.3 & 16.3 \\
19214+1556 & 30 & 0.09 & 0.01 & 0.08 & 0.01 & 0.54 & 0.06 & 1.37 & 0.14 & 2.47 & 0.25 & 65.9 & 242.5 & U & 2.7 & 15.3 \\
19255+1531 & 41 & 0.30 & 0.03 & 0.26 & 0.03 & 1.44 & 0.15 & 3.64 & 0.37 & 3.78 & 0.39 & 63.3 & 156.6 & U & 2.2 & 14.3 \\
19256+1705 & 22 & 0.07 & 0.01 & 0.07 & 0.01 & 0.36 & 0.04 & 0.90 & 0.09 & 2.13 & 0.22 & 28.4 & 88.1 & F & 0.1 & 9.7 \\
19248+1730 & 32 & 0.24 & 0.02 & 0.17 & 0.02 & 0.81 & 0.08 & 2.02 & 0.21 & 4.99 & 0.51 & 31.5 & 108.6 & U & 1.9 & 7.0 \\
19247+1829 & 23 & 0.08 & 0.01 & 0.07 & 0.01 & 0.43 & 0.04 & 1.10 & 0.11 & 0.86 & 0.09 & 17.1 & 51.0 & F & 0.01 & 6.9 \\
19260+1821 & 23 & 0.09 & 0.01 & 0.09 & 0.01 & 0.48 & 0.05 & 1.16 & 0.12 & 2.20 & 0.23 & 17.7 & 54.2 & T & 2.1 & 2.1 \\
19266+1926 & 60 & 0.42 & 0.04 & 0.30 & 0.03 & 2.09 & 0.21 & 5.64 & 0.58 & 10.63 & 1.09 & 75.1 & 230.1 & U & 2.5 & 17.5 \\
19330+1956 & 50 & 0.86 & 0.09 & 0.89 & 0.09 & 2.45 & 0.25 & 5.35 & 0.55 & 2.60 & 0.27 & 53.2 & 104.2 & F & 1.7 & 10.2 \\
\enddata
\tablecomments{Photometry and luminosity of the objects listed as blobs or shells. Aperture refers to the aperture radius in arcseconds. Measured \emph{Spitzer} flux densities are listed with $1 \sigma$ total uncertainty. For 60 $\micron$ and 100 $\micron$, the \emph{IRAS} PSC value is listed. The flags are the same as in Table~\ref{blobsshells}. The near and far luminosities are calculated using the near and far distances in Table~\ref{blobsshells}.}
\end{deluxetable}

\begin{deluxetable}{ccccccrrcccc}
\tablecolumns{12}
\tabletypesize{\scriptsize}
\tablewidth{0pc}
\rotate
\tablecaption{IM SFRs - Molecular material from the GRS.\label{moleculartable}}
\tablehead{
\colhead{\emph{IRAS} name} & \colhead{$I_{\mathrm{peak}}$} & \colhead{$N(\mathrm{H_2})_{\mathrm{peak}}$} & \colhead{$N(M)_{\mathrm{peak}}$} & \colhead{$\Omega$} & \colhead{Flag} & \colhead{$M_{\mathrm{LTE}}^{\mathrm{near}}$} & \colhead{$M_{\mathrm{LTE}}^{\mathrm{far}}$} & \colhead{Cloud} & \colhead{Clump} & \colhead{$M_{\mathrm{Cloud}}$} & \colhead{Note} \\
\colhead{} & \colhead{(K~km~s$^{-1}$)} & \colhead{($10^{21}$ cm$^{-2}$)} & \colhead{(g~cm$^{-2}$)} & \colhead{($10^{-7}$ sr)} & \colhead{} & \colhead{(M$_{\sun}$)} & \colhead{(M$_{\sun}$)} & \colhead{GRSMC} & \colhead{} & \colhead{(M$_{\sun}$)} & \colhead{}
}
\startdata
18131-1606 & 28.5 & 11.7 & 0.05 & 11.2 & N & 500 & 10100 & \nodata & \nodata & \nodata &  \\
18180-1342 & 9.7 & 4.0 & 0.02 & 6.1 & N & 40 & 1800 & \nodata & \nodata & \nodata &  \\
18241-1320 & 14.9 & 6.1 & 0.03 & 19.5 & N & 800 & 8700 & G018.39-00.41 & c7 & 46500 & N \\
18224-1228 & 41.3 & 17.0 & 0.08 & 47.8 & N & 6100 & 56200 & G018.89+00.04 & c1 & 238500 & F (!) \\
18253-1210 & 16.5 & 6.8 & 0.03 & 32.1 & N & 2800 & 17000 & G019.54-00.46 & c1 & 61600 & F (!) \\
18322-0956 & 10.5 & 4.3 & 0.02 & 20.4 & N & 200 & 3800 & none & none & \nodata &  \\
18308-0741 & 25.3 & 10.4 & 0.05 & 19.4 & U & 4300 & 11500 & G024.09+00.44 & c1 & 225100 & N \\
18370-0607 & 14.7 & 6.0 & 0.03 & 10.8 & T & 1700 & 1700 & G025.94-00.11 & c3 & 156300 & T \\
18412-0440 & 29.1 & 11.9 & 0.05 & 41.2 & N & 2300 & 34000 & G027.99-0.46 & c2 & 17000 & N \\
18441-0118 & 15.2 & 6.2 & 0.03 & 20.0 & U & 2100 & 4900 & none & none & \nodata &  \\
18463-0052 & 17.0 & 7.0 & 0.03 & 8.9 & T & 1600 & 1600 & G031.79+00.14 & c1 & 56300 & T \\
18502-0018 & 21.2 & 8.7 & 0.04 & 4.8 & F & 200 & 2600 & G032.89-00.46 & c3 & 54500 & F \\
18537+0145 & 9.6 & 3.9 & 0.02 & 3.7 & U & 100 & 900 & none & none & \nodata &  \\
18564+0145 & 22.0 & 9.0 & 0.04 & 63.1 & N & 1000 & 21600 & G035.59-00.91 & c1 & 3600 & N \\
19014+0451 & 19.2 & 7.9 & 0.04 & 3.8 & U & 200 & 800 & G038.79-00.51 & c1 & 109000 & F \\
19012+0505-N & 41.2 & 16.9 & 0.08 & 16.2 & N & 700 & 12200 & G038.94-00.46 & c2 & 308200 & F (!) \\
19012+0505-S & 35.8 & 14.7 & 0.07 & 25.8 & N & 1400 & 21200 & G038.89-00.26 & c1 & 1800 & N \\
19023+0545 & 3.4 & 1.4 & 0.01 & 2.0 & F & 5 & 200 & none & none & \nodata &  \\
19023+0601 & 17.3 & 7.1 & 0.03 & 4.3 & N & 200 & 900 & G039.89-00.21 & c6 & \nodata & \nodata \\
19027+0656 & 12.9 & 5.3 & 0.02 & 11.3 & F & 40 & 3100 & none & none & \nodata &  \\
19072+0649 & 11.5 & 4.7 & 0.02 & 9.7 & U & 700 & 1700 & G041.14-00.86 & c2 & \nodata & \nodata \\
19049+0712 & 26.6 & 10.9 & 0.05 & 6.2 & N & 400 & 2000 & G041.19-00.21 & c4 & 218900 & F (!) \\
19062+0758 & 16.2 & 6.6 & 0.03 & 16.4 & N & 1600 & 2600 & G042.14-00.61 & c2 & 211700 & N \\
19105+0852 & 16.2 & 6.6 & 0.03 & 7.8 & U & 500 & 1800 & G043.39-00.66 & c2 & 1800 & F \\
19139+1045 & 14.0 & 5.7 & 0.03 & 6.1 & N & 300 & 600 & none & none & \nodata &  \\
19138+1055 & 23.3 & 9.5 & 0.04 & 12.2 & U & 1000 & 2600 & none & none & \nodata &  \\
19157+1319 & 16.1 & 6.6 & 0.03 & 16.9 & T & 1400 & 1400 & none & none & \nodata &  \\
19207+1329 & 17.2 & 7.1 & 0.03 & 7.0 & U & 500 & 1200 & G048.84-00.51 & c4 & 23200 & F \\
19207+1348 & 31.7 & 13.0 & 0.06 & 19.1 & T & 4300 & 4300 & G049.04-00.46 & c3 & 22300 & T \\
19156+1441 & 8.7 & 3.6 & 0.02 & 9.2 & U & 300 & 600 & none & none & \nodata &  \\
19193+1443 & 7.6 & 3.1 & 0.01 & 13.1 & T & 600 & 600 & G049.44-00.06 & none & 162600 & T \\
19233+1413 & 23.0 & 9.4 & 0.04 & 15.6 & U & 600 & 3200 & G049.54-00.91 & c1 & 37500 & F \\
19205+1447 & 32.4 & 13.3 & 0.06 & 22.0 & U & 2000 & 6700 & G049.74-00.01 & c1 & 5400 & N \\
19221+1456 & 14.4 & 5.9 & 0.03 & 16.0 & T & 1300 & 1300 & G050.44-00.41 & c1 & 12500 & T \\
19214+1556 & 25.5 & 10.4 & 0.05 & 18.7 & U & 1000 & 5400 & G050.84+00.24 & c2 & 12500 & N \\
19255+1531 & 11.9 & 4.9 & 0.02 & 19.9 & U & 300 & 1800 & G050.94-00.61 & c2 & 6300 & N \\
19256+1705 & 7.7 & 3.2 & 0.01 & 4.4 & F & 10 & 600 & none & none & \nodata &  \\
19248+1730 & 6.1 & 2.5 & 0.01 & 8.2 & U & 100 & 500 & none & none & \nodata &  \\
19247+1829 & 7.1 & 2.9 & 0.01 & 10.9 & F & 1 & 1300 & none & none & \nodata &  \\
19260+1821 & 16.5 & 6.7 & 0.03 & 36.1 & T & 2000 & 2000 & G053.49+00.49 & c1 & 1800 & T \\
19266+1926 & 14.6 & 6.0 & 0.03 & 17.4 & U & 400 & 2500 & G054.64+00.79 & c2 & 50900 & F \\
19330+1956 & 10.6 & 4.4 & 0.02 & 7.5 & F & 100 & 900 & G055.34+00.19 & c2 & 15200 & F \\
\enddata
\tablecomments{Molecular material associated with the objects listed as blobs or shells. Flags are taken from Table~\ref{blobsshells}. Cloud masses are calculated using $L(^{13}$CO) from \citet{roman-duval} and conversion factors from \citet{simon01} and references therein. The last column tells what distance \citet{roman-duval} put the cloud, (!) indicates where that distance determination differs from the standard ruler method used in this paper.}
\end{deluxetable}

\clearpage

\begin{deluxetable}{ccccccccc}
\tablecolumns{9}
\tabletypesize{\scriptsize}
\tablewidth{0pc}
\rotate
\tablecaption{IM SFRs - Molecular material from the BGPS.\label{bgpstable}}
\tablehead{
\colhead{\emph{IRAS} name} & \colhead{Source} & \colhead{$F_{\mathrm{peak}}$} & \colhead{$N(\mathrm{H_2})_{\mathrm{peak}}$} & \colhead{$N(M)_{\mathrm{peak}}$} & \colhead{$S$} & \colhead{Flag} & \colhead{$M_{\mathrm{BGPS}}^{\mathrm{near}}$} & \colhead{$M_{\mathrm{BGPS}}^{\mathrm{far}}$} \\
\colhead{} & \colhead{BGPS} & \colhead{(Jy~beam$^{-1}$)} & \colhead{($10^{21}$ cm$^{-2}$)} & \colhead{(g~cm$^{-2}$)} & \colhead{(Jy)} & \colhead{} & \colhead{(M$_{\sun}$)} & \colhead{(M$_{\sun}$)}
}
\startdata
18131-1606 & G014.634+00.308 & 0.92 & 62.4 & 0.29 & 2.3 & N & 800 & 16500 \\
18180-1342 & none & \nodata & \nodata & \nodata & \nodata & N & \nodata & \nodata \\
18241-1320 & \nodata & \nodata & \nodata & \nodata & \nodata & N & \nodata & \nodata \\
18224-1228 & G018.890+00.045 & 0.47 & 31.6 & 0.15 & 2.9 & N & 1800 & 16100 \\
18253-1210 & G019.542-00.457 & 0.36 & 24.6 & 0.12 & 1.7 & N & 1300 & 8200 \\
18322-0956 & \nodata & \nodata & \nodata & \nodata & \nodata & N & \nodata & \nodata \\
18308-0741 & G024.100+00.456 & 0.28 & 18.9 & 0.09 & 0.6 & U & 800 & 2200 \\
18370-0607 & G026.190-00.179 & 0.31 & 21.1 & 0.10 & 0.6 & T & 1300 & 1300 \\
18367-0529 & G026.722+00.173 & 0.24 & 16.0 & 0.07 & 0.8 & A & \nodata & \nodata \\
18362-0517 & G026.827+00.391 & 0.16 & 11.0 & 0.05 & 0.2 & A & \nodata & \nodata \\
18412-0440 & G027.972-00.422 & 0.52 & 35.5 & 0.17 & 1.6 & N & 600 & 8800 \\
18411-0312 & G029.241+00.251 & 0.26 & 17.7 & 0.08 & 0.6 & A & \nodata & \nodata \\
18433-0327 & G029.285-00.333 & 0.43 & 29.0 & 0.14 & 2.2 & A & \nodata & \nodata \\
18441-0118 & G031.283+00.495 & 0.20 & 13.8 & 0.06 & 0.5 & U & 700 & 1500 \\
18463-0052 & G031.911+00.189 & 0.17 & 11.6 & 0.05 & 0.7 & T & 1400 & 1400 \\
18502-0018 & G032.861-00.411 & 0.33 & 22.6 & 0.11 & 1.4 & F & 600 & 6700 \\
18504+0025 & G033.544-00.133 & 0.16 & 10.7 & 0.05 & 0.4 & A & \nodata & \nodata \\
18537+0145 & G035.102-00.246 & 0.23 & 15.7 & 0.07 & 0.3 & U & 100 & 1300 \\
18527+0203 & G035.268+00.118 & 0.22 & 15.2 & 0.07 & 1.0 & A & \nodata & \nodata \\
18564+0145 & \nodata & \nodata & \nodata & \nodata & \nodata & N & \nodata & \nodata \\
19014+0451 & G038.758-00.530 & 0.22 & 14.9 & 0.07 & 0.4 & U & 300 & 1100 \\
19012+0505-N & G038.920-00.352 & 1.53 & 103.7 & 0.49 & 6.4 & N & 1700 & 27600 \\
19012+0505-S & G038.959-00.468 & 0.73 & 49.2 & 0.23 & 2.8 & N & 800 & 11700 \\
19023+0545 & none & \nodata & \nodata & \nodata & \nodata & F & \nodata & \nodata \\
19023+0601 & G039.878-00.169 & 0.18 & 12.2 & 0.06 & 0.3 & N & 200 & 1100 \\
19027+0656 & G040.741+00.161 & 0.19 & 12.6 & 0.06 & 0.2 & F & 10 & 1000 \\
19060+0657 & none & \nodata & \nodata & \nodata & \nodata & A & \nodata & \nodata \\
19072+0649 & \nodata & \nodata & \nodata & \nodata & \nodata & U & \nodata & \nodata \\
19049+0712 & G041.228-00.205 & 0.30 & 20.0 & 0.09 & 1.0 & N & 600 & 2800 \\
19062+0758 & G042.051-00.123 & 0.21 & 14.5 & 0.07 & 0.4 & N & 500 & 800 \\
19105+0852 & \nodata & \nodata & \nodata & \nodata & \nodata & U & \nodata & \nodata \\
19056+0947 & \nodata & \nodata & \nodata & \nodata & \nodata & No CO & \nodata & \nodata \\
19139+1045 & none & \nodata & \nodata & \nodata & \nodata & N & \nodata & \nodata \\
19138+1055 & G045.524-00.363 & 0.16 & 11.1 & 0.05 & 0.4 & U & 300 & 700 \\
19157+1319 & G047.871+00.313 & 0.32 & 21.5 & 0.10 & 0.8 & T & 1000 & 1000 \\
19207+1329 & \nodata & \nodata & \nodata & \nodata & \nodata & U & \nodata & \nodata \\
19207+1348 & G048.872-00.512 & 0.28 & 18.9 & 0.09 & 0.6 & T & 700 & 700 \\
19156+1441 & \nodata & \nodata & \nodata & \nodata & \nodata & U & \nodata & \nodata \\
19193+1443 & none & \nodata & \nodata & \nodata & \nodata & T & \nodata & \nodata \\
19233+1413 & \nodata & \nodata & \nodata & \nodata & \nodata & U & \nodata & \nodata \\
19205+1447 & G049.722-00.012 & 0.66 & 44.7 & 0.21 & 1.5 & U & 900 & 3000 \\
19221+1456 & none & \nodata & \nodata & \nodata & \nodata & T & \nodata & \nodata \\
19214+1556 & G050.846+00.333 & 0.33 & 22.4 & 0.10 & 1.3 & U & 500 & 2900 \\
19255+1531 & \nodata & \nodata & \nodata & \nodata & \nodata & U & \nodata & \nodata \\
19256+1705 & none & \nodata & \nodata & \nodata & \nodata & F & \nodata & \nodata \\
19248+1730 & none & \nodata & \nodata & \nodata & \nodata & U & \nodata & \nodata \\
19247+1829 & \nodata & \nodata & \nodata & \nodata & \nodata & F & \nodata & \nodata \\
19260+1821 & none & \nodata & \nodata & \nodata & \nodata & T & \nodata & \nodata \\
19266+1926 & \nodata & \nodata & \nodata & \nodata & \nodata & U & \nodata & \nodata \\
19330+1956 & none & \nodata & \nodata & \nodata & \nodata & F & \nodata & \nodata \\
\enddata
\tablecomments{BGPS sources associated with the objects listed as blobs or shells. $S$ is the integrated flux density from the BGPS catalog \citep{bgpscatalog}. Flags are taken from Table~\ref{blobsshells}.}
\end{deluxetable}

\clearpage

\clearpage

\end{document}